\documentclass[useAMS,usenatbib,usegraphicx]{mn2e} 
\usepackage{amssymb} 
\usepackage{times} 
\usepackage{aas_macros} 

\pdfoutput=1 
\usepackage[usenames]{color}

\newcommand{\dsct}{$\delta$~Sct }
\newcommand{\gdor}{$\gamma$~Dor }
\newcommand{\Kepler}{\textit{Kepler} }
\newcommand{\bcep}{$\beta$~Cep }

\title[Amplitude modulation in \Kepler \dsct stars]{Amplitude modulation in \dsct stars: statistics from an ensemble study of \Kepler targets}

\author[D. M. Bowman et al.]{Dominic M. Bowman,$^{1}$\thanks{E-mail: dmbowman@uclan.ac.uk (DMB)}
Donald W. Kurtz,$^{1}$
Michel Breger,$^{2}$
Simon J. Murphy,$^{3,4}$
\newauthor{and Daniel L. Holdsworth$^{1}$}
\\
$^{1}$Jeremiah Horrocks Institute, University of Central Lancashire, Preston PR1 2HE, UK \\
$^{2}$Department of Astronomy, University of Texas, Austin, TX~78712, USA \\
$^{3}$Sydney Institute for Astronomy (SIfA), School of Physics, The University of Sydney, NSW 2006, Australia \\
$^{4}$Stellar Astrophysics Centre, Department of Physics and Astronomy, Aarhus University, Ny Munkegade 120, DK-8000 Aarhus C, Denmark \\
}
 
\date{Accepted 2016 May 11. Received 2016 May 11; in original form 2016 April 8}

\pubyear{2016}

\begin{document}
\label{firstpage}
\pagerange{\pageref{firstpage}--\pageref{lastpage}}
\maketitle

\begin{abstract}
We present the results of a search for amplitude modulation of pulsation mode frequencies in {983} \dsct stars, which have effective temperatures between 6400 $\leq T_{\rm eff} \leq$ 10\,000~K in the \Kepler Input Catalogue and were continuously observed by the \Kepler Space Telescope for 4~yr. We demonstrate the diversity in pulsational behaviour observed, in particular nonlinearity, which is predicted for \dsct stars. We analyse and discuss examples of \dsct stars with constant amplitudes and phases; those that exhibit amplitude modulation caused by beating of close-frequency pulsation modes; those that exhibit {pure} amplitude modulation (with no associated phase variation); those that exhibit phase modulation caused by binarity; and those that exhibit amplitude modulation caused by {nonlinearity}. Using models and examples of individual stars, we demonstrate that observations of the changes in amplitude and phase of pulsation modes can be used to distinguish among the different scenarios. We find that {603} \dsct stars ({61.3} per~cent) exhibit at least one pulsation mode that varies significantly in amplitude over 4~yr. Conversely, many \dsct stars have constant pulsation amplitudes so short-length observations can be used to determine precise frequencies, amplitudes and phases for the most coherent and periodic \dsct stars. It is shown that amplitude modulation is not restricted to a small region on the HR~diagram, therefore not necessarily dependent on stellar parameters such as $T_{\rm eff}$ or $\log\,g$. Our catalogue of {983} \dsct stars will be useful for comparisons to similar stars observed by K2 and TESS, because the length of the 4-yr \Kepler data set will not be surpassed for some time.
\end{abstract}



\section{Introduction} 
\label{section: Introduction}

Almost a century ago, Sir Arthur Eddington proposed that a star could act as a heat engine and become unstable to pulsation \citep{Eddington1917b, Eddington1926}. For mono-periodic radial pulsators such as classical Cepheid variables and RR~Lyrae stars (e.g. RRab stars using the classification of \citealt{Bailey1902}), a piston-like driving mechanism in which layers of gas in the stellar atmosphere periodically expand and contract about an equilibrium point is relatively simple to envisage. Conversely, multi-periodic pulsators with many excited pulsation modes can be more complicated. The amplitude spectra of multi-periodic pulsators are often forest-like with a greater likelihood of mode interaction among the pulsation mode frequencies (e.g. \citealt{Chapellier2012}). The reward of studying a multi-periodic pulsator is a greater insight into a star's interior, because different pulsation mode frequencies probe different depths within a star (e.g. \citealt{Kurtz2014}). The process of identifying and modelling pulsation mode frequencies is called asteroseismology, with an in-depth review given by \citet{ASTERO_BOOK}.

	\subsection{Delta Sct stars}
	\label{subsection: DSct stars}
	
	The multi-periodic pulsators known as delta Scuti ($\delta$~Sct) stars are the most common group of variable A and F stars, and are found at the intersection of the classical instability strip and main-sequence on the Hertzsprung--Russell (HR) diagram. On the main-sequence, \dsct stars typically range from A2 to F2 in spectral type \citep{Rod2001} and within the effective temperature range of $6300 \leq T_{\rm eff} \leq 8600$~K \citep{Uytterhoeven2011}. The \dsct stars can be considered intermediate-mass stars, as they lie in a transition-region from radiative cores and thick convective envelopes in low-mass stars ($M \lesssim 1\,{\rm M}_{\rm \odot}$), to large convective cores and thin convective envelopes in high-mass stars ($M \gtrsim 2\,{\rm M}_{\rm \odot}$). Concurrently, they also represent a transition from the high-amplitude radial pulsators, such as Cepheid variables, and the non-radial multi-periodic pulsators within the classical instability strip \citep{Breger2000a}. 
		
	Pulsations in \dsct stars are excited by the $\kappa$-mechanism operating in the \mbox{He\,{\sc ii}} ionisation zone at $T \sim 50\,000$~K \citep{Cox1963, Chevalier1971a} producing low-order pressure (p) modes. Typical pulsation periods observed in \dsct stars are of order a few hours \citep{Breger2000a}, but can be as short as 15~min \citep{Holdsworth2014a}. Hotter \dsct stars generally have shorter pulsation periods (i.e. higher pulsation mode frequencies) than cooler \dsct stars. Pulsational instability is a balance between driving and damping within a star; for example, the depth of the convective envelope is predicted to be large enough at the red edge of the classical instability strip to damp the \dsct pulsations \citep{C-D2000, Houdek2000, Griga2010a}. Thorough reviews of \dsct stars are provided by \citet{Breger2000a}, \citet{ASTERO_BOOK} and \citet{Murphy_PhD}. 
	
	The evolutionary state of a star, specifically if it is near the zero-age main-sequence (ZAMS) or terminal-age main-sequence (TAMS), can influence its pulsational characteristics. The \dsct stars are interesting as they lie on and beyond the main-sequence and so experience large changes in their interiors in a relatively short period of time after the hydrogen in their cores is exhausted. Evolved stars often are observed to show a form of mode interaction called mixed modes, which are pulsation modes that exhibit gravity (g) mode characteristics near the core and p-mode characteristics near the surface \citep{Osaki1975}. Mixed modes are often observed in evolved stars because the large density gradient outside the core couples the g- and p-mode pulsation cavities (e.g. \citealt{Lenz2010}). For a given $T_{\rm eff}$, a more-evolved \dsct star near the TAMS will generally have lower pulsation mode frequencies than its ZAMS counterpart. 
	
	Another strong influence on stellar pulsations is rotation. The Kraft Break \citep{Kraft1967d} divides the main-sequence into slowly rotating low-mass stars and fast-rotating high-mass stars, with the boundary occurring at approximately spectral type F5 ($M \simeq 1.3$~M$_\odot$). If the chemically peculiar Am and Ap stars are excluded, A and F stars lie above the Kraft Break and have typical values between $150 < v\sin\,i < 200$~km~s$^{-1}$ \citep{Zorec2012}. From a high-resolution spectroscopic study of bright \Kepler A and F stars, a mean $v\sin\,i$ value of 134~km~s$^{-1}$ was obtained by \citet{Niemczura2015}. Therefore, \dsct stars are generally considered moderate, and often, fast rotators \citep{Breger2000a}. Rotation lifts the degeneracy of a non-radial ($\ell > 0$) pulsation mode into its $2\ell + 1$ components, which are observed as a multiplet with nearly exact splitting among its component frequencies \citep{Pamyat2003}.
	
	The relationship between rotation and mode density is demonstrated by the high-amplitude \dsct (HADS) stars. The HADS stars were first classified by \citet{McNamara2000a}, who defined a sub-group of slowly-rotating \dsct stars with peak-to-peak light amplitude variations of more than 0.3~mag. These HADS stars have few pulsation mode frequencies in their amplitude spectra, with the dominant light variation usually being associated with the fundamental radial mode \citep{McNamara2000a}. The slow rotation seems to be a requirement for high-amplitude pulsations \citep{Breger2000a}. The HADS stars are found in a very narrow region within the classical instability strip with effective temperatures between $7000 \leq T_{\rm eff} \leq 8000$~K \citep{McNamara2000a}, and offer an opportunity to study nonlinearity in high-amplitude pulsations for which mode identification is relatively simple. We explore the nonlinearity of HADS stars observed by \Kepler in subsection~\ref{subsection: HADS stars}.
	
	\subsection{Hybrid stars}
	\label{subsection: hybrid stars}

	Near the red edge of the classical instability strip and the main-sequence on the HR~diagram, is another group of variable stars called gamma Doradus ($\gamma$~Dor) stars, which pulsate in high-order low-degree non-radial g~modes driven by the convective flux blocking mechanism operating at the base of the convective zone \citep{Guzik2000a, Dupret2005}. Typical g-mode pulsation frequencies in a \gdor star lie between $0.3 < \nu < 3$~d$^{-1}$ \citep{Uytterhoeven2011}. Reviews of \gdor stars are provided by \citet{Balona1994}, \citet{Kaye2000} and \citet{ASTERO_BOOK}.
	
	The theoretical instability regions of the \dsct and \gdor stars have been shown to overlap on the HR diagram \citep{Dupret2004}. Hybrid pulsators from the \dsct and \gdor pulsational excitation mechanisms occurring simultaneously within a star were first predicted by \citet{Dupret2005}, but only expected to comprise a small minority of A and F stars. The \Kepler mission data revealed that many \dsct stars are in fact hybrid pulsators \citep{Uytterhoeven2011, Balona2011e}, exhibiting both p and g~modes. Although it is common for the amplitude spectra of \dsct stars to contain low-frequency peaks, it is not established whether these frequencies are always caused by pulsation, the effects of rotation or have some other cause. Often low-frequency peaks in \dsct stars can be associated with combination frequencies of high-frequency pulsation modes. 
	
	Understanding the multi-periodic hybrid stars is an exciting prospect for asteroseismology as one can gain insight into sub-surface conditions, such as measurements of the rotation profile in main-sequence stars \citep{Kurtz2014, Saio2015b, Schmid2015, Keen2015, Triana2015, Murphy2016a}. Ideally, the hybrid stars could be used to study the pulsation excitation mechanisms directly, particularly the possible exchange of energy between pulsation modes excited by the different mechanisms. This idea was explored by \citet{Chapellier2012}, who studied the interaction between 180 g-mode and 59 p-mode independent pulsation frequencies in the CoRoT hybrid star ID~105733033. The authors demonstrated that the p- and g-mode frequencies originated in the same star and that a coupling mechanism must exist to explain the observed mode interaction \citep{Chapellier2012}.
	
	\subsection{The {\it Kepler} mission}
	\label{subsection: Kepler}
	
	The \Kepler Space Telescope has revolutionised our understanding of pulsating stars, including \dsct and \gdor stars \citep{Uytterhoeven2011, Balona2014a}. The \Kepler spacecraft was launched in 2009~March into a 372.5-d Earth-trailing orbit with a primary goal to locate Earth-like exoplanets orbiting solar-like stars using the transit method \citep{Borucki2010}. A high photometric precision of order a few $\umu$mag, a high duty-cycle \citep{Koch2010} and data spanning 1470.5~d (4~yr) for more than 150\,000 stars, allow us to probe the structure of stars with a significantly higher precision than any ground-based telescope. Observations were made using a 29.5-min long cadence (LC) and 58.5-s short cadence (SC) \citep{Gilliland2010}. 
	
	Approximately 200\,000 targets stars were observed by {\it Kepler}, many of which were characterised with values of $T_{\rm eff}$, $\log\,g$ and $[{\rm Fe} / {\rm H} ]$ using {\it griz} and 2MASS $JHK$ broadband photometry prior to the launch of the telescope. These parameters were collated into the \Kepler Input Catalogue (KIC; \citealt{Brown2011}) and allowed stars to be placed on the HR~diagram. Since the end of the nominal \Kepler mission, \citet{Huber2014} revised the stellar parameters for the $\sim$200\,000 \Kepler targets and concluded that a colour-dependent offset exists compared to other sources of photometry (e.g. Sloan). This resulted in KIC temperatures for stars hotter than $T_{\rm eff} > 6500$~K being, on average, 200~K lower than temperatures obtained from Sloan photometry or the infrared flux method \citep{Pinsonneault2012}. Also, $\log\,g$ values for hot stars were overestimated by up to 0.2~dex. \citet{Huber2014} stresses that the $\log\,g$ and $[{\rm Fe} / {\rm H} ]$ values and their respective uncertainties should not be used for a detailed analysis on a star-by-star basis, as they are only accurate in a statistical sense.
	
	The 4-yr \Kepler observations provide a Rayleigh resolution criterion in frequency of $1/\Delta T = 0.00068$~d$^{-1}$ (8~nHz). Using signal-to-noise ratios of the most stable frequencies, an amplitude precision of 1~$\umu$mag is achievable with \Kepler data \citep{Kurtz2014}. This unprecedented frequency and amplitude precision has been utilised to study frequency and amplitude modulation in \dsct stars (e.g. \citealt{Bowman2014, Bowman2015a, Barcelo2015}). In this paper, we extend the search for amplitude modulation to a large number of \dsct stars observed by the \Kepler mission, and use case studies to demonstrate how to distinguish among different physical causes.
	
	In this paper we discuss the various causes of amplitude modulation in section~\ref{section: causes of AM} and our method of identifying the \dsct stars in our ensemble that exhibit amplitude modulation in section~\ref{section: method}. We present our catalogue of \dsct stars and demonstrate the diversity in pulsational behaviour using case studies of individual stars in section~\ref{section: results}. We discuss beating models in section~\ref{section: beating model} and coupling models in section~\ref{section: coupling model}. Finally, we discuss statistics from the ensemble study in section~\ref{section: ensemble} and our conclusions in section~\ref{section: conclusions}.
	

\section{Causes of amplitude modulation}
\label{section: causes of AM}

The various causes of why \dsct stars exhibit variable pulsation amplitudes (and/or frequencies) can be loosely grouped as intrinsic and extrinsic, i.e. those physical and interior to the star and those caused by external effects, respectively. In the following subsections, we discuss examples of the different mechanisms that can cause variable pulsation amplitudes and/or frequencies.

	\subsection{Intrinsic: Beating}
	\label{subsection: Beating}
	
	A study of seven well-known \dsct stars by \citet{Breger2002d} found that pairs of close-frequency pulsation modes, with spacings less than $0.01$~d$^{-1}$, were not uncommon. Moreover, these pairs of close-frequency modes were found near the expected frequencies of radial modes in these stars \citep{Breger2002d}. Such close-frequency pulsation modes are unlikely to be explained by rotational effects as most \dsct stars are fast rotators. For example, even the slowly-rotating \dsct star 44~Tau (HD~26322) with $v\sin i = 3 \pm 2$~km~s$^{-1}$ \citep{Zima2006c} would produce a rotational splitting of approximately 0.02~d$^{-1}$ \citep{Breger2009a}. The study by \citet{Breger2002d}, however, only included a small sample of \dsct stars limited by the frequency precision obtained from intermittent ground-based data spanning a few decades. 
	
	Later, it was shown that pulsation mode frequencies in \dsct stars are not distributed at random, and that many non-radial modes had frequencies near radial mode frequencies \citep{Breger2009a}. These regularities in the amplitude spectra were explained by mode trapping in the stellar envelope \citep{Dziembowski1990a}, which was clearly demonstrated as the cause of regularities in the amplitude spectrum of the \dsct star FG~Vir (HD~106384) by \citet{Breger2009a}.
	
	The variability of the \dsct star 4~CVn (HD~107904) was first discovered by \citet{Jones1966}, and the star has been extensively studied since, with 26 independent pulsation mode frequencies and many more combination frequencies discovered \citep{Breger1990c, Breger1999, Breger2000b, Breger2009b, Schmid2014, Breger2016a*}. This makes it one of the longest- and best-studied \dsct stars. Many of the pulsation mode frequencies show frequency and amplitude variations, some of which can be explained by a mode coupling mechanism \citep{Breger2000b} or the beating of two close frequencies \citep{Breger2009b}. The two pulsation mode frequencies, $6.1007$~d$^{-1}$ and $6.1170$~d$^{-1}$, were highly variable in amplitude over the observations \citep{Breger2010}. The changes in frequency and amplitude of these two pulsation mode frequencies were used to construct beating models, which were matched to the observations of amplitude modulation in 4~CVn \citep{Breger2010}. 
	
	These investigations of \dsct stars may not contain a large number of stars, but they do demonstrate the diverse pulsational behaviour. In our study, we use the definition from \citet{Breger2002d} of close frequencies having a separation of less than $0.01$~d$^{-1}$. We emphasise that it is only studying the {\it changes} in amplitude and frequency (i.e. phase at fixed frequency) of pulsation modes that allows one to construct beating models of close-frequency pulsation modes and {pure} amplitude modulation of a single pulsation mode \citep{Breger2002d}. This cannot be achieved from simple inspection of the light curve or amplitude spectrum of a pulsating star, regardless of the data precision, because the convolved amplitude and frequency modulation signals cannot be disentangled. 
	
	The beating of a pair of close and resolved pulsation mode frequencies appears as periodic amplitude modulation, with a characteristic sharp change in phase at the epoch of minimum amplitude for each frequency in the pair (e.g. \citealt{Breger2006a}). The simplest scenario is the example of two similar frequency cosinusoids with equal amplitude, each of the form

\begin{equation}
	y = A \cos ( 2\pi \nu t + \phi) ~ ,
\label{equation: line}
\end{equation}

\noindent where $A$ is the amplitude, $\nu$ is the frequency and $\phi$ is the phase. Using the sum rule in trigonometry for two equal-amplitude cosinusoids with frequencies $\nu_1$ and $\nu_2$, each with a phase of 0.0~rad, a summation is given by
	
	\begin{equation}
		\begin{array}{ll} \displaystyle
		y_1 + y_2 &= A \cos 2\pi\nu_{1}t + A \cos2\pi\nu_{2}t \\
		\cr \displaystyle
				&= 2A\cos2\pi \frac{\nu_1 + \nu_2}{2} t \cos2\pi \frac{\nu_1 - \nu_2}{2} t ~ ,
		\end{array}
		\label{equation: beating 1}
	\end{equation}
	
	\noindent from which, the beat frequency is defined by
	
	\begin{equation}
		\nu_{\rm beat} = | \nu_1 - \nu_2 | ~ ,
		\label{equation: beating 2}
	\end{equation}
	
	\noindent which is the absolute difference in frequency of the two cosinusoids. 
	
	The characteristic behaviour of beating is most easily recognised with two frequencies of equal amplitude. In such a case the visible (and assumed single) frequency will vary sinusoidally in amplitude with a period equal to the beat period, but also vary in phase: a half cycle (i.e. $\pi$~rad) change in phase will occur at the epoch of minimum amplitude \citep{Breger2006a}, and the amplitude will modulate between $2A$ and $0$. If the cosinusoids have increasingly different amplitudes, the amplitude and phase changes get progressively smaller. However, the amplitude and phase must always vary synchronously with a phase lag (shift) close to $\pi/2$~rad, such that the epoch of minimum amplitude in the beat cycle occurs at the time of average and most rapid change in phase \citep{Breger2006a}. We construct beating models for \dsct stars observed by \Kepler in section~\ref{section: beating model}.
	
	\subsection{Intrinsic: nonlinearity and mode coupling}
	\label{subsection: resonance}
	
	There are different non-linear effects that can create combination frequencies in the amplitude spectrum of a pulsating star. Combination frequencies are mathematical sum and difference frequencies of pulsation mode frequencies, $\nu_{\rm i}$ and $\nu_{\rm j}$, that have the form $n \nu_{\rm i} \pm m \nu_{\rm j}$, in which $n$ and $m$ are integers. Possible mechanisms to explain combination frequencies in variable DA and DB white dwarfs discussed by \citet{Brickhill1992a} were that the stellar medium does not respond linearly to the pulsation wave, or that the dependence of emergent flux variation is not a linear transformation from the temperature variation $(F = \sigma T^4)$, which are often grouped into what is termed a non-linear distortion model (e.g. \citealt{Degroote2009a}).
	
Combination frequencies are common in \dsct stars, for example KIC~11754974 \citep{Murphy2013b} and KIC~8054146 \citep{Breger2012b, Breger2014}, but also in SPB, Be and \gdor stars \citep{Kurtz2015b, VanReeth2015b}. Identifying which peaks are combination frequencies and which are real pulsation mode frequencies is important, so that the amplitude spectra of pulsating stars can be greatly simplified (e.g. \citealt{Papics2012b, Kurtz2015b}); the real pulsation mode frequencies are the main parameters used for asteroseismic modelling. One method of identifying combination frequencies is by mathematically generating all the possible combination terms from a small number of parent frequencies, fitting them by least-squares and removing them by pre-whitening. Alternatively, iterative pre-whitening can be used to extract all statistically-significant frequencies and then exclude those that satisfy a combination frequency relation (e.g. \citealt{VanReeth2015b}). However, the physical mechanism that causes these frequencies is not immediately obvious.

At this juncture, it is important to note that there is a subtle difference between combination frequencies and coupled frequencies. Combination frequencies and harmonics occur due to the mathematical representation of the summation of sine and cosine terms when calculating the Fourier transform caused by pulsational nonlinearity. This effect differs to families of pulsation mode frequencies that are resonantly excited due to the coupling of modes inside the star (e.g. \citealt{Breger2014}). Mode coupling through the resonant interaction of pulsation modes has been discussed theoretically in detail by \citet{Dziembowski1982} and \citet{Buchler1997a}. This form of pulsational nonlinearity gives rise to variable frequencies and amplitudes in pulsation modes over time \citep{Buchler1997a}, which will appear as a cluster of unresolved peaks in the amplitude spectrum if the length of observations is shorter than the modulation cycle. From the unresolved behaviour, it is difficult to determine if a cluster of peaks represents a single pulsation mode with frequency and/or amplitude variability, or multiple independent close-frequency pulsation modes.
	 
	Theoretical models of \dsct stars often predict much larger pulsation mode amplitudes than are observed, which suggests that an amplitude limitation mechanism or limit cycle is required \citep{Breger2000a}. An example of how nonlinearity can act as an amplitude limitation mechanism in \dsct stars is the parametric resonance instability \citep{Dziembowski1982}, which states that two linearly unstable low-frequency modes (i.e. parent modes) can damp a high-frequency unstable mode (i.e. child mode) once it reaches a critical amplitude. Resonant mode coupling has been suggested as the amplitude limitation mechanism operating in \dsct stars but not in HADS stars, which explains the large difference in pulsation mode amplitudes between the two subgroups \citep{Dziembowski1985a}. 
	
	Coupled frequencies are grouped into families of child and parent modes (e.g. \citealt{Breger2014}), and this coupling can facilitate the exchange of energy between different members of the family \citep{Dziembowski1982, Buchler1997a, Nowakowski2005}. Coupled child and parent modes must satisfy the resonance condition of 
	
	\begin{equation}
		\nu_1 \simeq \nu_2 \pm \nu_3 ~ ,
	\label{equation: freq coupling}
	\end{equation} 
	
	\noindent where $\nu_1$ is the child mode, and $\nu_2$ and $\nu_3$ are the parent modes. The \dsct star KIC~8054146 was found to exhibit several families of pulsation modes \citep{Breger2012a}, some of which were later suggested to be caused by resonant mode coupling \citep{Breger2014}. The authors commented that it is difficult to distinguish physically coupled modes from combination frequencies, emphasising the need to study the frequency, amplitude and phase of the members within each family.
		
	However, the frequency resonance criterion given in eq.~\ref{equation: freq coupling} does not solely distinguish which frequency within a family is a combination or coupled mode frequency of the other two. To make this distinction, \citet{Breger2014} modelled the amplitude of a child mode as a product of the two parent mode amplitudes using
	
	\begin{equation}
		A_1 = \mu_{\rm c} ( A_2  A_3 ) ~ , 
	\label{equation: amp coupling}
	\end{equation} 
	
	\noindent and the linear combination of the parent phases
	
	\begin{equation}
		\phi_1 = \phi_2 \pm \phi_3 ~ ,
	\label{equation: phase coupling}
	\end{equation}
	
	\noindent where $A_i$ and $\phi_i$ represent amplitude and phase of the child and parent modes, respectively, and $\mu_{\rm c}$ is defined as the coupling factor. For combination frequencies arising from a non-linear distortion model, small values of $\mu_{\rm c}$ are expected and thus the amplitude and/or phase variability in combination frequencies will simply mimic the parent modes that produce them \citep{Brickhill1992a, Wu2001c, Breger2008c}. However, for resonant mode coupling, one expects the amplitudes of the three modes to be similar and values of $\mu_{\rm c}$ to be larger because mode energy is physically being exchanged between the child and parent modes. In the case of KIC~8054146, $\mu_{\rm c}$ was of the order $1000$ for coupled modes for parent mode amplitudes of order 0.1~mmag \citep{Breger2014}.
	
	For the case of resonant mode coupling discussed by \citet{Dziembowski1985a}, the most likely outcome in \dsct stars is two linearly damped g-mode parents coupling with an unstable p-mode child. The coupling of these modes would cause the growth and decay of the child p~mode in anti-correlation with the parent modes, as energy is exchanged among the family members \citep{Dziembowski1985a}. Furthermore, the parent g~modes may not be excited to observable amplitudes at the surface of the star and are thus undetectable. This was suggested as a plausible mechanism for the observed amplitude modulation in the \dsct star KIC~7106205 by \citet{Bowman2014}, who showed that a single p-mode frequency decreased significantly in amplitude over 4~yr with no change in amplitude or phase in any other visible pulsation mode. Further work by \citet{Bowman2015a} extended this study back to 2007 using archive data from the Wide Angle Search for Planets (WASP; \citealt{Pollacco2006}). It is possible that the observed amplitude modulation in KIC~7106205 could be caused by an amplitude limitation mechanism, i.e. resonant mode coupling, transferring energy from the child p~mode to undetectable parent g~modes in the core of the star; otherwise, the p-mode pulsation energy could be lost to an unknown damping region. It was concluded for the case of KIC~7106205 that the visible pulsation energy was not conserved \citep{Bowman2014}.
	
	In their analysis of KIC~8054146, \citet{Breger2012b} stated that even if three frequencies within a family obey the frequency, amplitude and phase relations outlined in eqs~\ref{equation: freq coupling}, \ref{equation: amp coupling} and \ref{equation: phase coupling}, it does not {\it prove} that they are combination frequencies, but merely behave like combination frequencies, thus variable modes can be interpreted as being caused by nonlinearity from a non-linear distortion model or resonant mode coupling. The coupling coefficient $\mu_{\rm c}$ in eq.~\ref{equation: amp coupling} represents the {\it strength} of nonlinearity in a star and thus how much coupling exists among the different members within a family. Therefore, the testable prediction for resonant mode coupling between a child and two parent modes in \dsct stars is large-scale amplitude modulation in three similar-amplitude modes with large values of $\mu_{\rm c}$ \citep{Breger2014}. Using a similar approach, we test the coupling hypothesis in section~\ref{section: coupling model} for the \dsct star KIC~4733344, which contains families of pulsation modes satisfying eqs~\ref{equation: freq coupling}, \ref{equation: amp coupling} and \ref{equation: phase coupling}. 
	
	\subsection{Extrinsic: Binarity and multiple systems}
	\label{subsection: Binarity}
		
	A spectroscopic study of 4~CVn by \citet{Schmid2014} revealed that it is in an eccentric binary with $P_{\rm orb} = 124.44 \pm 0.03$~d and $e = 0.311 \pm 0.003$. After removing the binary signature, further amplitude and phase variability on timescales of the order 1~yr remained in pulsation mode frequencies. These could not be explained by the beating of two (or more) close-frequency modes, because the beating cycle was unresolved in the length of the observations \citep{Schmid2014}. 
	
	Binarity in a system can also be tested with photometric data using the Frequency Modulation technique (FM; \citealt{Shibahashi2012, Shibahashi2015}) and the Phase Modulation technique (PM; \citealt{Murphy2014, Murphy2015b}). The FM technique uses the fact that the motion of a pulsating star about the Barycentre of a binary (or multiple) system will perturb a pulsation mode frequency throughout the orbit and introduce a small frequency shift. If observations are longer than the orbital period, then the perturbation on a pulsation mode frequency is resolved and produces sidelobes on either side of the pulsation mode frequency in the amplitude spectrum \citep{Shibahashi2012}. The orbital period, can be directly measured as the inverse of the separation in frequency between the central peak and the sidelobes of a multiplet in the amplitude spectrum. 
	
	Similarly, the PM technique uses the fact that there will be a difference in the light travel time across the orbit in a multiple system and thus the phases of pulsation mode frequencies will vary on the same timescale as the orbit \citep{Murphy2014}. The amplitude of the observed phase modulation  is a function of frequency when expressed as light travel time delays (see equation~3 from \citealt{Murphy2014}). If all the pulsation modes in a star vary in phase with the same period, this can be explained by the Doppler effect modulating the signal throughout the orbital period \citep{Murphy2014}. The significance of the FM and PM techniques is made evident as not only can the orbital period be determined, but also $e$, $a\sin\,i$, $f(m)$ and argument of periastron without obtaining radial velocity measurements. An excellent example of the PM technique used to study hybrid pulsators using \Kepler data was that of \citet{Schmid2015} and \citet{Keen2015}, who demonstrated that KIC~10080943 is an eccentric binary system containing two hybrid pulsators with masses $M_1 = 2.0 \pm 0.1$~M$_{\odot}$ and $M_2 = 1.9 \pm 0.1$~M$_{\odot}$. From the common phase modulation, \citet{Schmid2015} were able to identify which pulsation frequencies originated from each hybrid star and confirm the orbital period of $P_{\rm orb} = 15.3364 \pm 0.0003$~d for KIC~10080943.
	

\section{Method}
\label{section: method}

To study amplitude modulation using a statistical ensemble, the time series for all \Kepler targets with effective temperatures between $6400 \leq T_{\rm eff} \leq 10\,000$~K in the KIC \citep{Brown2011} were downloaded. We used the publicly available (msMAP) PDC data \citep{Stumpe2012, Smith2012}, which can be downloaded from the Mikulski Archive for Space Telescopes (MAST)\footnote{MAST website: {http://archive.stsci.edu/kepler/}}. The extracted time series were stored locally in the format of reduced Barycentric Julian Date (${\rm BJD} - 2\,400\,000$) and magnitudes, which were normalized to be zero in the mean, for each quarter of LC and month of SC data. Amplitude spectra for each quarter of LC data were calculated using the methodology described by \citet{Deeming1975}, which produced a data catalogue for all of the $\sim$10\,400 stars in this $T_{\rm eff}$ range.

	\subsection{Creating an amplitude modulation catalogue}
	\label{subsection: creating catalogue}

	To maximise the outputs from this study, the final ensemble of stars comprised the targets that met {\it all} of the following criteria:
	
	\begin{enumerate}
	\item Characterised by $6400 \leq T_{\rm eff} \leq 10\,000$~K in the KIC;
	\vspace{0.2cm}
	\item Observed continuously in LC from Q0 (or Q1) to Q17;
	\vspace{0.2cm}
	\item Contain peaks in the amplitude spectrum in the sub-Nyquist p-mode frequency regime ($4 \leq \nu \leq 24$~d$^{-1}$) with amplitudes greater than 0.10~mmag;
	\vspace{0.2cm}
	\end{enumerate}
	
	\noindent which resulted in {983} \dsct and hybrid stars. In the following paragraphs, we justify the motivation for each of these criteria.
	
	We chose the lower $T_{\rm eff}$ limit of $6400$~K for our search for amplitude modulation, because this is the observational red edge of the classical instability strip for \dsct stars \citep{Rod2001}. The ZAMS red edge was calculated to be approximately $6900$~K by \citet{Dupret2004}, but cooler high-luminosity \dsct stars are found below $6900$~K and so we chose a lower limit of $6400$~K to include these targets. Only about ten \dsct stars were found in the KIC effective temperature range of $6400 \leq T_{\rm eff} \leq 6500$, supporting the observational red edge defined by \citet{Rod2001}. An upper limit of $T_{\rm eff} \leq 10\,000$~K was chosen to exclude pulsators that do not lie within the classical instability strip, such as SPB and \bcep stars (e.g. see \citealt{Balona2011b} and \citealt{McNamara2012}).
		
	We chose to use LC data as this gives the largest number of stars in our ensemble observed over the largest possible time span of 4~yr. Only a small fraction of intermediate (and high) mass stars were observed in SC and even fewer for many consecutive SC months. We do not include any \dsct stars that lie on module 3 because of the lower duty-cycle. Module 3 of the \Kepler CCD failed during the 4-yr mission, thus these stars have every fourth data quarter missing in their light curves and have complex window patterns in their amplitude spectra. This choice is motivated by previously studied \dsct stars that have amplitude modulation of the order of years and decades \citep{Breger1999, Breger2000b, Breger2016a*}, and so complete data coverage over the maximum of 4~yr is most useful. 

	We selected stars that contain pulsation mode frequencies within $4 \leq \nu \leq 24$~d$^{-1}$, because the LC Nyquist frequency is $\nu_{\rm Nyq} = 24.4$~d$^{-1}$. Even though \dsct stars can pulsate at higher frequencies than $\nu_{\rm Nyq}$ (e.g. \citealt{Holdsworth2014a}), we made this selection because alias peaks are subject to frequency (phase) variations and amplitude suppression \citep{Murphy2013a}. \Kepler data were sampled at a regular cadence on board the spacecraft, but Barycentric time stamp corrections were made resulting in a non-constant cadence. Therefore, real and alias frequencies can be identified without the need to calculate an amplitude spectrum beyond the LC Nyquist frequency. Note that if a star pulsates with frequencies that lie above and below the LC Nyquist frequency, it was included in our sample as some of its extracted frequencies will not be super-Nyquist aliases. We chose an amplitude cut-off of {0.10}~mmag, which is much higher than the typical noise level in \Kepler data of order a few $\umu$mag. This choice is so that reasonable phase uncertainties are generated as they are dependent on the amplitude signal-to-noise ratio \citep{Montgomery1999}.
	
	\subsection{Identifying pulsation modes with variable amplitudes}
	\label{subsection: finding amod}
	
	After identifying the ensemble of {983} \dsct stars for this project, an automated tracking routine was used to determine amplitude and phase variations (at fixed frequency) for the 12 highest amplitude peaks in the amplitude spectrum using the 4-yr data set for each star. The choice of tracking specifically 12 peaks is somewhat arbitrary. The main motivation was to identify the dominant (changes in) pulsational behaviour in \dsct stars. For a star that pulsates in only a few modes, selecting 12 peaks was more than sufficient. On the other hand, for a star that pulsates in dozens of modes and has hundreds of combination frequencies in its amplitude spectrum, 12 frequencies may not be enough to fully disentangle the star, but does provide vital information on the most dominant behaviour. Note that only peaks with amplitudes greater than $0.10$~mmag were extracted in our analysis, so fewer than the maximum number of 12 frequencies can be extracted for a star.

	After extracting the appropriate number of frequencies by sequentially pre-whitening a star's amplitude spectrum, the frequencies were optimized using a simultaneous multi-frequency non-linear least-squares fit to the 4-yr data set, ensuring the highest possible frequency and amplitude precision were obtained. We then divided the data set into {30} time bins, each {100}~d in length (except the last bin) with a {50-d} overlap, and optimized amplitude and phase at fixed frequency using linear least-squares in each bin for each frequency. This approach has previously been used to study the amplitude and phase variability in the \dsct star KIC~7106205 \citep{Bowman2014}. The amplitudes and phases of each time bin for the frequencies were plotted against time in what we term tracking plots, which allowed us to investigate similar variability in different frequencies within each star. 
	
	\subsection{Defining significant amplitude modulation}
	\label{subsection: defining amod}
	
	\begin{figure}
		\centering
		\includegraphics[width=\columnwidth]{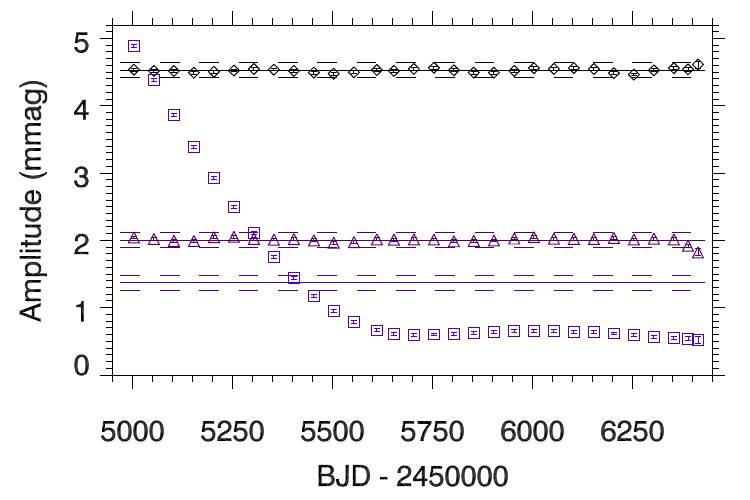}
		\caption{Amplitude tracking plot for KIC~7106205 showing pulsation mode frequencies $\nu_1 = 10.032366$~d$^{-1}$, $\nu_2 = 10.727317$~d$^{-1}$ and $\nu_3 = 13.394175$~d$^{-1}$ as diamonds, triangles and squares, respectively. The solid line for each frequency is the mean of 30 time bins and the dashed lines represent the {$\pm 5\sigma$} amplitude significance interval from the mean. A frequency is flagged as exhibiting significant amplitude modulation if {15} (half) of its bins lie outside the $\pm5\sigma$ amplitude from the mean.}
		\label{figure: significant amod}		
	\end{figure}
	
	Classifying a significant change in amplitude for each frequency in each star by visual inspection is straightforward, but time consuming for a large sample. Therefore we automated this process using the following methodology. A mean amplitude was calculated from the time bins\footnote{For clarity, the last time bin is often excluded from the tracking plots, as it contains fewer data points.} for each frequency and any bins that were more than {$\pm5\sigma$} in amplitude away from this mean were flagged. An example of this is shown in Fig.~\ref{figure: significant amod} using a solid line as the mean value and dashed lines as the {$\pm5\sigma$} significance interval, for the three highest-amplitude frequencies in KIC~7106205. We chose to define a frequency as exhibiting {\it significant} amplitude modulation if at least {15} (i.e half) of its amplitude bins were more than {$\pm5\sigma$} from its mean value. For example, only $\nu_3 = 13.394175$~d$^{-1}$ in KIC~7106205 satisfied the significant amplitude modulation criterion described, which is shown in Fig.~\ref{figure: significant amod}. All other frequencies were flagged as having constant amplitudes.
	
	This method was applied to each of the extracted frequencies and the number of amplitude-modulated frequencies was noted for each star in our ensemble. We use the abbreviations of AMod ({\bf A}mplitude {\bf Mod}ulated) to describe the \dsct stars that exhibit at least a single pulsation mode that is variable in amplitude over the 4-yr \Kepler data set, and NoMod ({\bf No} {\bf Mod}ulation) for those that do not.
	
	\subsection{Phase adjustment}
	\label{subsection: phase adjustment}
	
	For the purpose of calculating phases in the tracking plots, a zero-point in time was chosen as the centre of the 4-yr \Kepler data set, specifically $t_0 = 2\,455\,688.770$~BJD, in the cosinusoid function $y = A \cos ( 2\pi \nu (t-t_0) + \phi)$. Phase is defined in the interval $-\pi \leq \phi \leq \pi$~rad, and so can be adjusted by adding or subtracting integer values of $2\pi$~rad. If the difference between consecutive phase bins exceeded $5$~rad in a tracking plot, then a $2\pi$~rad phase adjustment was made. As previously discussed in section~\ref{subsection: Beating}, the phase change for beating cannot exceed $\pi$~rad so this phase adjustment did not remove any beating signals in pulsation mode frequencies.
	
	\subsection{Pulsation mode identification}
	\label{subsection: Q constants}
	
	For high-amplitude pulsators, the period ratios of pulsation modes can be used to identify modes. The period ratio of the first overtone to fundamental mode for \dsct stars is between ${P_1}/{P_0} = (0.756-0.787)$, and subsequent ratios of ${P_2}/{P_0} = (0.611-0.632)$ and ${P_3}/{P_0} = (0.500-0.525)$ for the second and third overtones, respectively \citep{Stellingwerf1979}. 
	
	Pulsation modes can also be identified by calculating pulsation constants using 
	
	\begin{equation}
	Q = P \sqrt{ \frac{\bar\rho}{\bar\rho_{\odot}} } ~ ,
	\label{equation: stellingwerf 2}
	\end{equation}
	
	\noindent where $Q$ is the pulsation constant in days, $P$ is the pulsation period in days, and $\bar\rho$ is the mean stellar density. Eq.~\ref{equation: stellingwerf 2} can be re-written as

	\begin{equation}
	\log~Q = \log\,P + \frac{1}{2}\log\,g +  \frac{1}{10}M_{\rm Bol} + \log\,T_{\rm eff} - 6.454 ~ ,
	\label{equation: stellingwerf}
	\end{equation}

	\noindent where $\log\,g$ is the surface gravity in cgs units, $M_{\rm Bol}$ is the Bolometric absolute magnitude and $T_{\rm eff}$ is the effective temperature in K. A value of $M_{\rm Bol}$ can be calculated using $T_{\rm eff}$ and $\log\,g$ values in comparison with the Pleiades main-sequence stars. Typical values of pulsation constants for the fundamental, first and second-overtone radial p~modes in \dsct stars lie between $0.022 \leq Q \leq 0.033$~d \citep{Breger1975}. The pulsation constant can be used to identify the order (overtone number, $n$) of radial modes \citep{Stellingwerf1979}, but the calculation using eq.~\ref{equation: stellingwerf} is very dependent on the stellar parameters used, particularly $\log\,g$. For example, \citet{Breger1990b} quotes fractional uncertainties in $Q$ values as high as 18~per~cent, which could cause a first overtone radial mode to be confused for the fundamental or second overtone radial mode. Therefore, caution is advised when applying this method of mode identification.


\section{Catalogue discussion}
\label{section: results}

The analysis of {983} \dsct stars have been collated into a single catalogue\footnote{The amplitude modulation catalogue containing the amplitude spectra and tracking plots of all 983 \dsct stars can be obtained from {http://uclandata.uclan.ac.uk/id/eprint/42} as a PDF.} In the following subsections, individual stars are used as case studies to demonstrate the diversity of pulsational behaviour in \dsct stars. For each star presented in this paper, Table~\ref{table: all stars} lists the stellar parameters from \citet{Huber2014}, the number of AMod and NoMod frequencies, and the pulsator type as either \dsct or hybrid based on its frequencies.

\begin{table*}
	\centering
	\caption[]{Stellar parameters for the \dsct stars discussed in this paper, as listed in \citet{Huber2014}. For each star, the number of constant-amplitude and variable-amplitude pulsation mode frequencies, ${\rm N_{\nu}}$, above our amplitude cut-off of {0.10}~mmag are listed under the columns {NoMod} and {AMod}, respectively. Super-Nyquist alias peaks are identified using sNa. Orbital periods obtained using the PM technique \citep{Murphy2014} are consistent with the phase modulation in the four example stars identified as binary systems. Each star is labelled as either a \dsct or a hybrid in the pulsator type column based on its frequencies. A version of this table for all {983} stars is given as online-only material as a PDF, with a machine-readable version available through CDS.}
	\small
	\begin{tabular}{c r r r r c c c l}
		\hline
		\multicolumn{1}{c}{KIC ID} & \multicolumn{1}{c}{T$_{\rm eff}$} & \multicolumn{1}{c}{$\log\,g$} & \multicolumn{1}{c}{[Fe/H]} & \multicolumn{1}{c}{Kp} & \multicolumn{1}{c}{Pulsator type} & \multicolumn{2}{c}{N$_{\nu}$} & \multicolumn{1}{c}{Comments} \\
		
		\multicolumn{1}{c}{ } & \multicolumn{1}{c}{(K)} & \multicolumn{1}{c}{(cgs)} & \multicolumn{1}{c}{(dex)} & \multicolumn{1}{c}{(mag)} & \multicolumn{1}{c}{ } & \multicolumn{1}{c}{NoMod} & \multicolumn{1}{c}{AMod} & \multicolumn{1}{c}{ } \\	
				
\hline
\multicolumn{9}{l}{NoMod (constant amplitude) stars:} \vspace{0.1cm} \\

1718594   &   $7800 \pm 270$   &   $3.97 \pm 0.22$   &   $0.07 \pm 0.30$   &	10.37	&	\dsct		&	8	&	0	&	$\nu_4$ is sNa	\\

2304168   &   $7220 \pm 270$   &   $3.67 \pm 0.19$   &   $-0.06 \pm 0.30$   &	12.41	&	\dsct		&	11	&	0	&	\citet{Balona2011g} \\

6613627   &   $7310 \pm 280$   &   $4.14 \pm 0.24$   &   $-0.02 \pm 0.31$   &	12.55	&	\dsct		&	5	&	0	&	\\

9353572   &   $7420 \pm 280$   &   $3.95 \pm 0.21$   &   $-0.40 \pm 0.31$   &	10.62	&	\dsct		&	5	&	0	&	\\	

\hline
\multicolumn{9}{l}{AMod explainable by the beating of close-frequency pulsation modes:} \vspace{0.1cm} \\
	
4641555   &   $7170 \pm 250$   &   $4.22 \pm 0.25$   &   $-0.12 \pm 0.32$   &	12.61	&	\dsct		&	8	&	1	&	$P_{\rm beat} = 1166 \pm 1$~d \\
		
8246833   &   $7330 \pm 270$   &   $3.96 \pm 0.22$   &   $-0.32 \pm 0.30$   &	11.87	&	\dsct		&	9	&	3	&	$P_{\rm beat} = 1002 \pm 1$~d \\	

\hline
\multicolumn{9}{l}{AMod explainable by {nonlinearity}:} \vspace{0.1cm} \\

4733344   &   $7210 \pm 260$   &   $3.50 \pm 0.23$   &   $-0.12 \pm 0.28$   &	10.08	&	hybrid 	&	3	&	9	&	\\
	
\hline
\multicolumn{9}{l}{Stars with {pure} AMod:} \vspace{0.1cm} \\

2303365   &   $7520 \pm 270$   &   $3.64 \pm 0.18$   &   $0.00 \pm 0.28$   &	11.14	&	\dsct		&	10	&	2	&	$\nu_{10}$ has a $\sim250$-d beat signal \\	

5476273   &   $7430 \pm 280$   &   $4.17 \pm 0.23$   &   $-0.22 \pm 0.35$   &	13.62	&	\dsct		&	11	&	1	&	$\nu_9$ is sNa	\\	

7685307   &   $7690 \pm 270$   &   $3.80 \pm 0.20$   &   $-0.20 \pm 0.31$   &	12.14	&	\dsct		&	5	&	2	&	\\	

8453431   &   $7180 \pm 270$   &   $3.63 \pm 0.19$   &   $-0.06 \pm 0.29$   &	12.53	&	\dsct		&	2	&	1	&	\\

\hline
\multicolumn{9}{l}{Stars with phase modulation explainable by binarity and confirmed using the PM technique:} \vspace{0.1cm} \\

3650057   &   $7320 \pm 280$   &   $4.06 \pm 0.22$   &   $-0.10 \pm 0.32$   &	13.92	&	\dsct		&	9	&	3	&	$P_{\rm orb} = 804.6 \pm 2.0$~d \\	
	
4456107   &   $7250 \pm 280$   &   $4.08 \pm 0.24$   &   $0.06 \pm 0.28$   &	13.83	&	\dsct		&	11	&	1	&	$P_{\rm orb} = 335.5 \pm 0.5$~d	\\	
		
5647514   &   $7410 \pm 280$   &   $4.13 \pm 0.23$   &   $-0.04 \pm 0.32$   &	12.43	&	\dsct 	&	10	&	2	&	$P_{\rm orb} = 1123 \pm 6$~d	\\	

9651065   &   $7010 \pm 150$   &   $3.83 \pm 0.13$   &   $-0.10 \pm 0.15$   &	11.07	&	hybrid	&	12	&	0	&	$P_{\rm orb} = 272.7 \pm 0.8$~d \\

\hline
\multicolumn{9}{l}{HADS stars:} \vspace{0.1cm} \\
		
5950759   &   $8040 \pm 270$   &   $4.05 \pm 0.22$   &   $-0.10 \pm 0.33$   &	13.96	&	HADS	&	{2}	&	{10}	&	$\nu_{4,6,9,10,12}$ are sNa	\\	
	
9408694   &   $6810 \pm 140$   &   $3.78 \pm 0.11$   &   $-0.08 \pm 0.15$   &	11.46	&	HADS	&	{7}	&	{5} 	&	\citet{Balona2012a}	\\	
	
\hline
\multicolumn{9}{l}{Other special cases:} \vspace{0.1cm} \\

7106205   &   $6900 \pm 140$   &   $3.70 \pm 0.13$   &   $0.32 \pm 0.13$   &	11.46	&	\dsct		&	8	&	1	&	\citet{Bowman2014}		\\
		
		\hline
		\end{tabular}
	\label{table: all stars}
\end{table*}

	\subsection{Super-Nyquist asteroseismology}
	\label{subsection: sNa peaks}
	
\begin{figure*}
	\centering
		\includegraphics[width=0.59\textwidth]{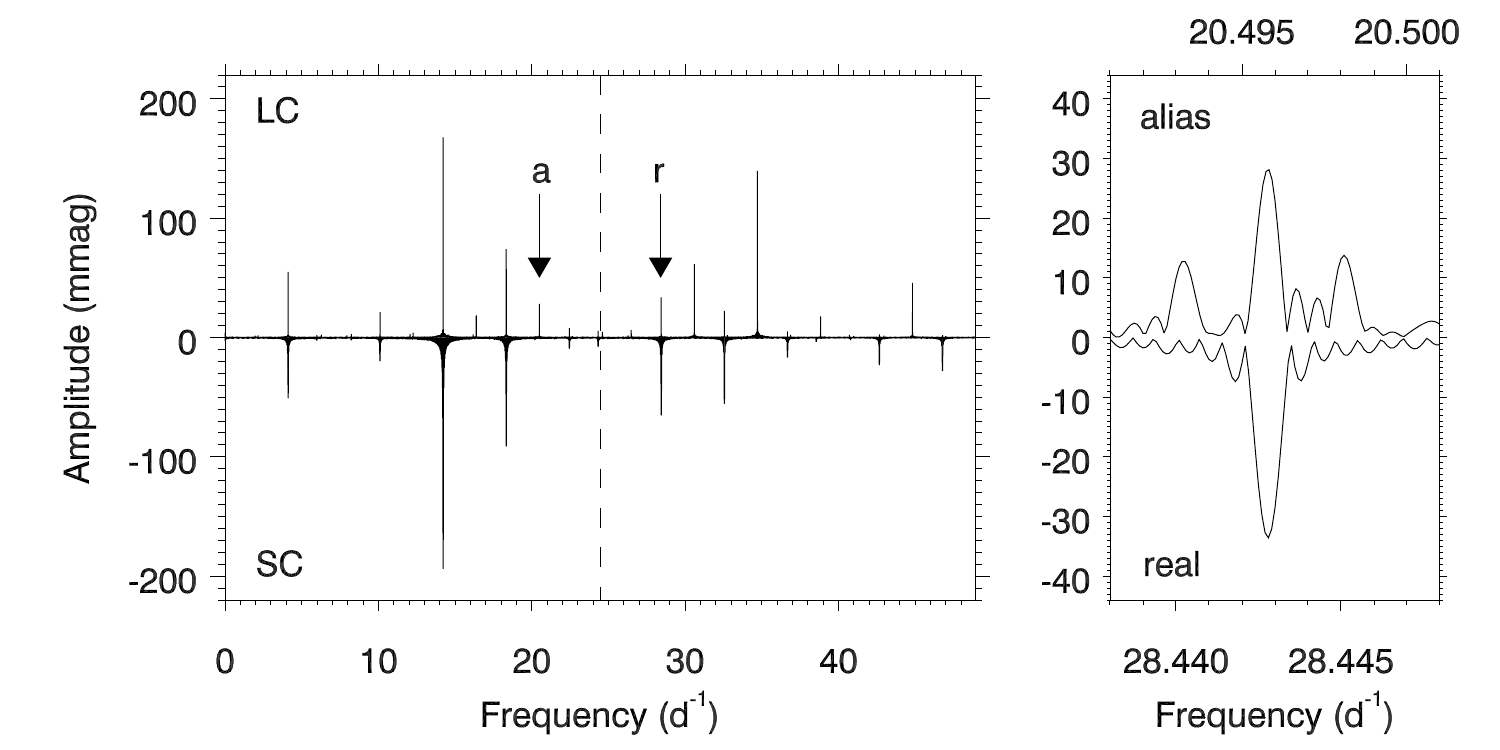}
		\includegraphics[width=0.4\textwidth]{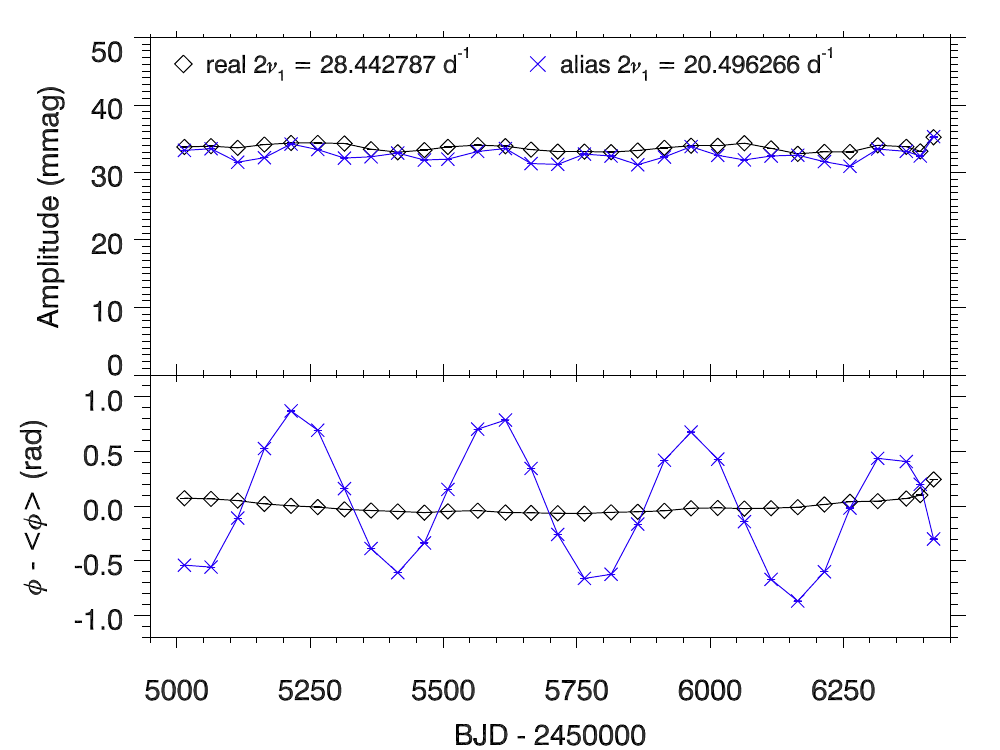}
		\caption{Demonstration of super-Nyquist asteroseismology with the HADS star KIC~5950759. Real and alias peaks associated with the harmonic of the fundamental radial mode are marked by `r' and `a', respectively, in the LC amplitude spectrum given in the left panel. The LC Nyquist frequency is indicated by the vertical dashed line and the SC amplitude spectrum is shown below for comparison. The middle panel contains inserts of the LC amplitude spectrum showing the real peak below and the alias peak above. The alias peak is easily identified as its multiplet structure is split by the \Kepler orbital frequency. The right panel shows the amplitude and phase tracking plot revealing the periodic phase modulation of the alias peak created from the Barycentric time-stamp corrections made to \Kepler data, whereas the real peak has approximately constant phase. Some peaks that exist in the SC amplitude spectrum do not appear in the LC amplitude spectrum as they lie close to the LC sampling frequency and are heavily suppressed in amplitude.}
		\label{figure: sNa peaks}	
	\end{figure*}

	To demonstrate the super-Nyquist asteroseismology (sNa) technique described by \citet{Murphy2013a}, we have applied our amplitude and phase tracking method to a real frequency peak and its super-Nyquist alias in the HADS star KIC~5950759. This HADS star acts as a useful example because of the high $S/N$ in its pulsation mode frequencies and because simultaneous LC and SC are available. The LC and SC amplitude spectra for KIC~5950759 are given in the left panel of Fig.~\ref{figure: sNa peaks}, which contain the fundamental radial mode at $\nu_1 = 14.221372$~d$^{-1}$, and its harmonic $2\nu_{\rm 1,r} = 28.442744$~d$^{-1}$ labelled with `r', which lies above the LC Nyquist frequency indicated by a vertical dashed line. The alias of the harmonic $\nu_{\rm 1,a} = 20.496203$~d$^{-1}$ is labelled `a'. The middle panel in Fig.~\ref{figure: sNa peaks} shows a zoom-in of the amplitude spectrum using LC data, showing the multiplet structure split by the \Kepler satellite's orbital frequency of the alias peak in the top panel, compared to the real peak shown below for comparison.
	
	The right panel of Fig.~\ref{figure: sNa peaks} shows the results of the tracking method for the real and alias harmonics of the fundamental radial mode frequency. The alias peak experiences phase modulation with a peak-to-peak amplitude of approximately $\pi / 2$~rad and a period equal to the \Kepler satellite orbital period (372.5~d). The Barycentric correction to the data time stamps creates a variable Nyquist frequency. The reflection of a peak across the variable Nyquist frequency causes a variable alias peak with sidelobes split by the \Kepler orbital frequency and reduced central peak amplitude \citep{Murphy2013a}. The example of KIC~5950759 in Fig.~\ref{figure: sNa peaks} acts as a useful case study for other stars with sNa peaks that are extracted using the method described in section~\ref{section: method}. Aliases of real frequencies are easily identified from the 372.5-d periodic phase modulation (see also, figure~2 from \citealt{Murphy2014}). Frequencies identified as super-Nyquist aliases are labelled as sNa in figure captions and in Table~\ref{table: all stars}.
	
	\subsection{Constant amplitudes and phases: NoMod stars}
	\label{subsection: constant stars}
	
\begin{figure*}
		\centering
		
		\includegraphics[width=0.49\textwidth]{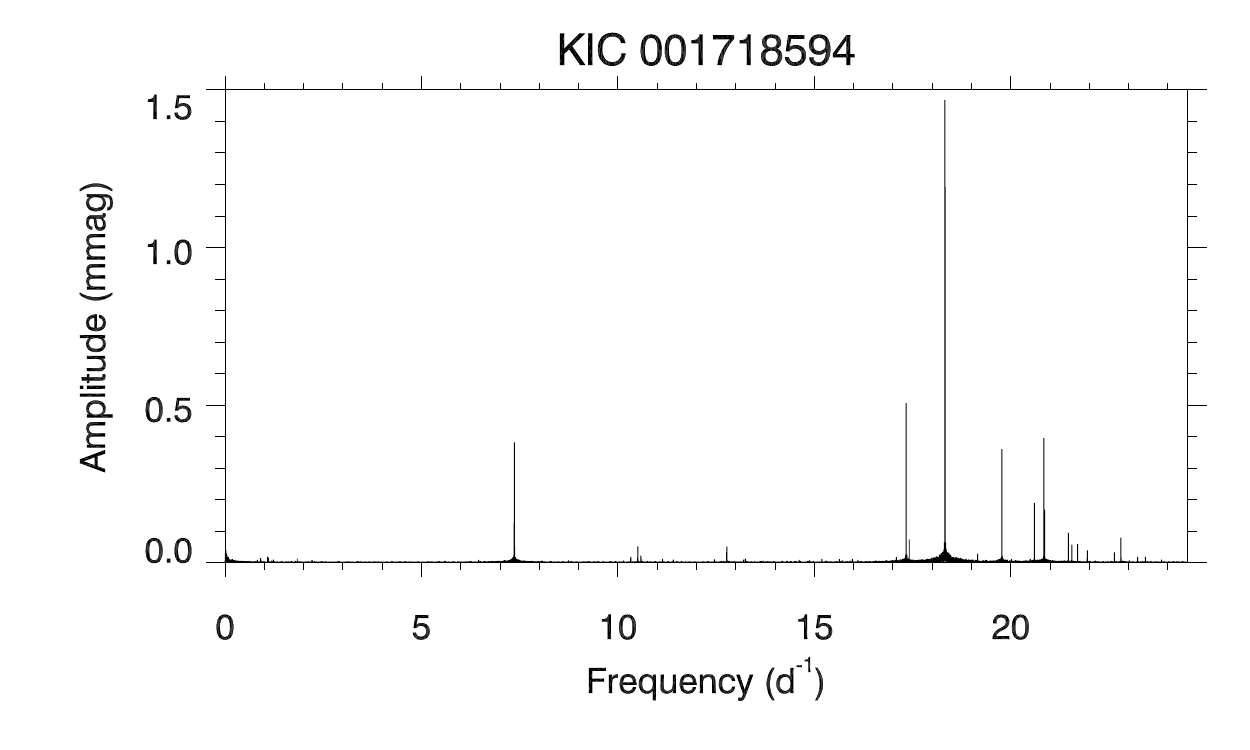}
		\includegraphics[width=0.49\textwidth]{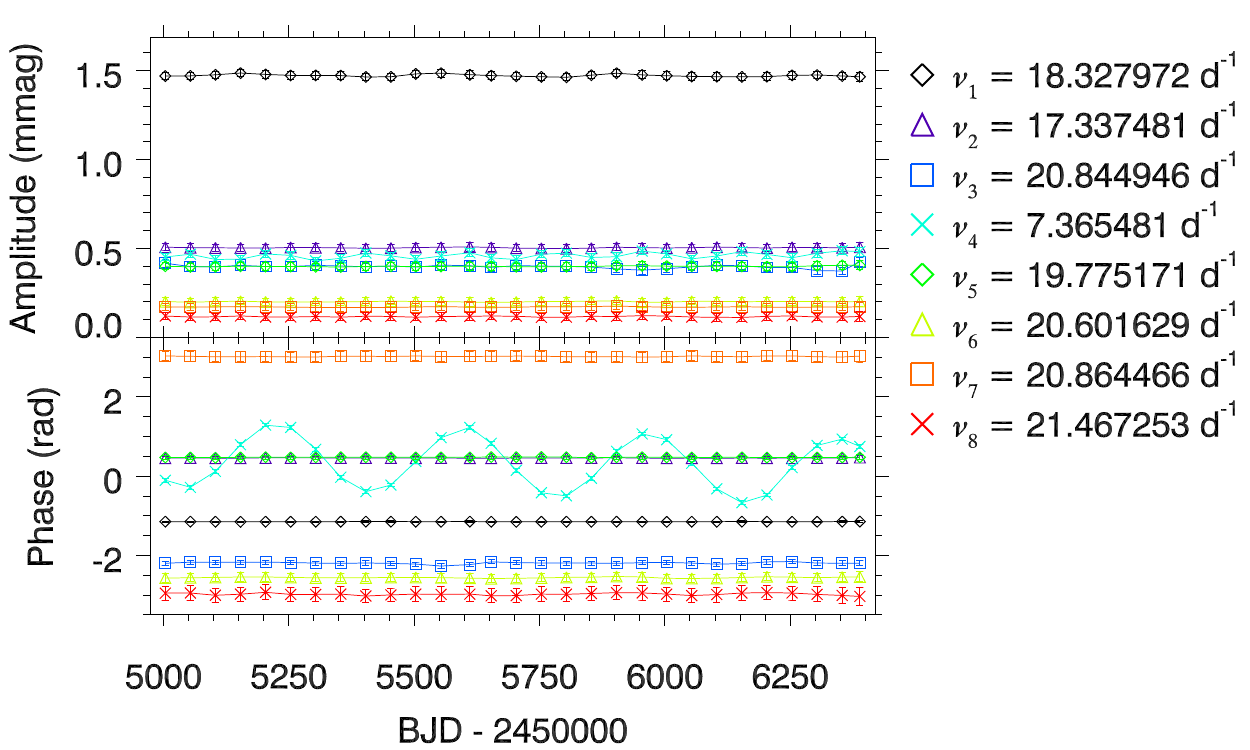}
		
		\vspace{0.5cm}	

		\includegraphics[width=0.49\textwidth]{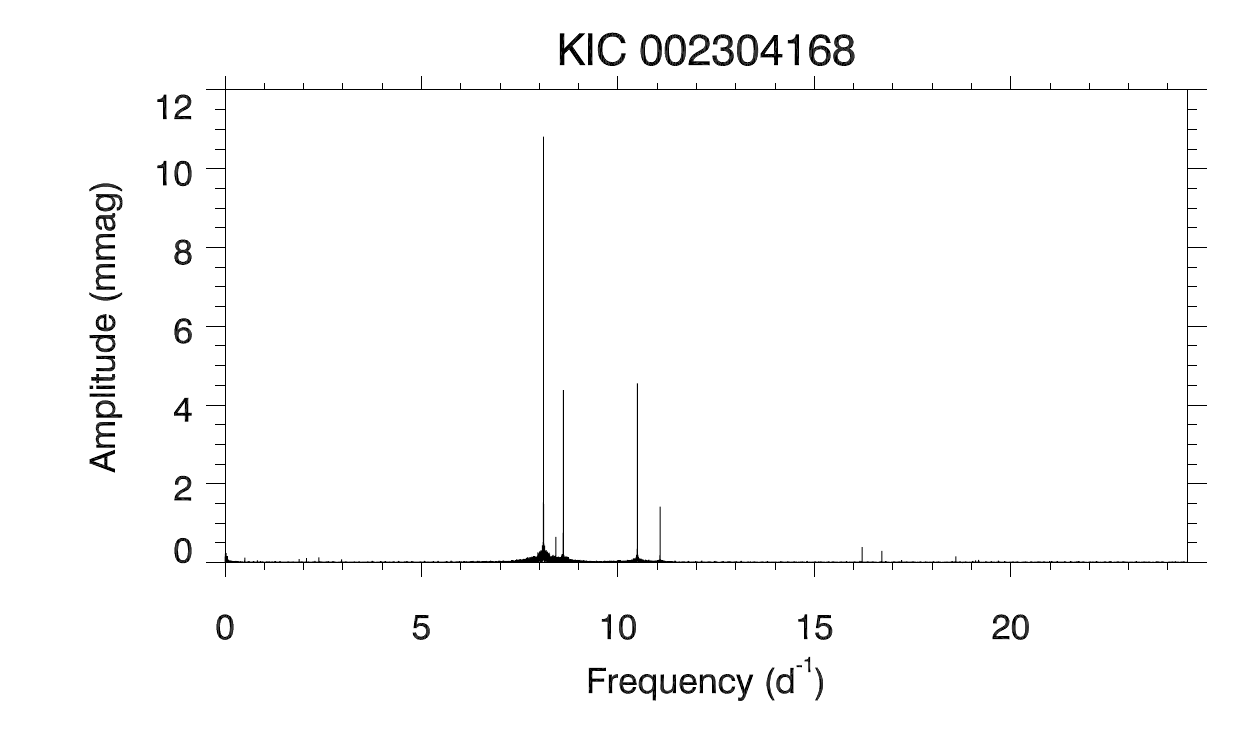}
		\includegraphics[width=0.49\textwidth]{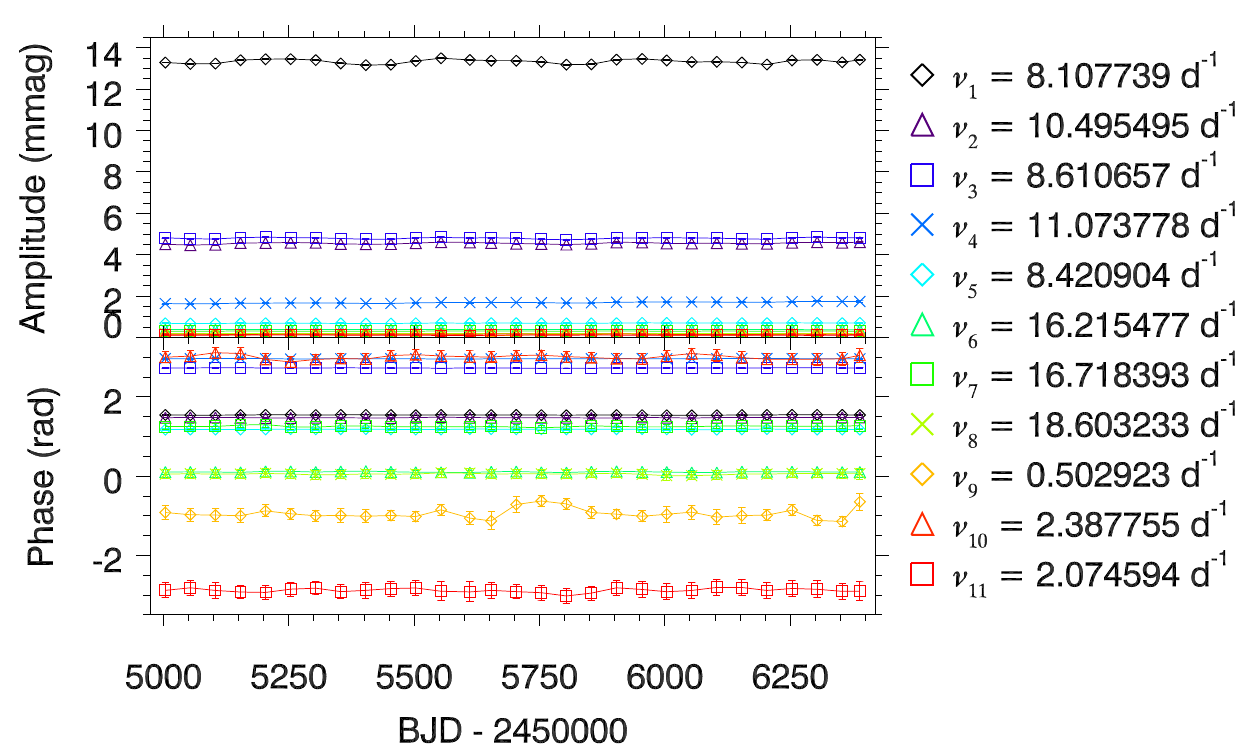}
		
		\vspace{0.5cm}	
		
		\includegraphics[width=0.49\textwidth]{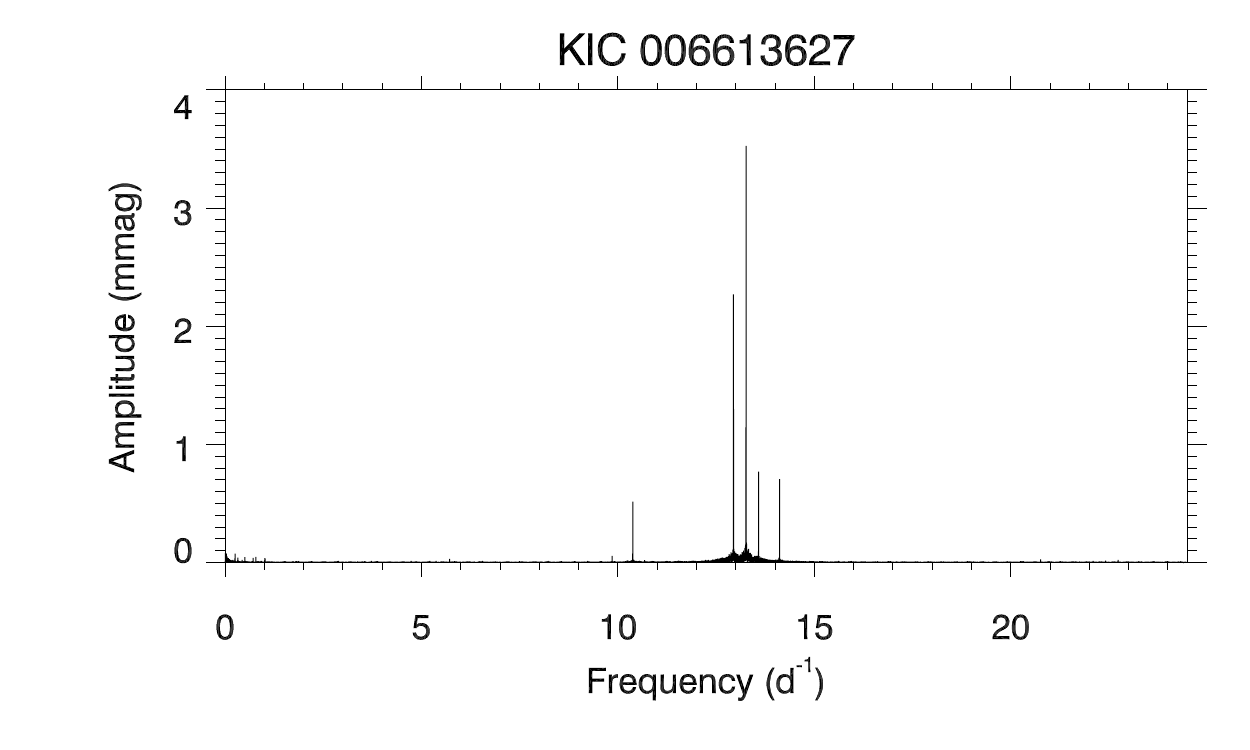}
		\includegraphics[width=0.49\textwidth]{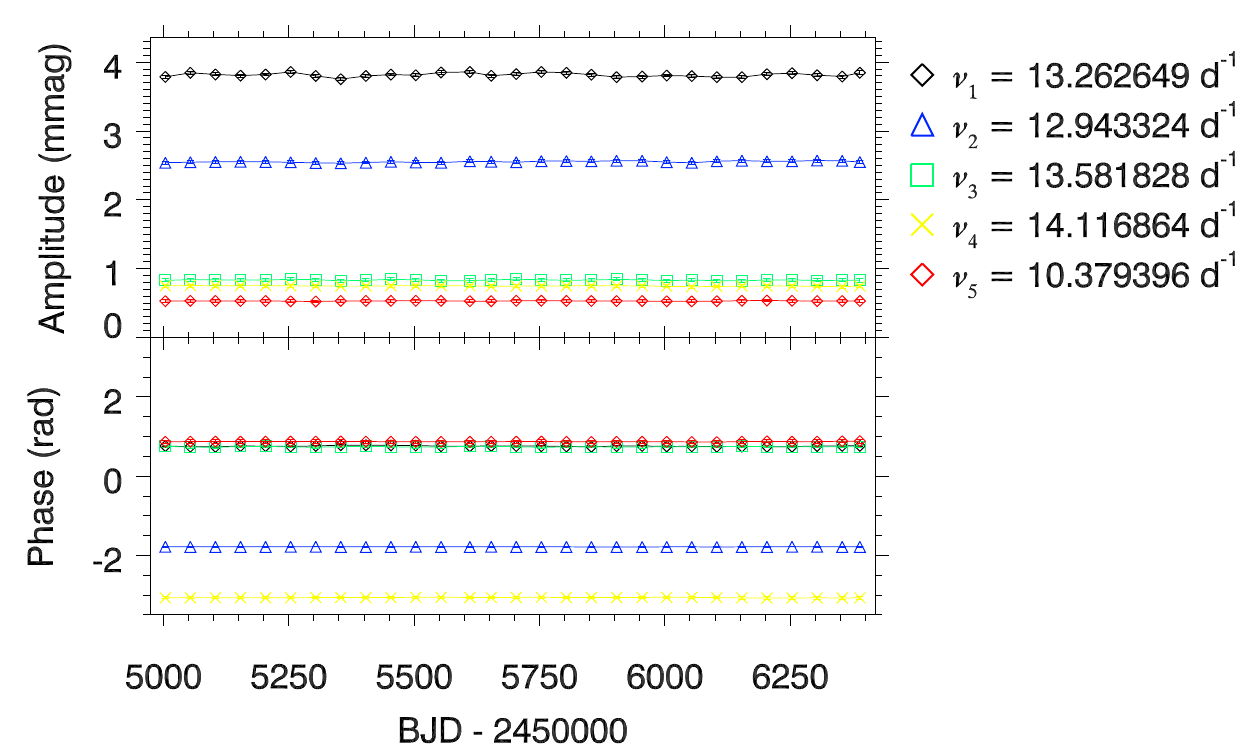}
		
		\vspace{0.5cm}
		
		\includegraphics[width=0.49\textwidth]{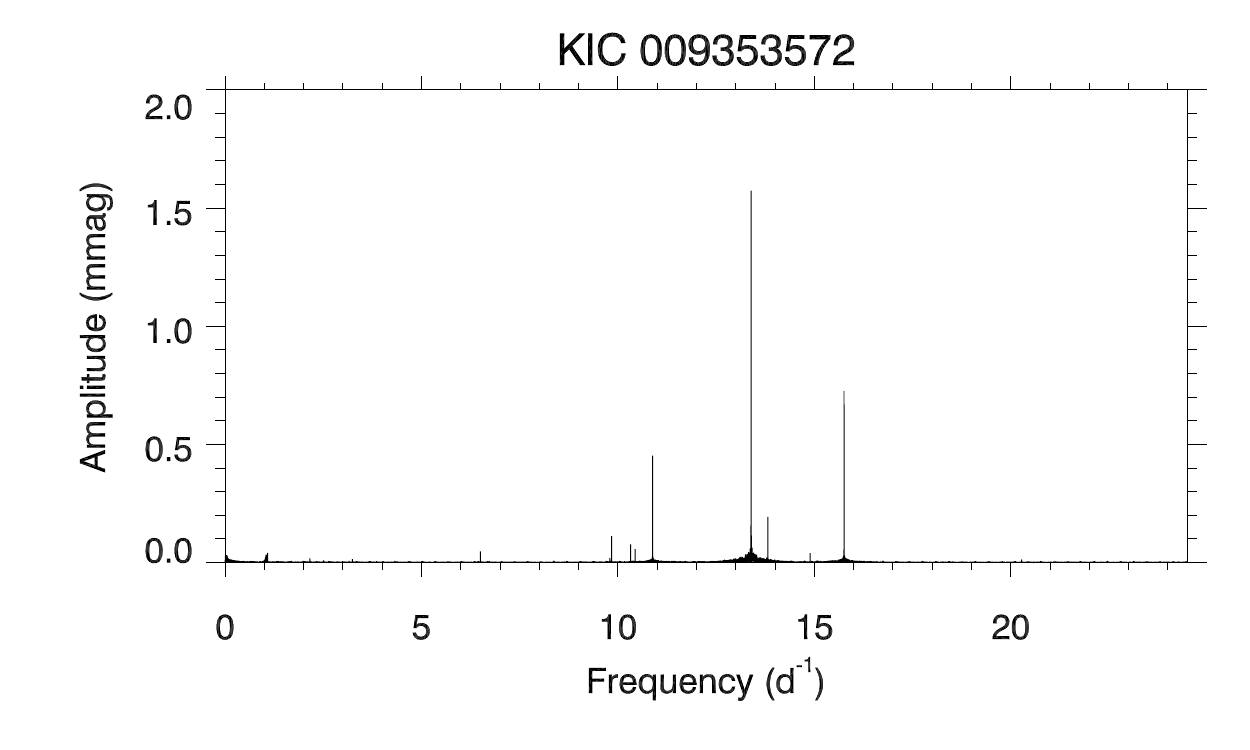}
		\includegraphics[width=0.49\textwidth]{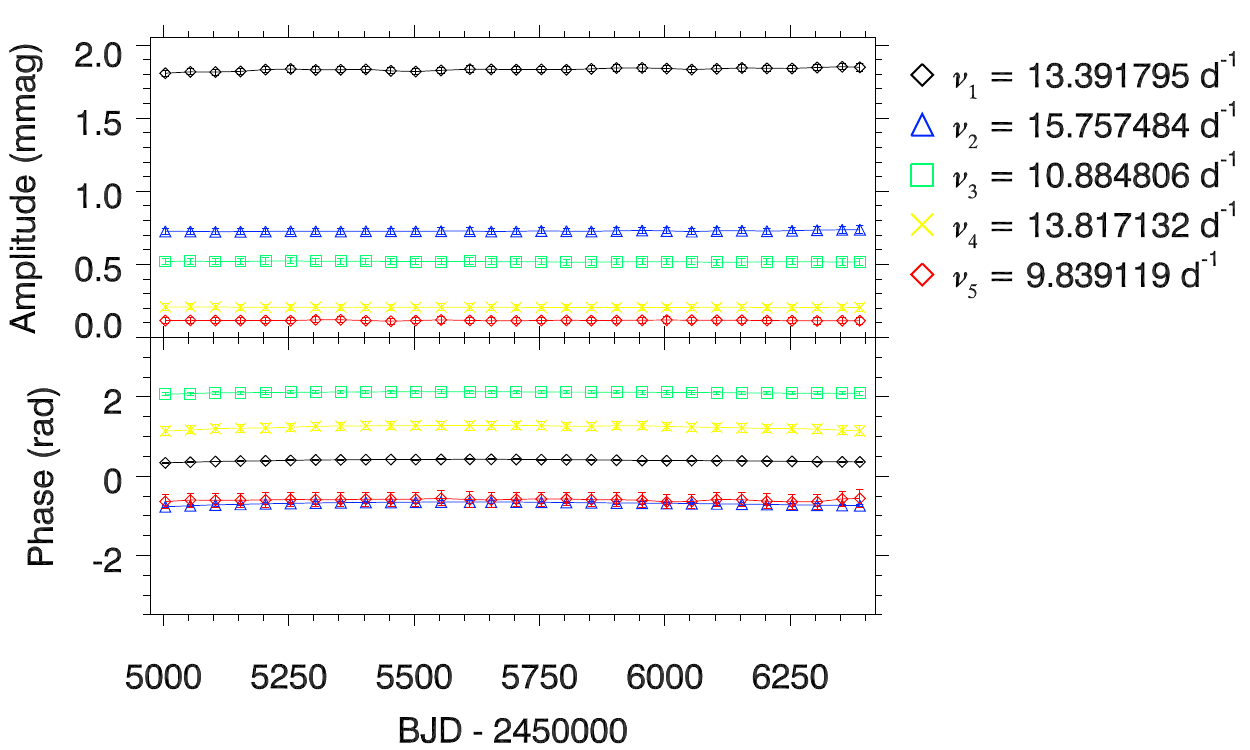}
		
		\caption{Four examples of NoMod \dsct stars in the KIC range $6400 \leq T_{\rm eff} \leq 10\,000$~K that show constant amplitudes and phases over the 4-yr \Kepler data set. From top to bottom: KIC~1718594 ($\nu_4$ is a super-Nyquist alias), KIC~2304168, KIC~6613627 and KIC~9353572. The left panels are the 4-yr amplitude spectra calculated out to the LC Nyquist frequency. The right panels show the amplitude and phase tracking plots which demonstrate the lack of variability in pulsation amplitudes and phases over 4~yr in each of these four NoMod stars.}
		\label{figure: constant stars}
		\end{figure*}

	There are {380} \dsct stars that have been classified as NoMod stars within our ensemble, thus {38.7}~per~cent of stars exhibit little or no change in their pulsation mode amplitudes. This subset supports the view that \dsct stars are coherent and periodic pulsators. Four examples of NoMod \dsct stars that exhibit constant-amplitude and constant-phase pulsation modes are shown in Fig.~\ref{figure: constant stars}.
	
	The \dsct star KIC~2304168 was studied by \citet{Balona2011g}, who used a subset of \Kepler data and asteroseismic modelling to identify the two principal pulsation mode frequencies (which they term $f_1 = 8.1055$~d$^{-1}$ and $f_2 = 10.4931$~d$^{-1}$) as the fundamental and first overtone radial modes, respectively. With a much longer data set available, we have re-analysed this star and calculated that the period ratio from 4~yr of \Kepler data for $\nu_1 = 8.107739$~d$^{-1}$ and $\nu_2 = 10.495495$~d$^{-1}$ is 0.7725, which is typically associated with the ratio of the first overtone and fundamental radial modes. We calculate pulsation constants for $\nu_1$ and $\nu_2$ as 0.028 and 0.022, respectively, which are consistent with the mode identification by \citet{Balona2011g} considering the typical uncertainties discussed by \citet{Breger1990b}. The first harmonic of $\nu_1$ is also present with significant amplitude, labelled as $\nu_6$ in the second row of Fig.~\ref{figure: constant stars}. KIC~2304168 is an excellent example of a NoMod \dsct star, as its amplitude spectrum has low mode density and mode identification is possible, but most importantly, all frequencies in its amplitude spectrum are completely stable over 4~yr. This raises the question: for stars of similar $T_{\rm eff}$, $\log\,g$ and $[{\rm Fe}/{\rm H}]$, what mechanism is driving amplitude modulation and nonlinearity in some \dsct stars, yet is absent in others?
	
		\subsection{Amplitude and phase modulation due to beating of close-frequency modes}
		\label{subsection: beating stars}
	
\begin{figure*}
		\centering
		
		\includegraphics[width=0.48\textwidth]{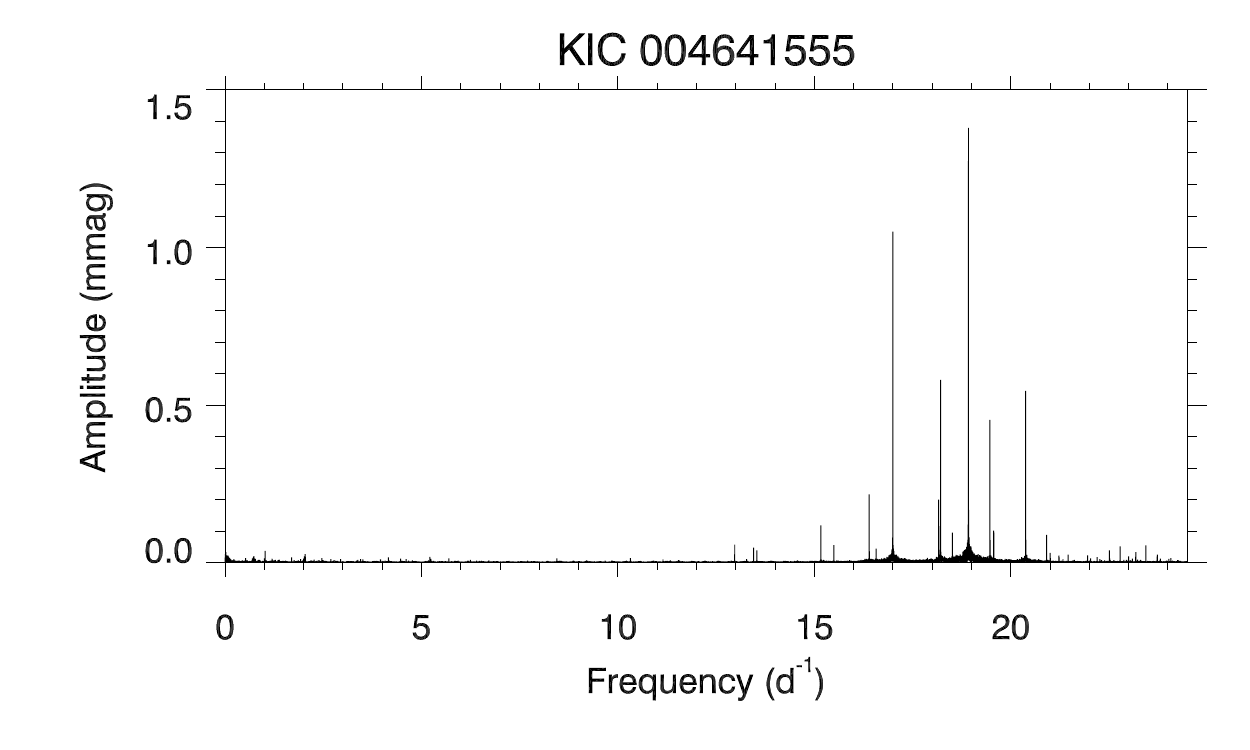}
		\hspace{0.5cm}
		\includegraphics[width=0.48\textwidth]{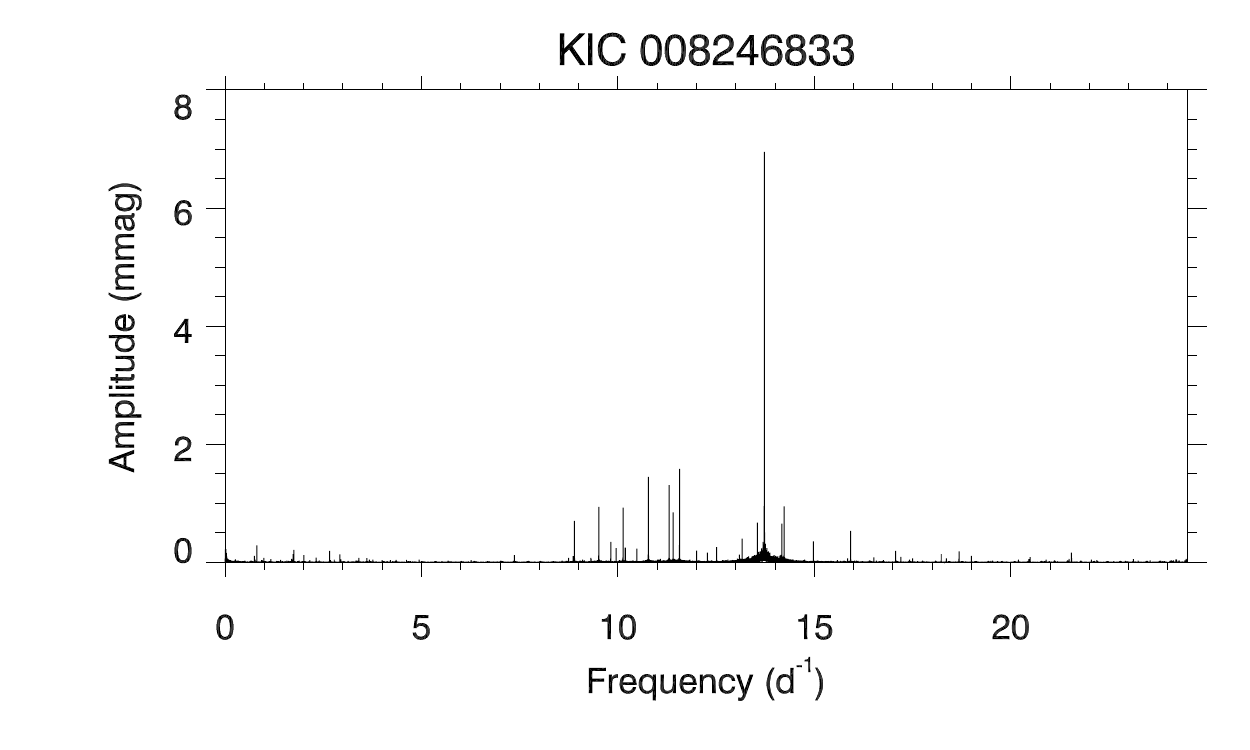}
		
		\includegraphics[width=0.48\textwidth]{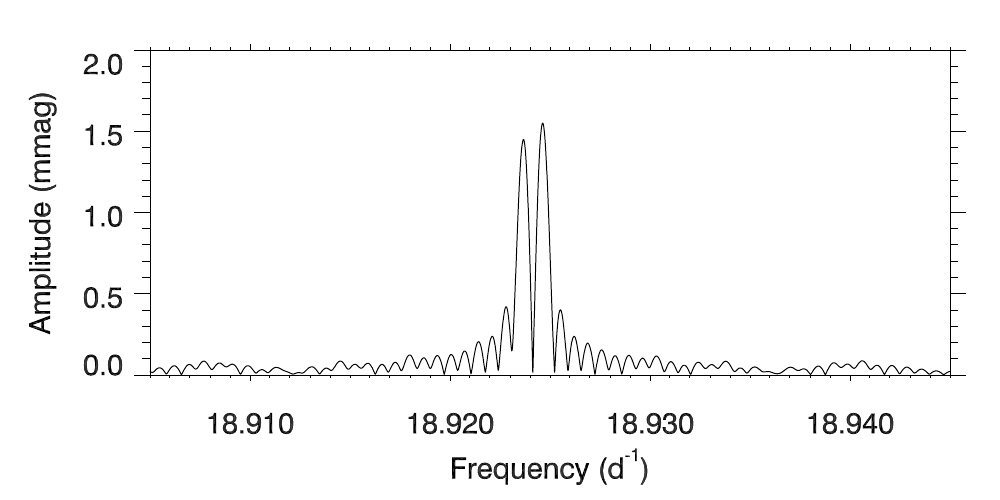}
		\hspace{0.5cm}
		\includegraphics[width=0.48\textwidth]{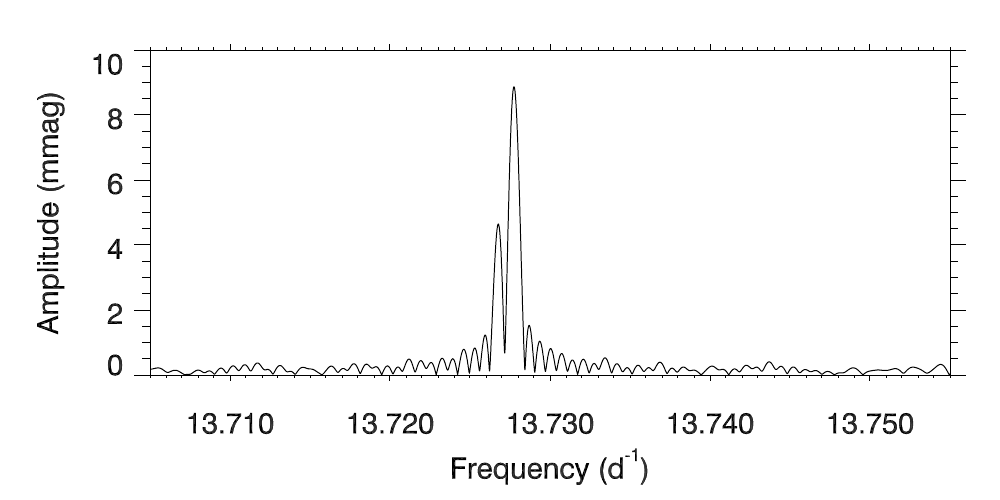}
		
		\vspace{0.5cm}		
		
		\includegraphics[width=0.46\textwidth]{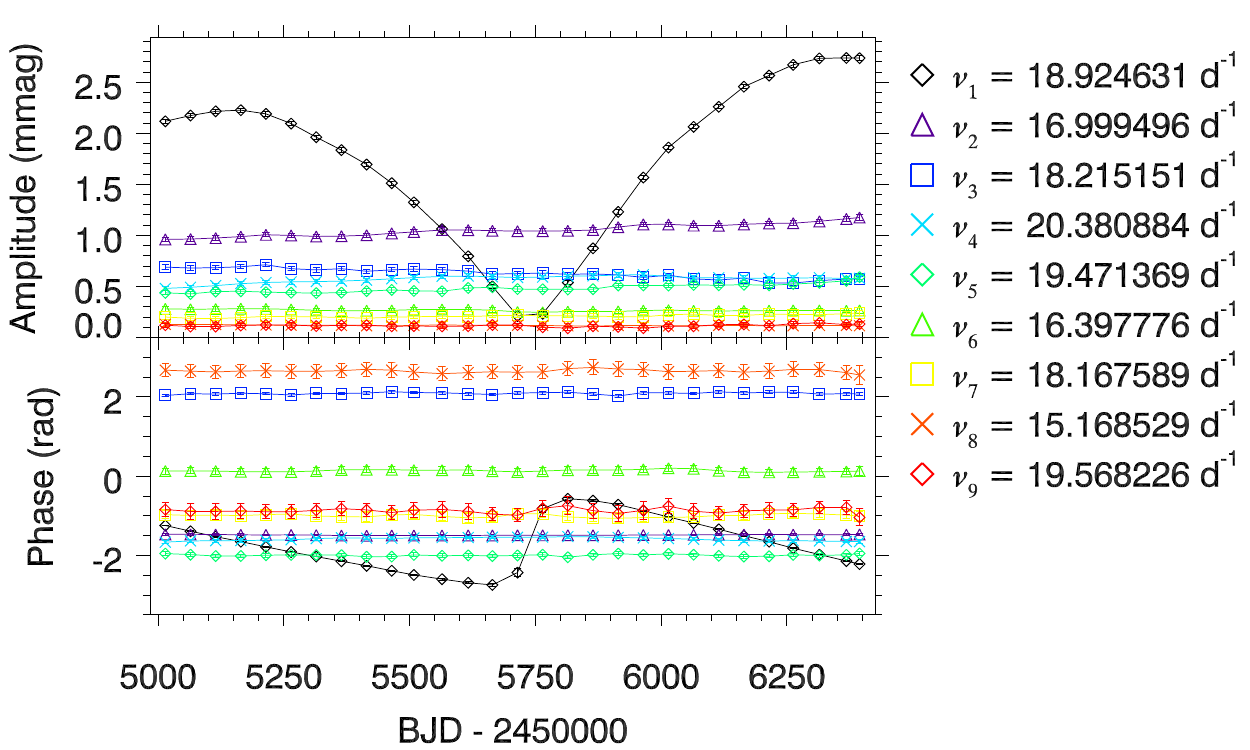}
		\hspace{1cm}
		\includegraphics[width=0.46\textwidth]{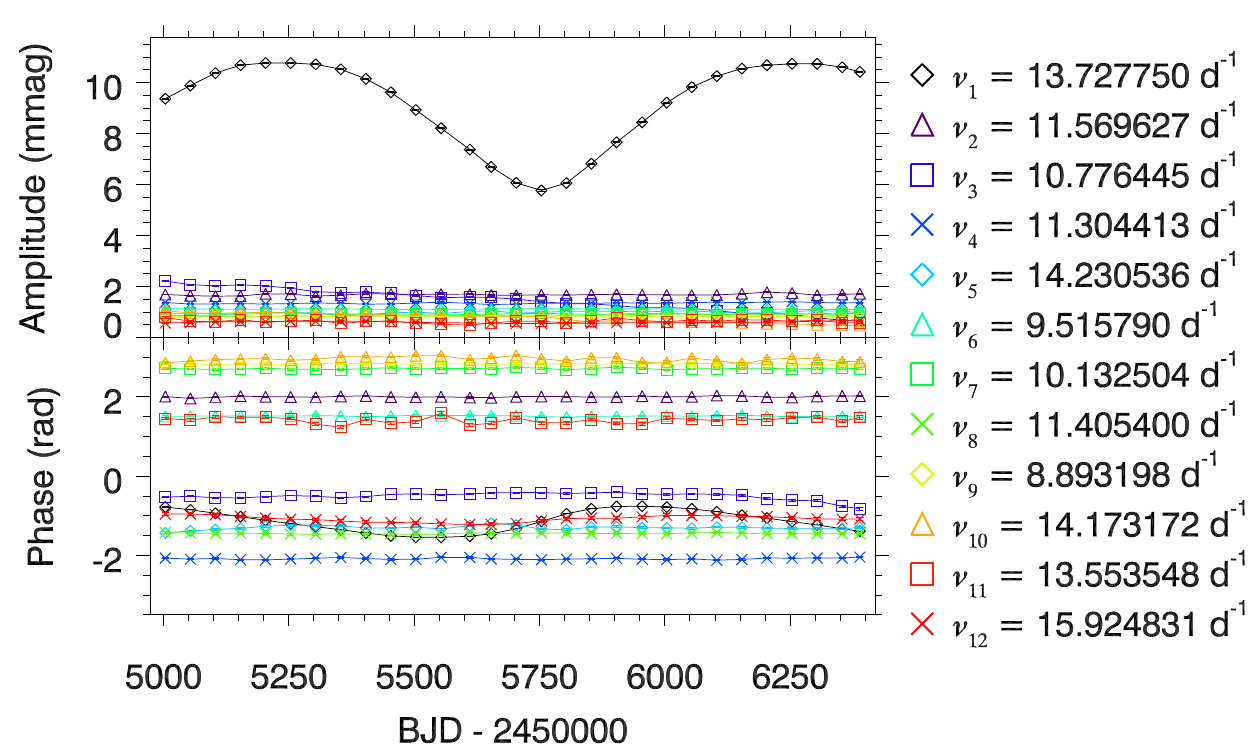}
		
		\vspace{0.5cm}
		
		\includegraphics[width=0.48\textwidth]{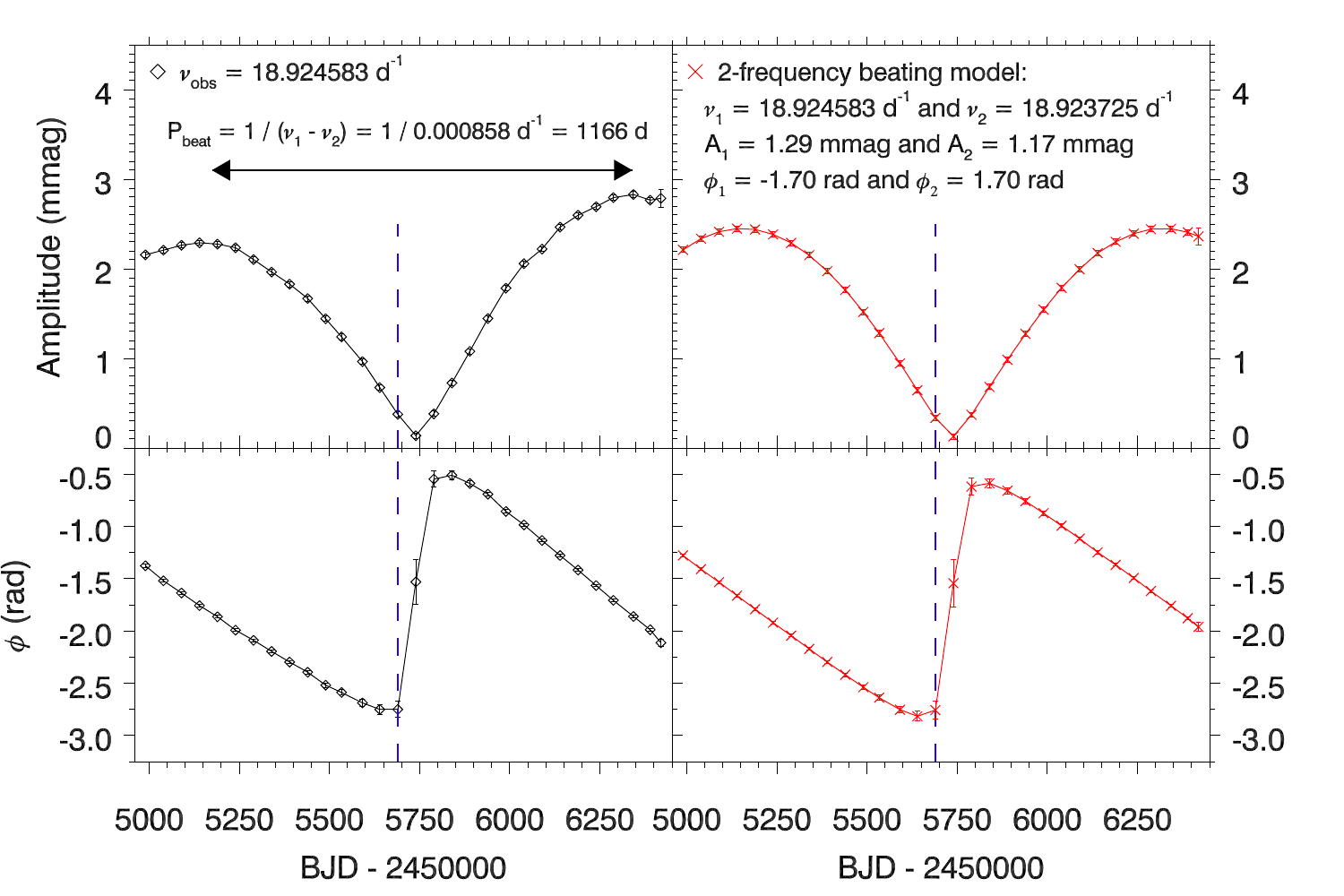}
		\hspace{0.5cm}
		\includegraphics[width=0.48\textwidth]{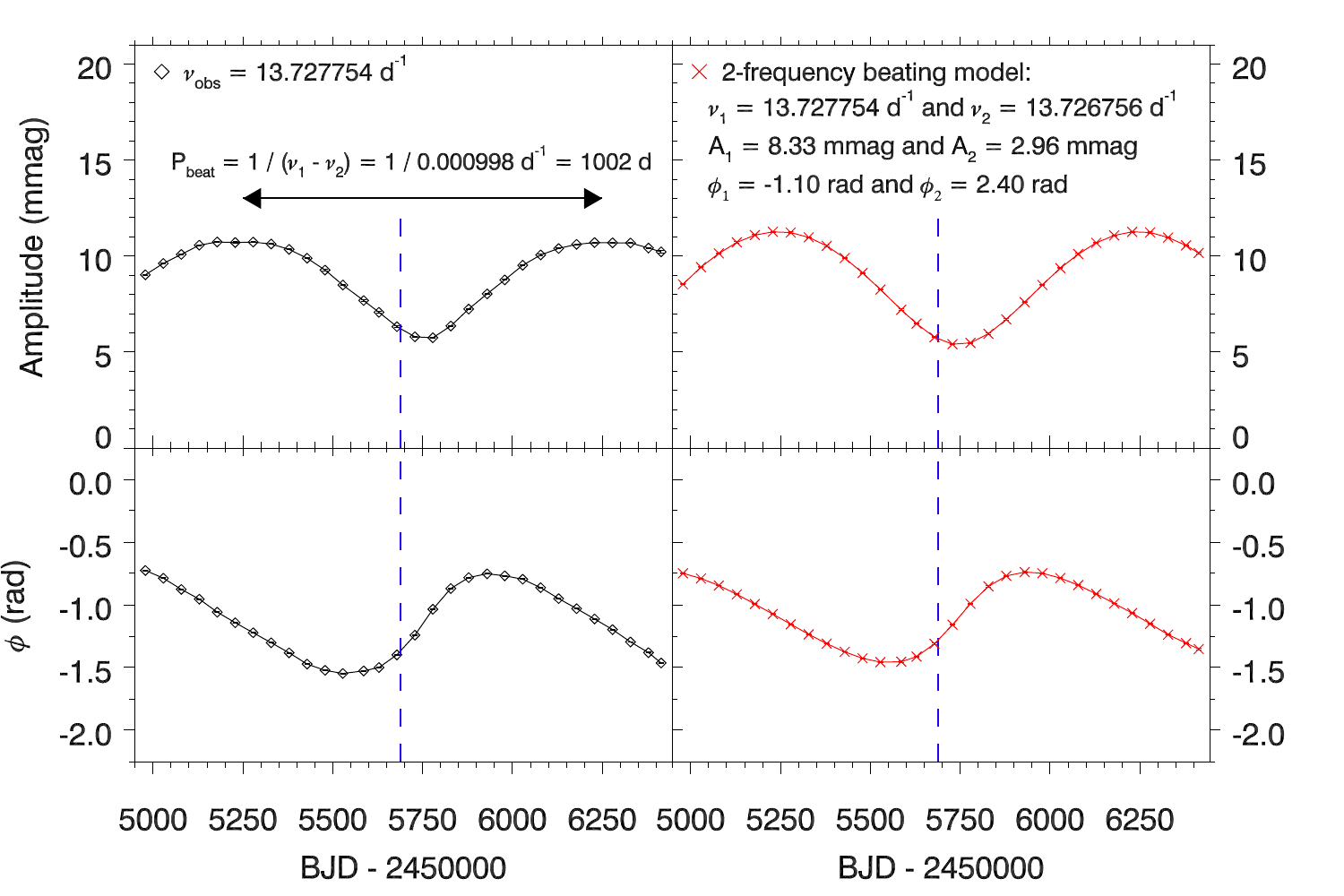}
		
		\caption{Two examples of AMod \dsct stars in the KIC range $6400 \leq T_{\rm eff} \leq 10\,000$~K that show variable pulsation amplitudes and phases over the 4-yr \Kepler data set from beating of extremely close-frequency modes. The left and right columns are KIC~4641555 and KIC~8246833, respectively. From top to bottom are the 4-yr amplitude spectrum calculated out to the LC Nyquist frequency; a zoom-in of the pair of close-frequency modes in the amplitude spectrum; the amplitude and phase tracking plots; and the accurate beating model (shown as crosses) matching the observed amplitude modulation (shown as diamonds). The dashed vertical line indicates the centre of the \Kepler data set that has been chosen as the zero point in time, specifically $t_0 = 2\,455\,688.770$~BJD. In each of these two example stars, KIC~4641555 (left column) and KIC~8246833 (right column), a pair of high-amplitude pulsation mode frequencies lie closer than 0.001~d$^{-1}$ in frequency, resulting in beat periods of $1166 \pm 1$~d and $1002 \pm 1$~d, respectively.}
		\label{figure: beating AMod stars}
		\end{figure*}

		We find two \dsct stars in our ensemble, KIC~4641555 and KIC~8246833, that are AMod from the beating of pulsation modes spaced closer than $0.001$~d$^{-1}$. This emphasises the superiority of \Kepler data purely because of the length: 4~yr is only just long enough to resolve these pulsation mode frequencies. The amplitude spectra and tracking plots for KIC~4641555 and KIC~8246833 are shown in the left and right columns of Fig.~\ref{figure: beating AMod stars}, respectively. We discuss how beating models of close-frequency pulsation modes can be used to distinguish between {pure} amplitude modulation and beating in section~\ref{section: beating model}.
		
		\subsection{{Pure} amplitude modulation with no phase variability}
		\label{subsection: pure AMod stars}

\begin{figure*}
		\centering

		\includegraphics[width=0.49\textwidth]{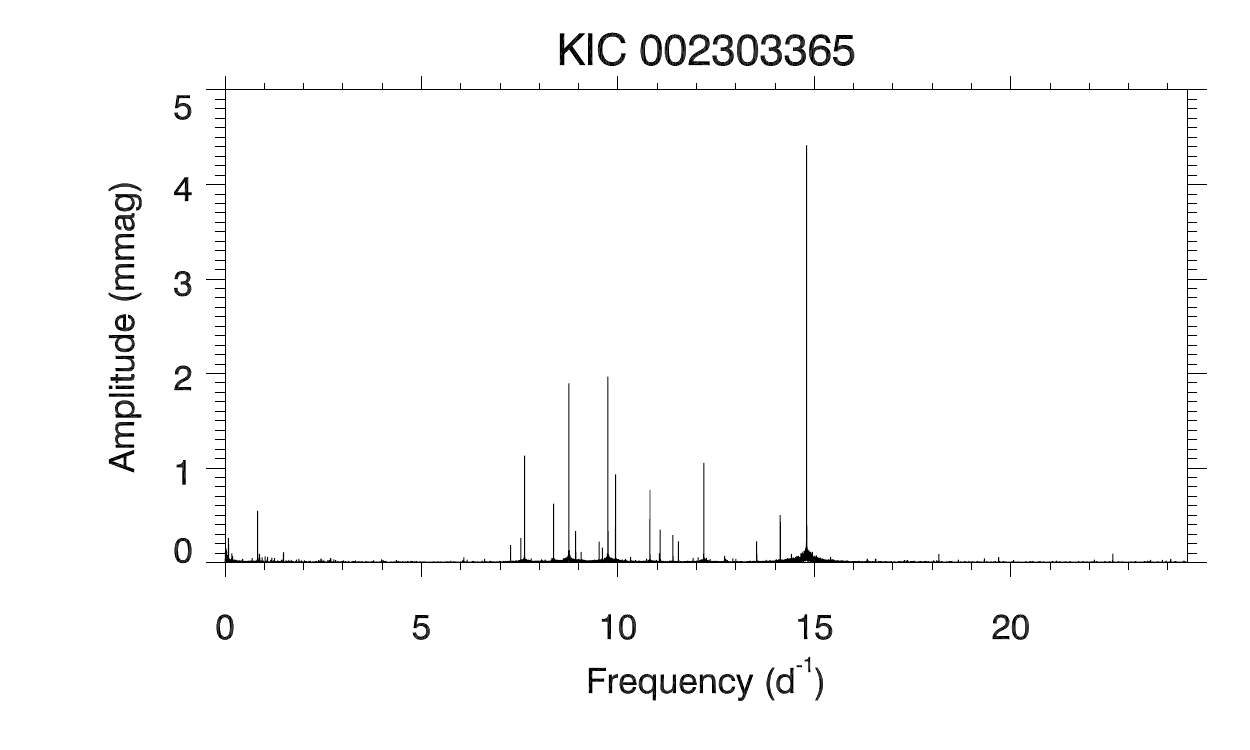}
		\includegraphics[width=0.49\textwidth]{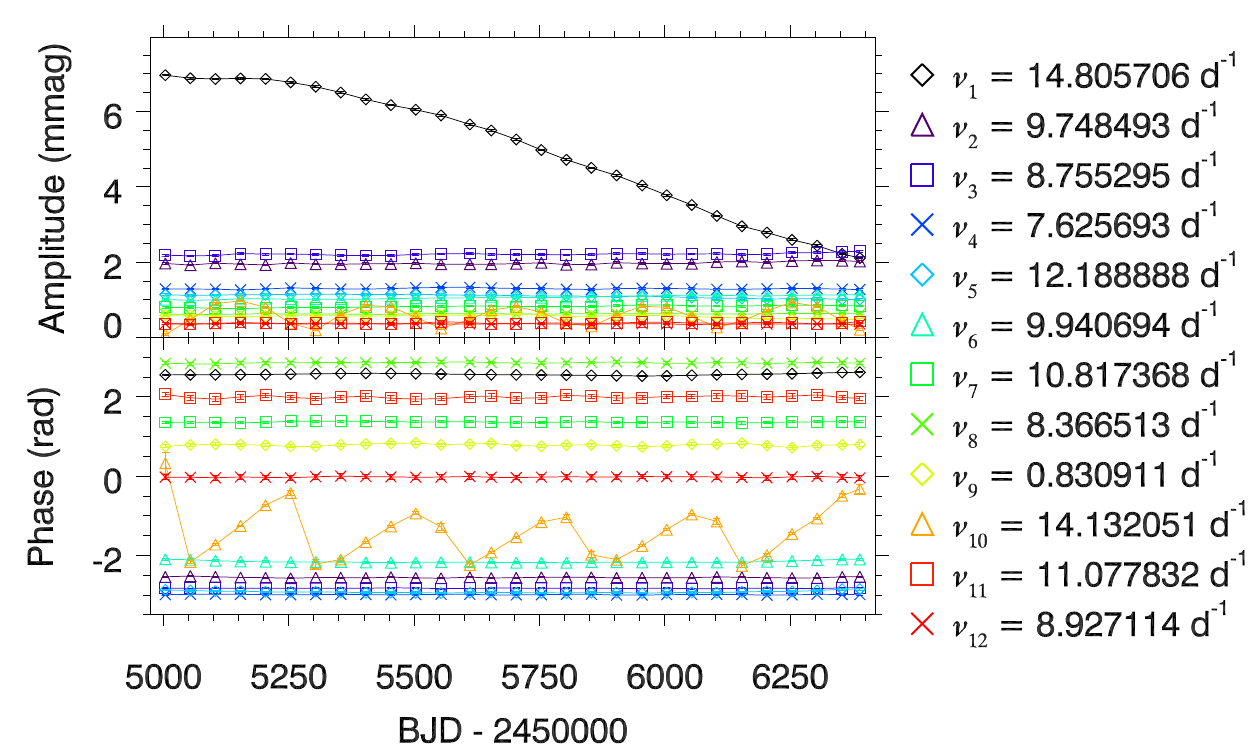}
		
		\vspace{0.4cm}	
				
		\includegraphics[width=0.49\textwidth]{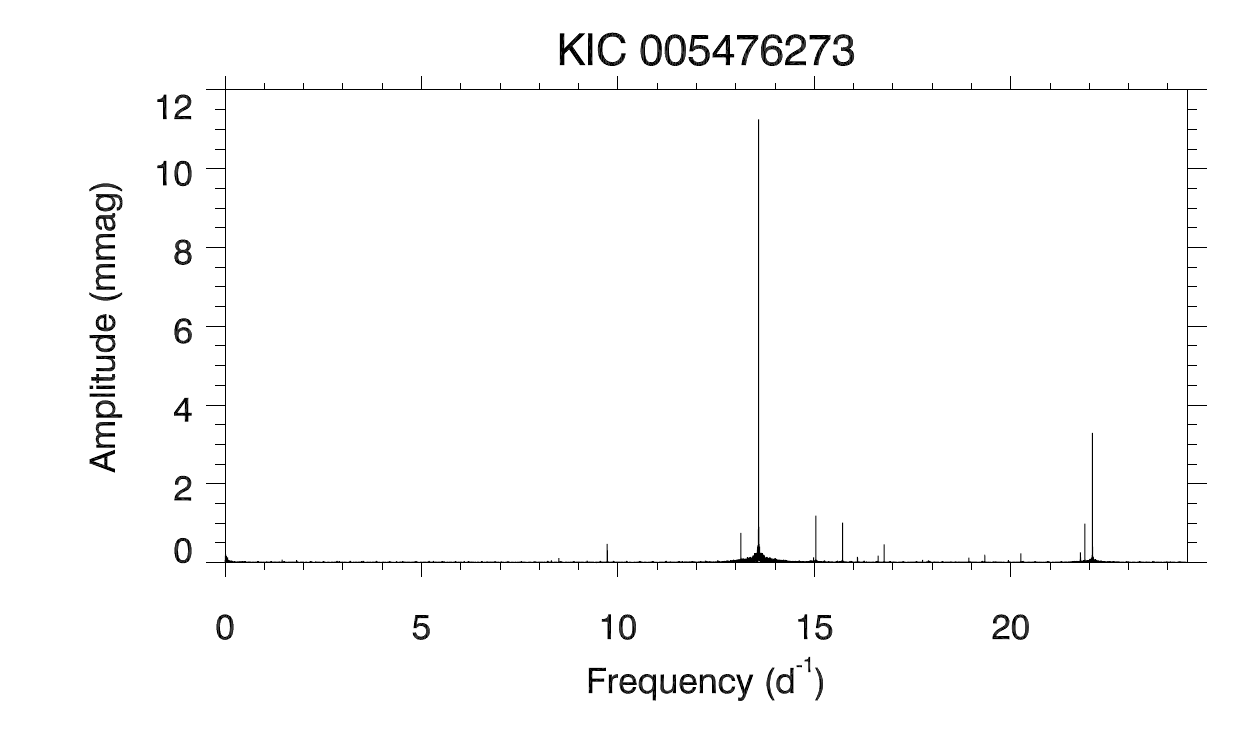}
		\includegraphics[width=0.49\textwidth]{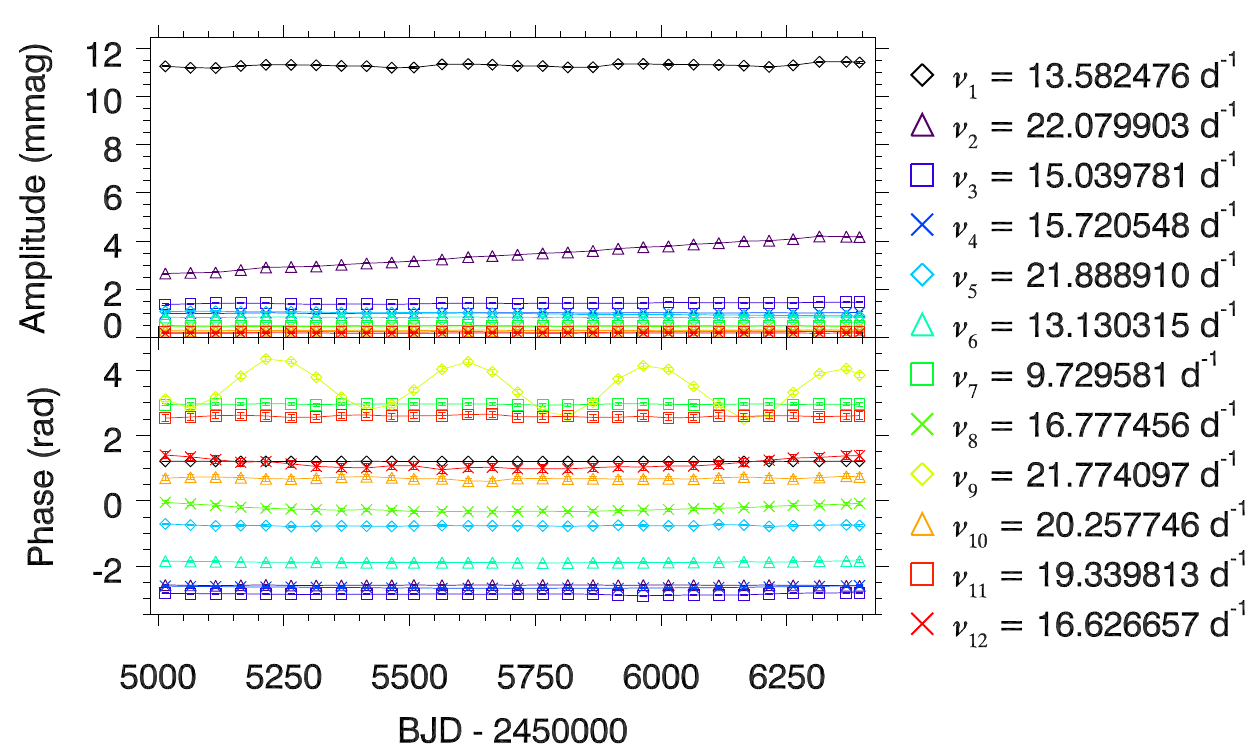}
		
		\vspace{0.4cm}	
		
		\includegraphics[width=0.49\textwidth]{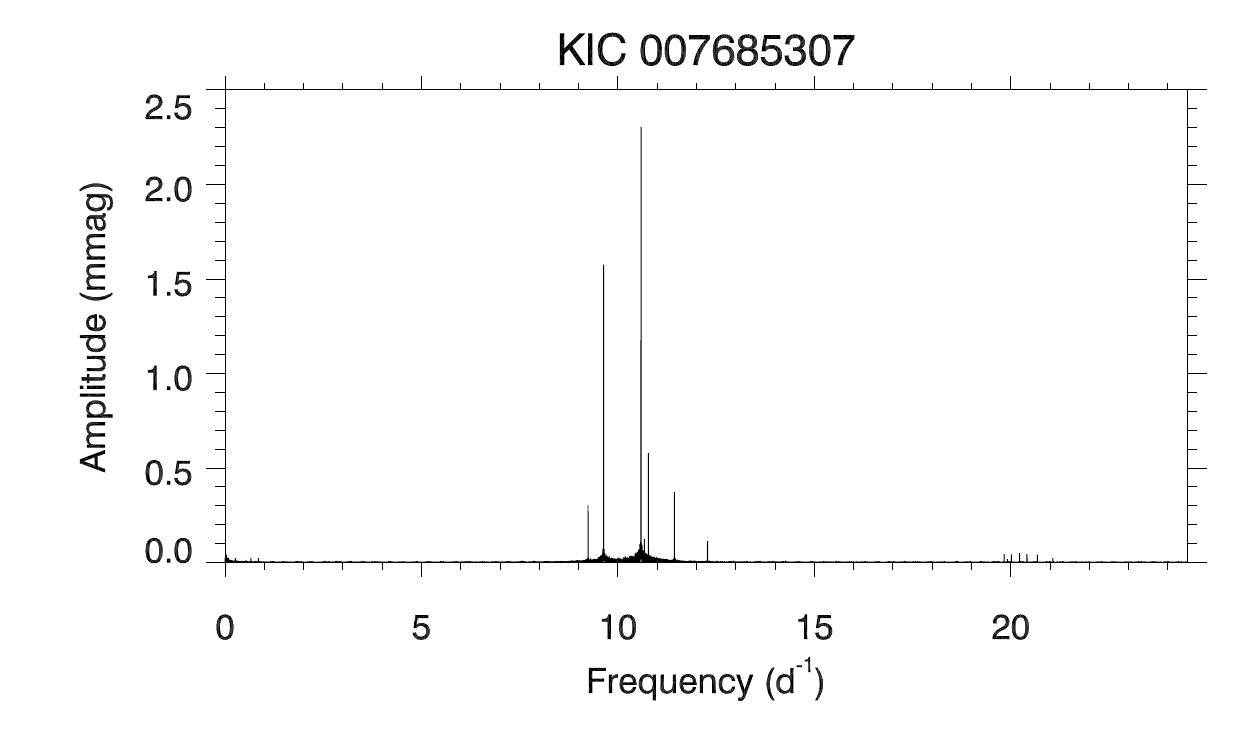}
		\includegraphics[width=0.49\textwidth]{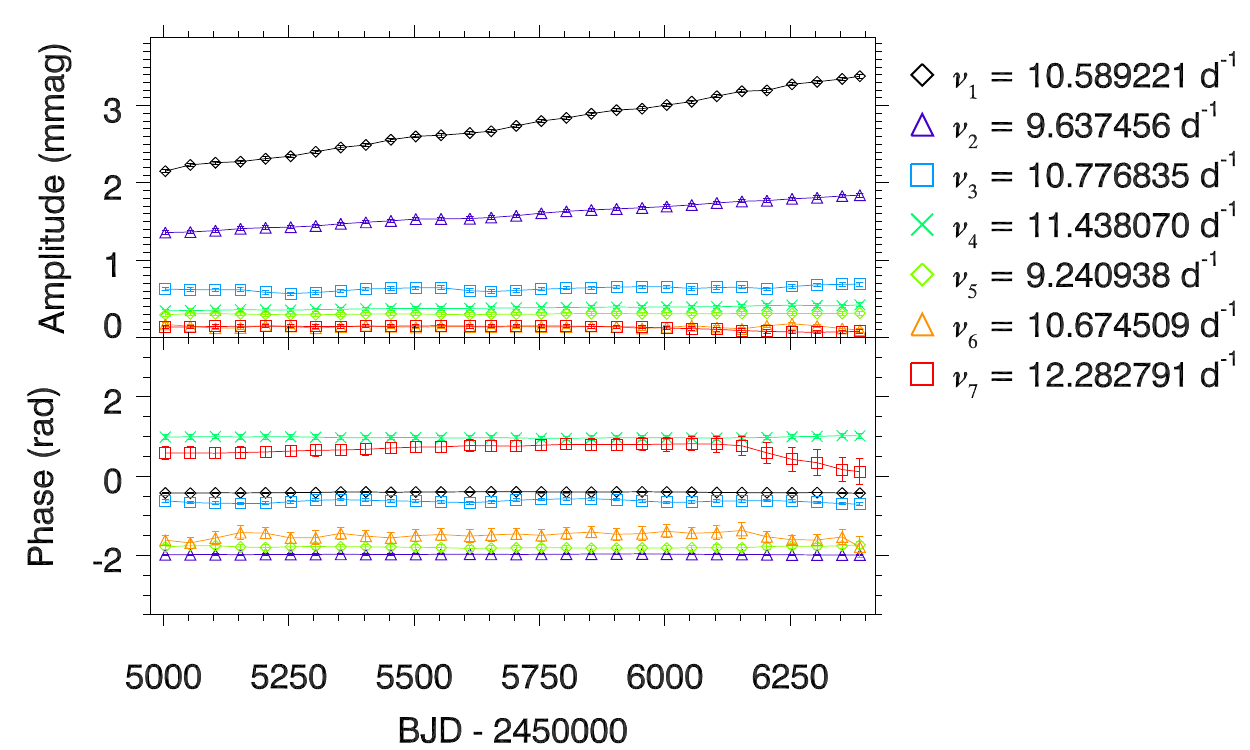}
		
		\vspace{0.4cm}	
		
		\includegraphics[width=0.49\textwidth]{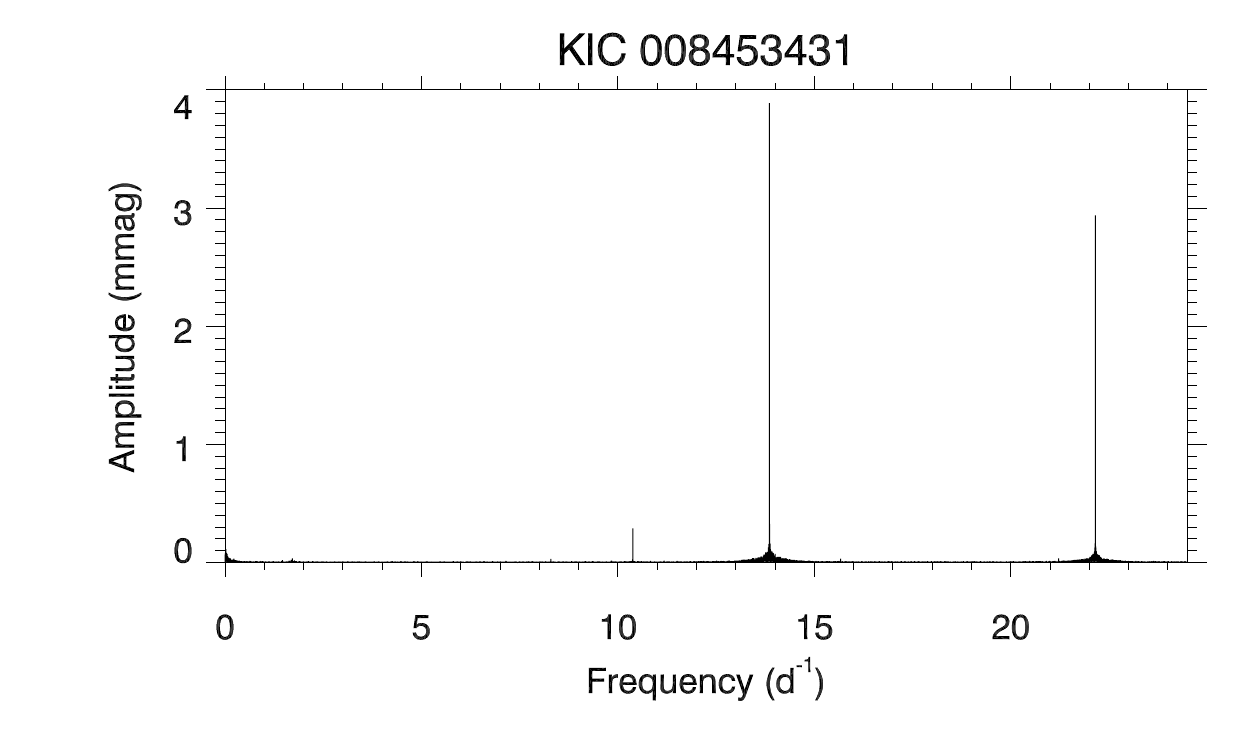}
		\includegraphics[width=0.49\textwidth]{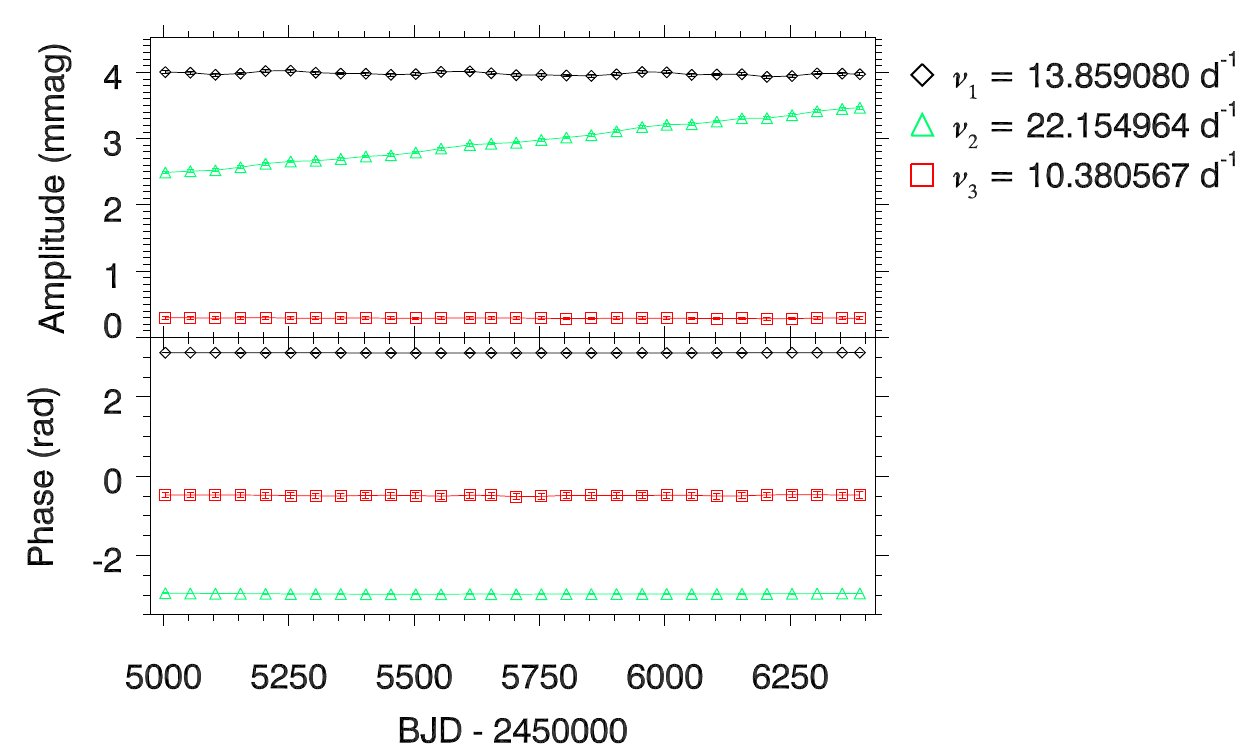}

		\caption{Four examples of pure AMod \dsct stars in the KIC range $6400 \leq T_{\rm eff} \leq 10\,000$~K that show variable pulsation amplitudes with constant phases over the 4-yr \Kepler data set. From top to bottom: KIC~23093365 ($\nu_1$ is pure AMod and $\nu_{10}$ contains a $\sim$250-d beat signal), KIC~5476273 ($\nu_2$ is pure AMod and $\nu_9$ is a super-Nyquist alias), KIC~7685307 ($\nu_1$ and $\nu_2$ are pure AMod) and KIC~8453431 ($\nu_2$ is a pure AMod frequency). The left panels are the 4-yr amplitude spectra calculated out to the LC Nyquist frequency. The right panels show the amplitude and phase tracking plots which demonstrate the modulation in pulsation amplitudes but constant phases over 4~yr. }
		\label{figure: pure AMod stars}
		\end{figure*}

		In this subsection, we present four examples of \dsct stars that exhibit amplitude modulation with no change in phase in at least one pulsation mode. This {pure} form of amplitude modulation is unlikely to be caused by beating of unresolved pulsation modes or mode coupling because no phase modulation is observed, which is required by both mechanisms. This subgroup contains perhaps the most interesting \dsct stars, with no obvious selection effect that determines why some stars do this and others do not. At this stage, we conjecture that {pure} amplitude modulation could be caused by variable driving and/or damping within a star. Therefore, it remains an unsolved problem why this occurs. Four examples of {pure} AMod \dsct stars are shown in Fig.~\ref{figure: pure AMod stars}. 
				
		A good example of a {pure} AMod star is KIC~8453431, which is shown in the bottom row of Fig.~\ref{figure: pure AMod stars}, as it only contains three pulsation mode frequencies with amplitudes greater than 0.01~mmag, specifically $\nu_1 = 13.859080$~d$^{-1}$, $\nu_2 = 22.154964$~d$^{-1}$ and $\nu_3 = 10.380567$~d$^{-1}$. The period ratio of $\nu_2$ and $\nu_1$ gives 0.6256 and the period ratio of $\nu_3$ and $\nu_1$ gives 0.7490, but calculation of pulsation constants using eq.~\ref{equation: stellingwerf} suggest that neither $\nu_3$ or $\nu_1$ is the fundamental radial mode. Therefore, we conclude that these three mode frequencies are likely low-overtone radial modes considering the typical uncertainties associated with calculating $Q$ values \citep{Breger1990b}. Regardless of the method for mode identification, only a single pulsation mode frequency, $\nu_2$, slowly increases in amplitude whilst staying at constant phase.
		
		\subsection{Phase modulation due to binarity}
		\label{subsection: binary stars}

		Binarity within a stellar system causes all the pulsation mode frequencies to be phase modulated with the same period, i.e. the orbital period of the star. A previously confirmed binary \dsct star included in our study is KIC~9651065, which was analysed using the PM technique by \citet{Murphy2014}, who calculated an orbital period of $P_{\rm orb} = 272.7 \pm 0.8$~d. Later, \citet{Shibahashi2015} used the FM technique to study KIC~9651065 and found pulsation mode frequencies with first, second and third FM sidelobes in the amplitude spectrum, meaning that the star is highly eccentric. An eccentricity of $e = 0.569 \pm 0.030$ was calculated from the amplitude ratio of the FM sidelobes \citep{Shibahashi2015}.
		
\begin{figure*}
		\centering

		\includegraphics[width=0.49\textwidth]{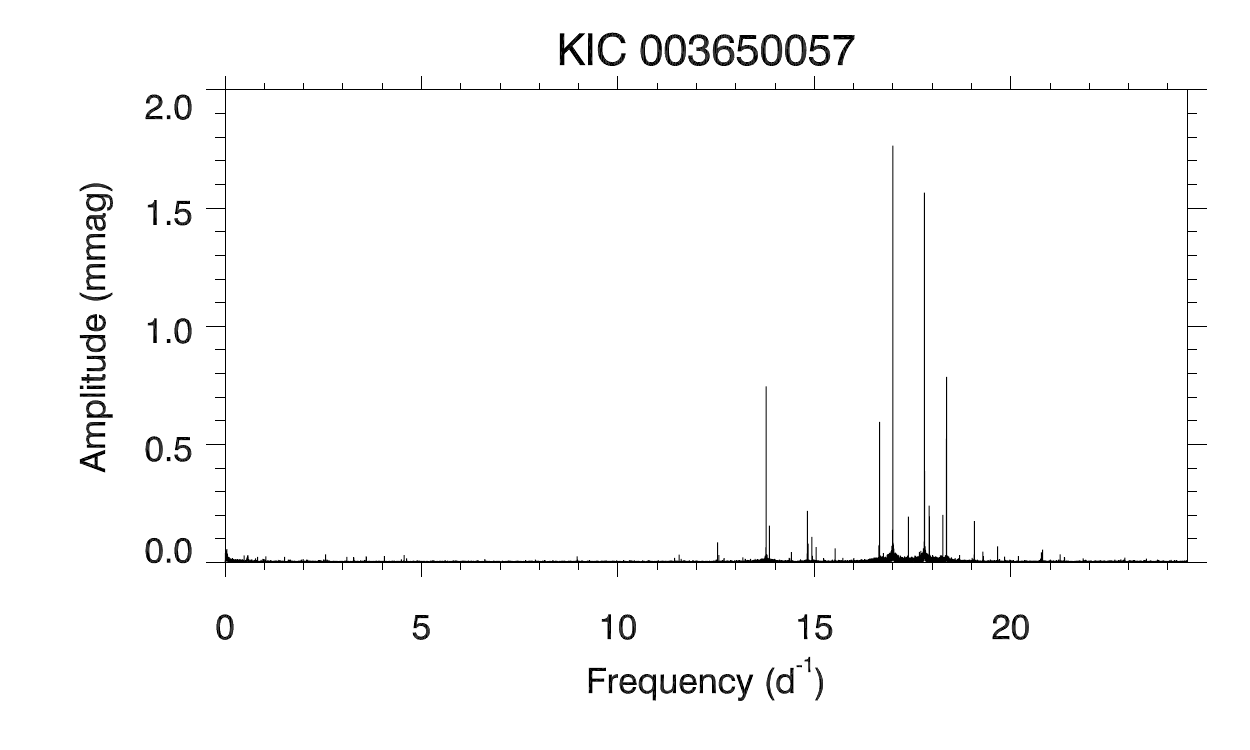}
		\includegraphics[width=0.49\textwidth]{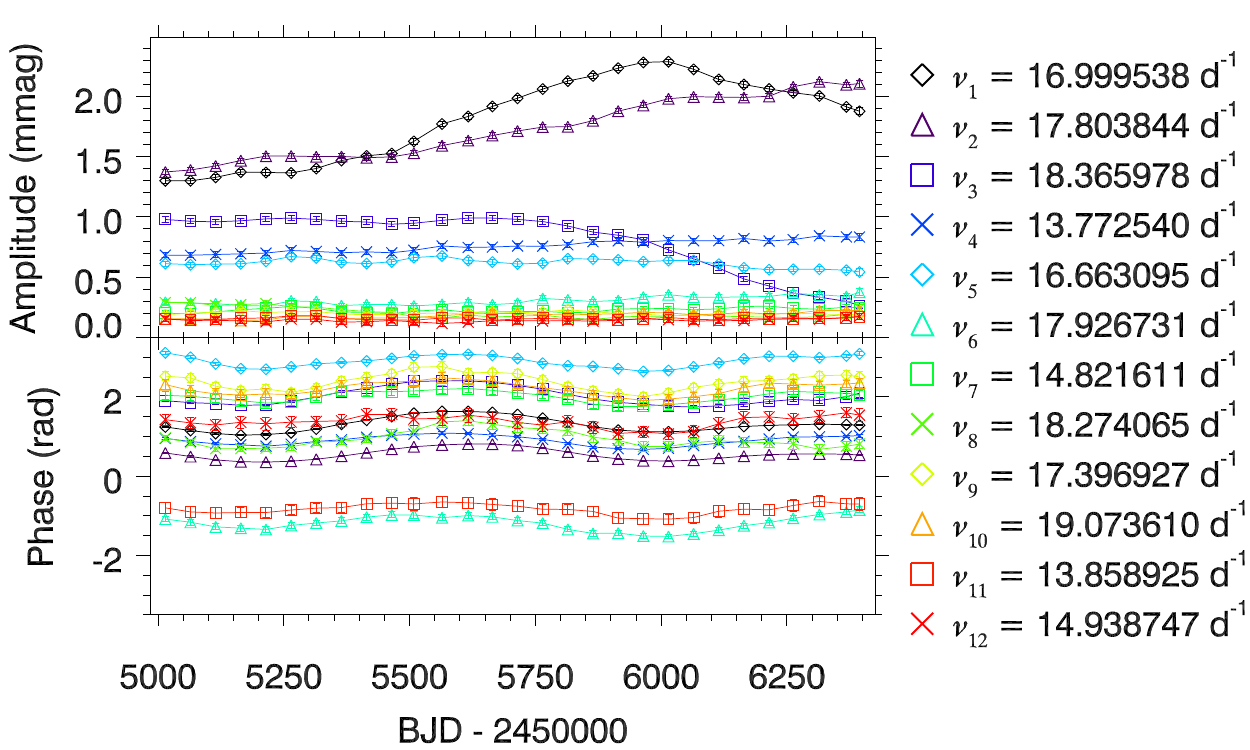}
		
		\vspace{0.4cm}	
		
		\includegraphics[width=0.49\textwidth]{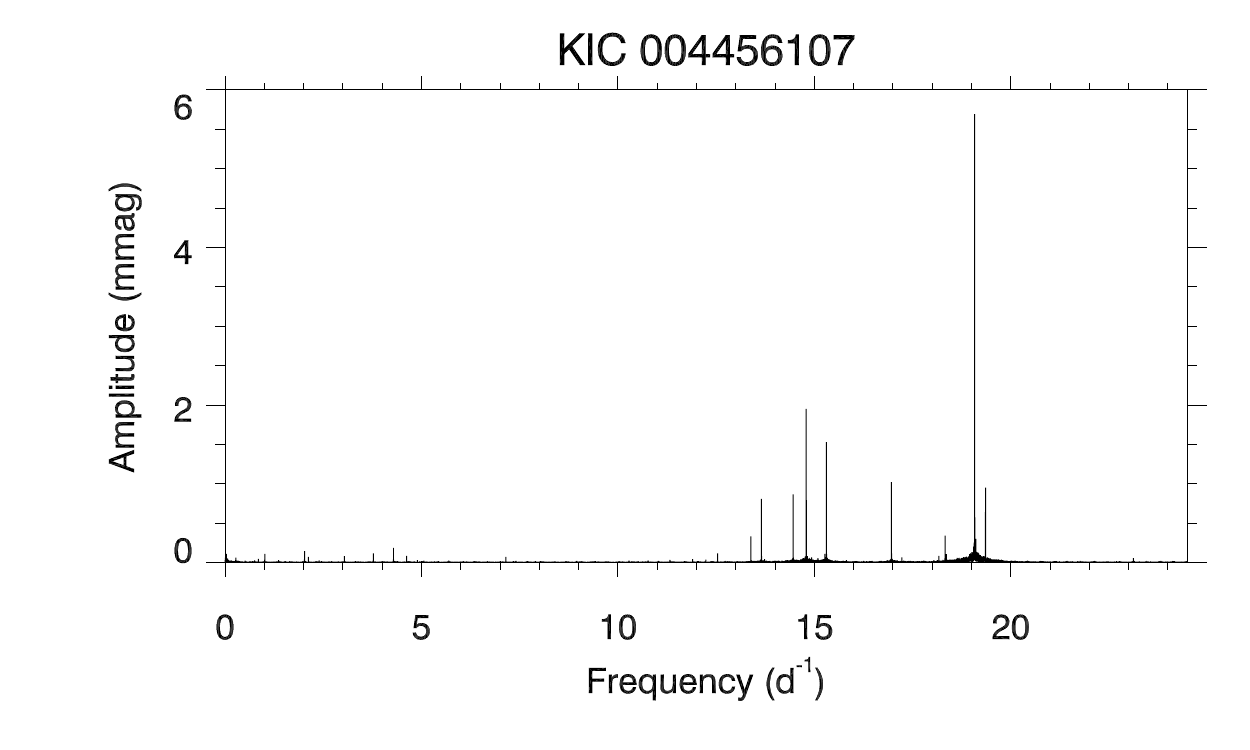}
		\includegraphics[width=0.49\textwidth]{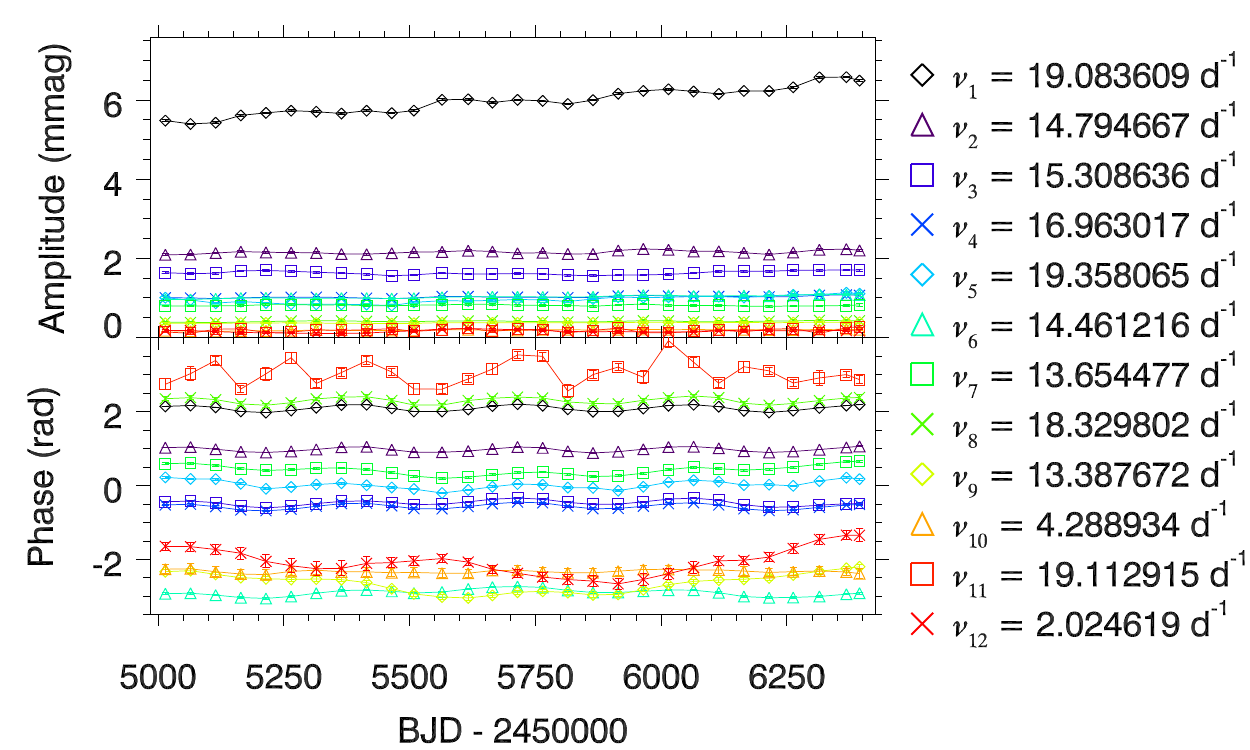}
		
		\vspace{0.4cm}	
		
		\includegraphics[width=0.49\textwidth]{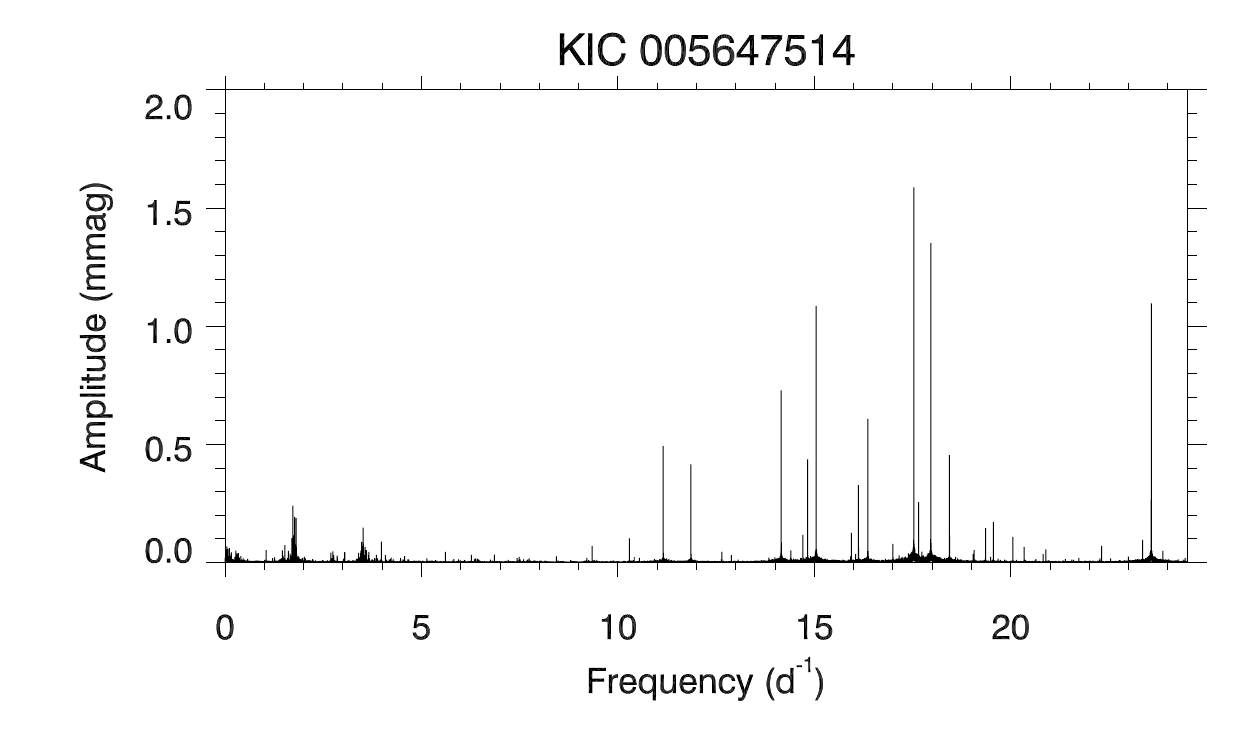}
		\includegraphics[width=0.49\textwidth]{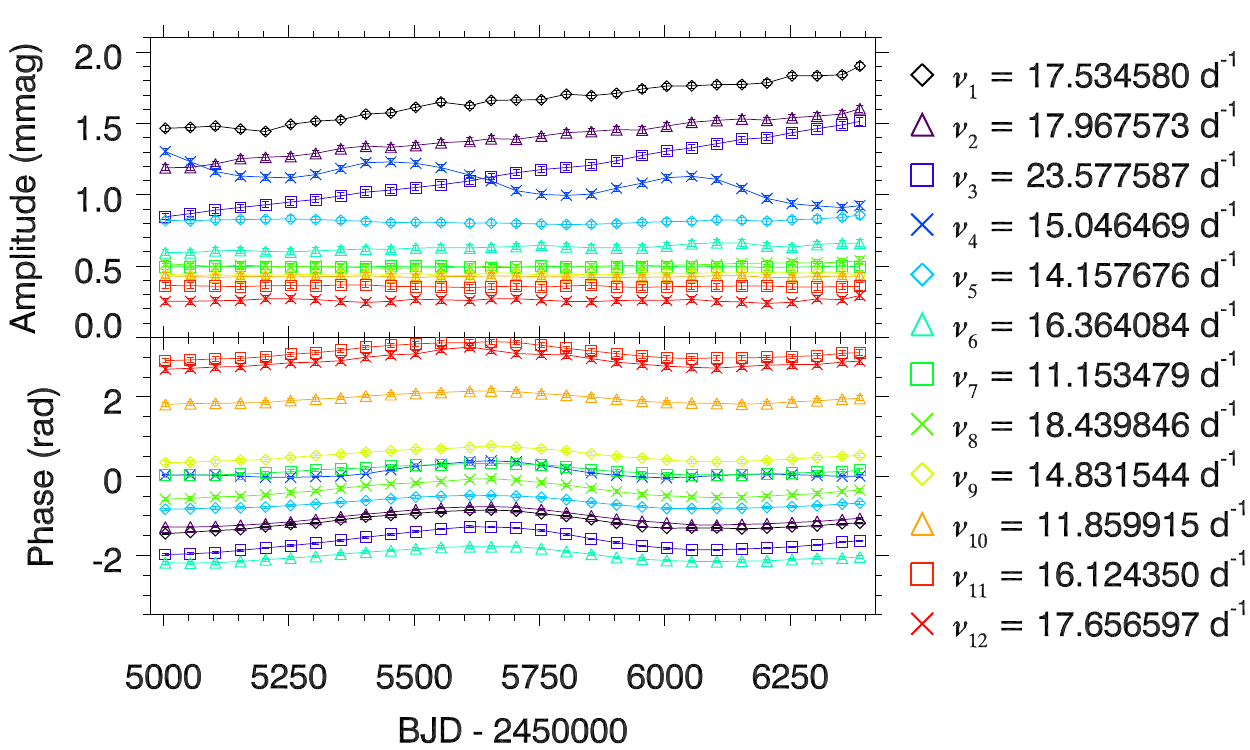}
		
		\vspace{0.4cm}	
		
		\includegraphics[width=0.49\textwidth]{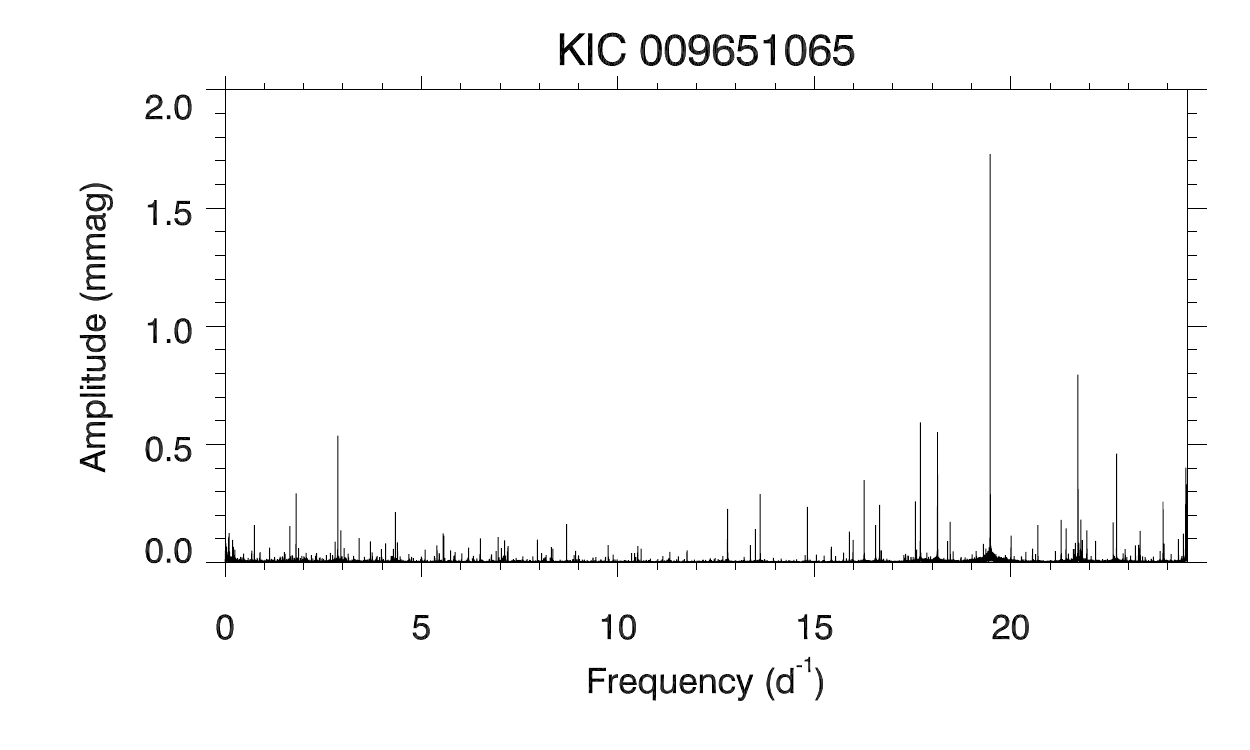}
		\includegraphics[width=0.49\textwidth]{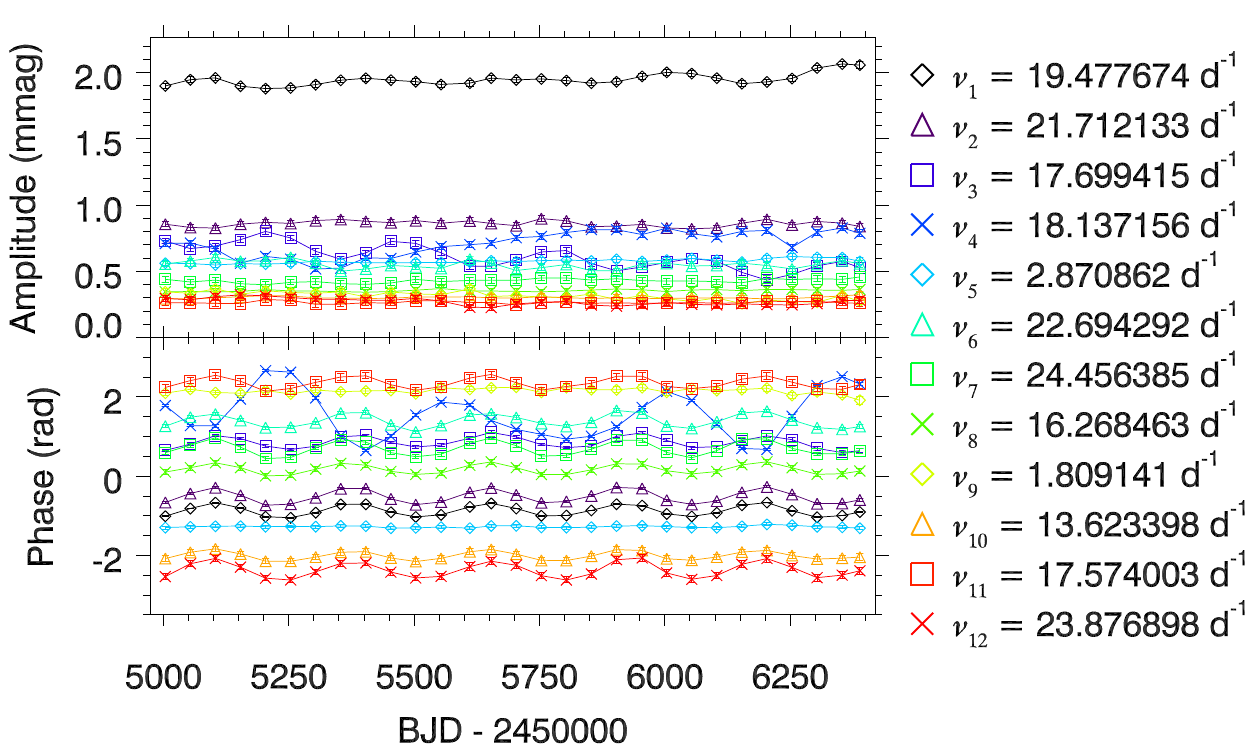}
		
		\caption{Four examples of \dsct stars in the KIC range $6400 \leq T_{\rm eff} \leq 10\,000$~K that show phase modulation over the 4-yr \Kepler data set because of binarity, but are also AMod stars. From top to bottom: KIC~3650057, KIC~4456107, KIC~5647514 and KIC~9651065 ($\nu_4$ is a super-Nyquist alias). The left panels are the 4-yr amplitude spectra calculated out to the LC Nyquist frequency. The right panels show the amplitude and phase tracking plots which demonstrate the modulation in pulsation amplitudes and phases over 4~yr. The orbital period is calculated from the period of the phase modulation in the tracking plot, and is consistent with the result from the PM technique given in Table~\ref{table: all stars} for each star.}
		\label{figure: binary stars}
		\end{figure*}

		Four examples of stars identified as binary systems from this study are shown in Fig.~\ref{figure: binary stars}. These stars are also AMod stars which appears unrelated to the phase modulation caused by the presence of a companion object. The bottom row of Fig.~\ref{figure: binary stars} shows the amplitude spectrum (left panel) and the tracking plot (right panel) of the well-studied star KIC~9651065 in which the period of the phase modulation is the orbital period of $P_{\rm orb} = 272.7 \pm 0.8$~d calculated by \citet{Murphy2014}. The search for binarity among the \dsct stars observed by \Kepler is not the goal of this study so we do not discuss it any further.
		
		\subsection{HADS stars}
		\label{subsection: HADS stars}

\begin{figure*}
		\centering
		
		\includegraphics[width=0.49\textwidth]{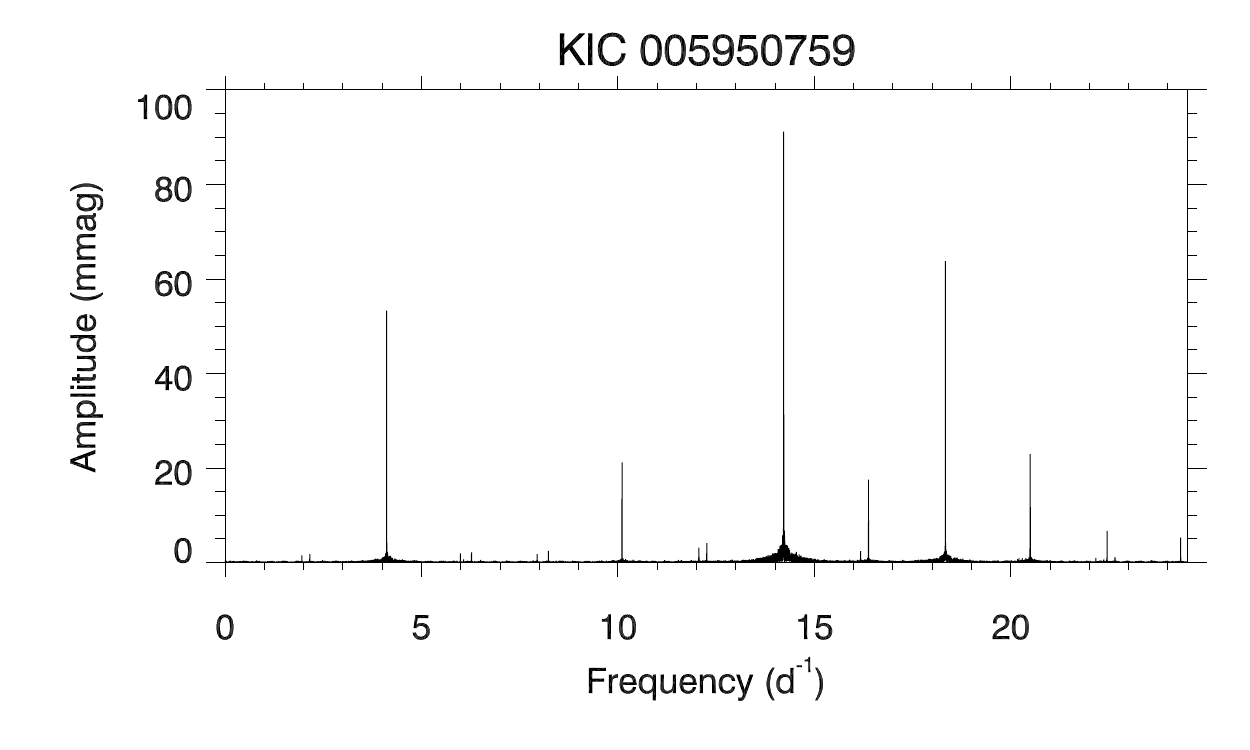}
		\includegraphics[width=0.49\textwidth]{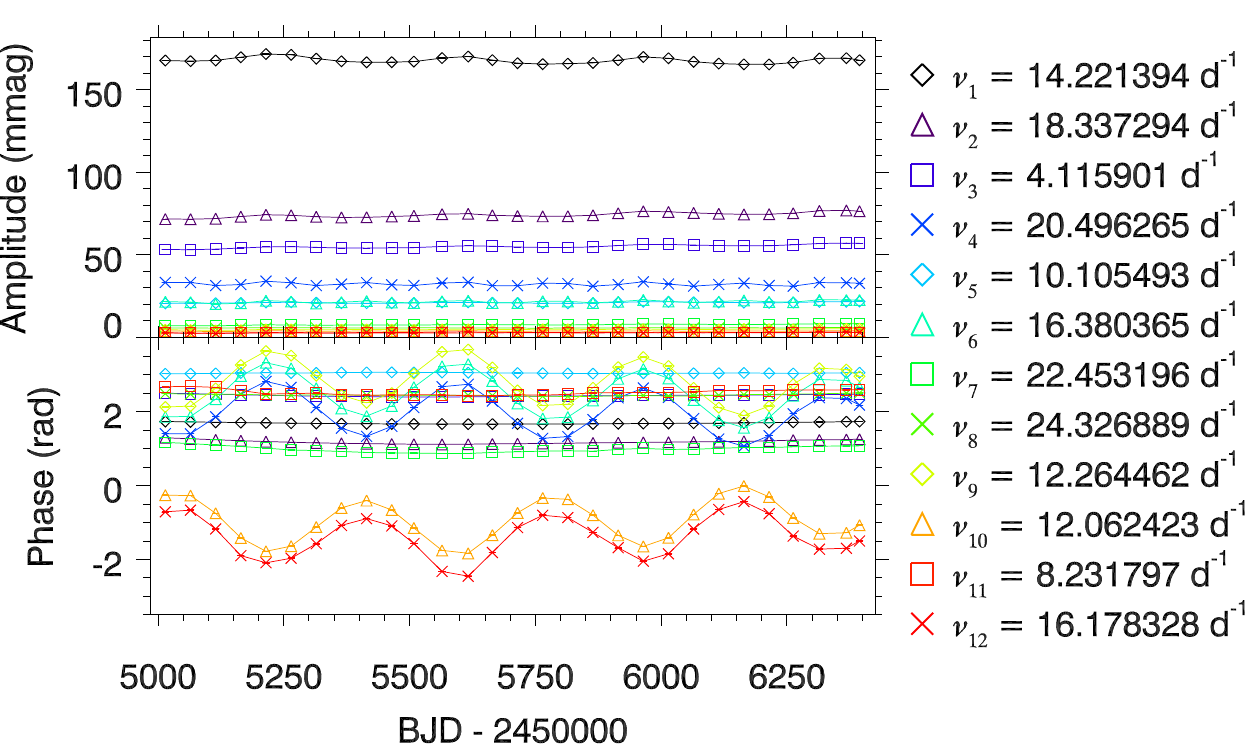}
		
		\vspace{0.5cm}	

		\includegraphics[width=0.49\textwidth]{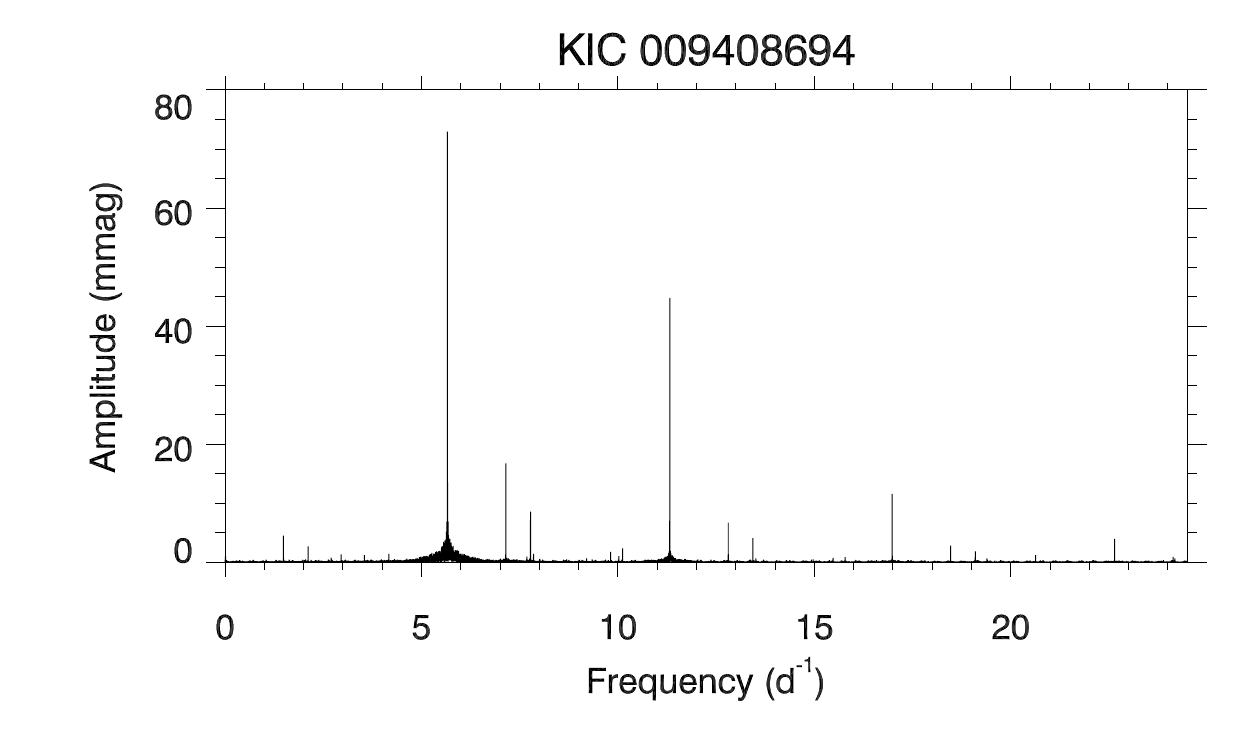}
		\includegraphics[width=0.49\textwidth]{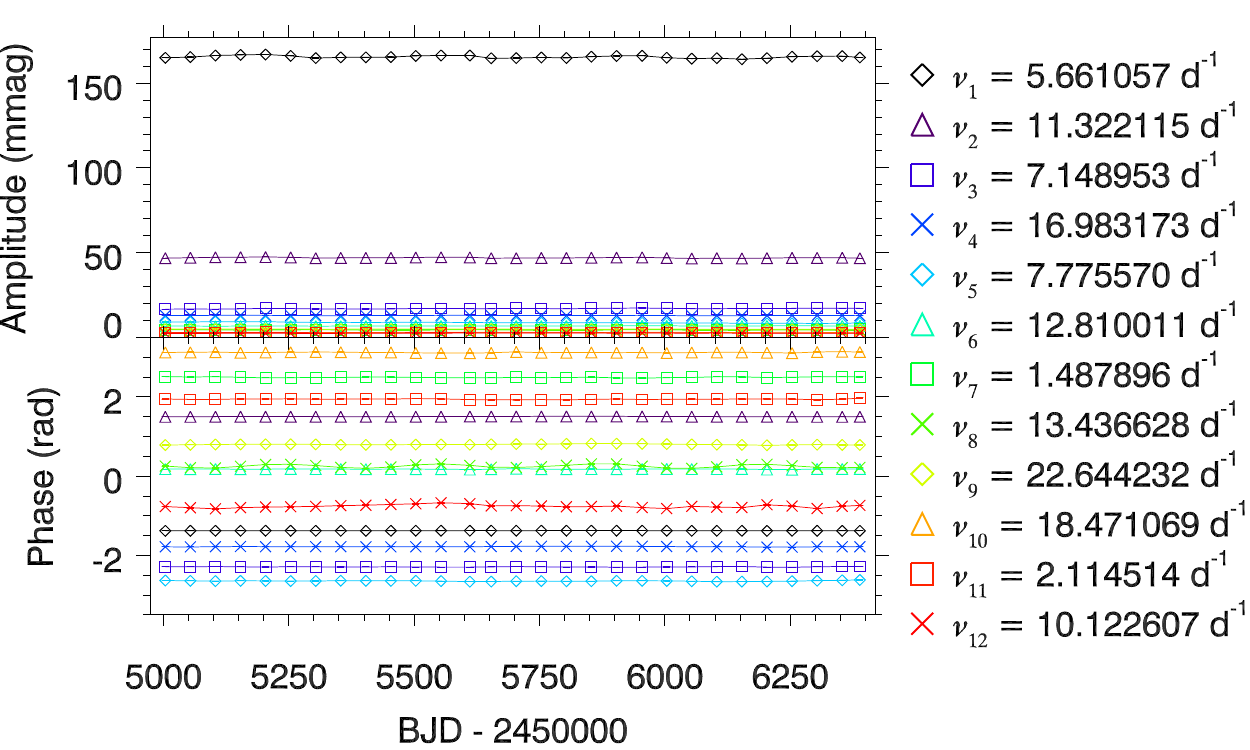}
		
		\caption{Two HADS stars in the KIC range $6400 \leq T_{\rm eff} \leq 10\,000$ ~K that show little or no variability in their high-amplitude radial pulsation modes over the 4-yr \Kepler data set. The top row is KIC~5950759 ($\nu_4$, $\nu_6$, $\nu_9$, $\nu_{10}$ and $\nu_{12}$ are super-Nyquist aliases) and the bottom row is KIC~9408694. The left panels are the 4-yr amplitude spectra calculated out to the LC Nyquist frequency. The right panels show the amplitude and phase tracking plots for these two HADS stars over 4~yr. There is little or no variability in the amplitudes and phases of the radial pulsation modes, if instrumental modulation caused by the \Kepler satellite is removed.}
		\label{figure: HADS stars}
		\end{figure*}

		There are few stars in our ensemble that meet the HADS definition from \citet{McNamara2000a} of peak-to-peak light amplitude variations greater than 0.3~mag. Some stars were only observed for limited subsets of LC data, and consequently have not been included in our ensemble as we only chose stars for which 4~yr of continuous \Kepler observations were available. We did not, however, preferentially exclude HADS stars from our sample. Since there are several thousand A and F stars in the \Kepler data set, the classical instability strip is well-sampled near the TAMS -- for example see \citet{Niemczura2015}, thus the implication from our study is that HADS stars are rare in \Kepler data. We find only two HADS stars within our ensemble, KIC~5950759 and KIC~9408694, which are shown in the top and bottom rows of Fig.~\ref{figure: HADS stars}, respectively. 

		For KIC~5950759, the period ratio of $\nu_1 = 14.221394$~d$^{-1}$ and $\nu_2 = 18.337294$~d$^{-1}$ gives 0.7755, which identifies these frequencies as the fundamental and first overtone radial modes, respectively. For KIC~9408694, the period ratio of $\nu_1 = 5.661057$~d$^{-1}$ and $\nu_3 = 7.148953$~d$^{-1}$ gives 0.7919, which is outside the expected range for the fundamental and first overtone radial modes. KIC~9408694 was studied by \citet{Balona2012a} who concluded that the fast rotation of KIC~9408694, which is unusual for a HADS star, perturbs the observed pulsation mode frequencies. A model including fast rotation successfully identified $\nu_1$ and $\nu_3$ as the fundamental and first overtone radial modes, respectively \citep{Balona2012a}. We find that both HADS stars in our ensemble exhibit fractional amplitude variability of order a few~per~cent with a period equal to the \Kepler orbital period -- each has several AMod frequencies using our $\pm5\sigma$ significance criterion. If this instrumental amplitude modulation is removed, both HADS stars have little or no variability in the amplitudes of their radial modes. This was also concluded by \citet{Balona2012a} for KIC~9408694.
		
		The same amplitude limitation mechanism predicted for the low-amplitude \dsct stars \citep{Breger2000a} does not seem to be at work within HADS stars, which pulsate at much higher amplitudes and yet do not continue to grow exponentially. It is interesting to note that high amplitude pulsations are typically associated with nonlinearity in the form of harmonics and combination frequencies, which are found in KIC~5950759 and KIC~9408694, but not normally associated with variable mode amplitudes. This implies that HADS stars are more similar to Cepheid variables\footnote{For example, \citet{Eggen1976} referred to \dsct stars as ultrashort-period Cepheids (USPC). They have also been called dwarf Cepheids.} than their low-amplitude \dsct star counterparts \citep{McNamara2000a}. 
		
	\subsection{Pure phase modulation}
	\label{subsection: phase modulation}
	
	Pulsation mode frequencies change with evolution of stellar structure, and the concomitant changes in the pulsation cavities of individual modes. Are these changes observable over the 4-yr \Kepler data set? In our study of 983 \dsct stars, we searched for pure phase modulation with no associated amplitude modulation of independent pulsation modes, and found no obvious cases. This, of course, excluded those stars which have phase modulation driven by an extrinsic cause, such as binarity, or because the frequencies are super-Nyquist aliases. However, there are cases of non-sinusoidal light variations that changed in shape of 4~yr, observed as phase modulation of harmonics of pulsation mode frequencies. For example, slight phase modulation is observed in the harmonic of the fundamental radial mode frequency, $2\nu_1 = 28.442787$~d$^{-1}$, in the HADS star KIC~5950759, which is shown in the tracking plot in the right panel of Fig.~\ref{figure: sNa peaks}.
				

	\subsection{Special case study stars}
	\label{subsection: special stars}
	
		\subsubsection{KIC~4733344}
		\label{subsubsection: KIC4733344}
		
		Frequency analysis of the \dsct star KIC~4733344 revealed that its pulsation mode frequencies $\nu_2 = 7.226764$~d$^{-1}$ and $\nu_3 = 9.412445$~d$^{-1}$ have a period ratio of 0.7678, which is typically associated with the fundamental and first overtone radial modes. Calculating pulsation constants using eq.~\ref{equation: stellingwerf} for $\nu_2$ and $\nu_3$ indicated they are likely the fundamental and the first overtone radial modes, respectively, considering the typical uncertainties associated with calculating $Q$ values \citep{Breger1990b}. The highest amplitude pulsation mode frequency, $\nu_1 = 8.462183$~d$^{-1}$, is not easily identifiable as it lies between $\nu_2$ and $\nu_3$, thus suggesting it is likely a non-radial mode. This is not surprising as non-radial modes can have higher amplitudes than radial modes (see figure~1.5 from \citealt{ASTERO_BOOK}). Our tracking routine revealed that these three pulsation mode frequencies are variable in amplitude and phase, and so KIC~4733344 pulsates with two variable low-overtone radial modes, $\nu_2$ and $\nu_3$. KIC~4733344 has a $\log\,g$ value of $3.50 \pm 0.23$ \citep{Huber2014}, suggesting that it is likely in a post-main-sequence state of evolution. 
		
		Previously, in section~\ref{subsection: constant stars} we discussed the example of the NoMod \dsct star KIC~2304168, which has similar $T_{\rm eff}$, $\log\,g$ and $[{\rm Fe}/{\rm H}]$ values to the AMod \dsct star KIC~4733344. Both of these \dsct stars have been shown to pulsate in low-overtone radial modes, in particular the fundamental and first overtone radial modes. This makes us ponder the possible differences in the driving and damping mechanisms at work within these two \dsct stars. In section~\ref{section: coupling model}, we discuss coupling models for families of pulsation mode frequencies in KIC~4733344.
		
		\subsubsection{KIC~7106205}
		\label{subsubsection: KIC7106205}

\begin{figure}
			\centering
			\includegraphics[width=0.49\textwidth]{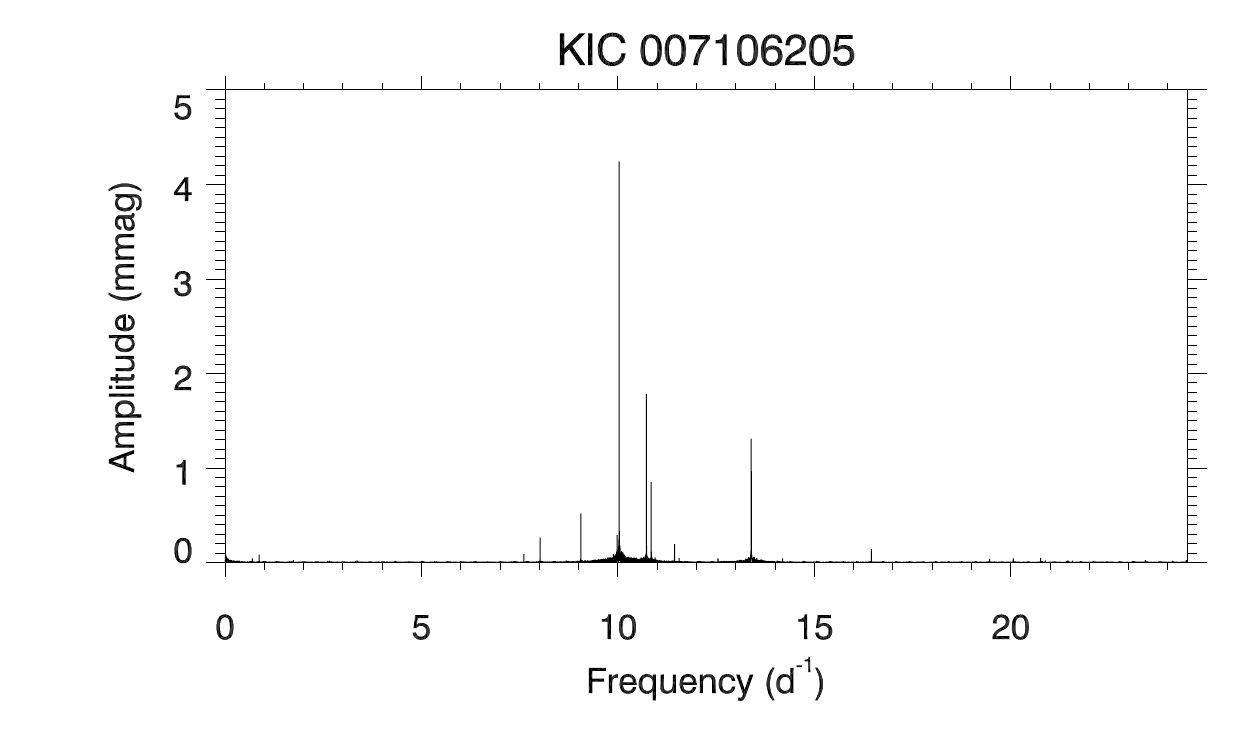}
			\includegraphics[width=0.49\textwidth]{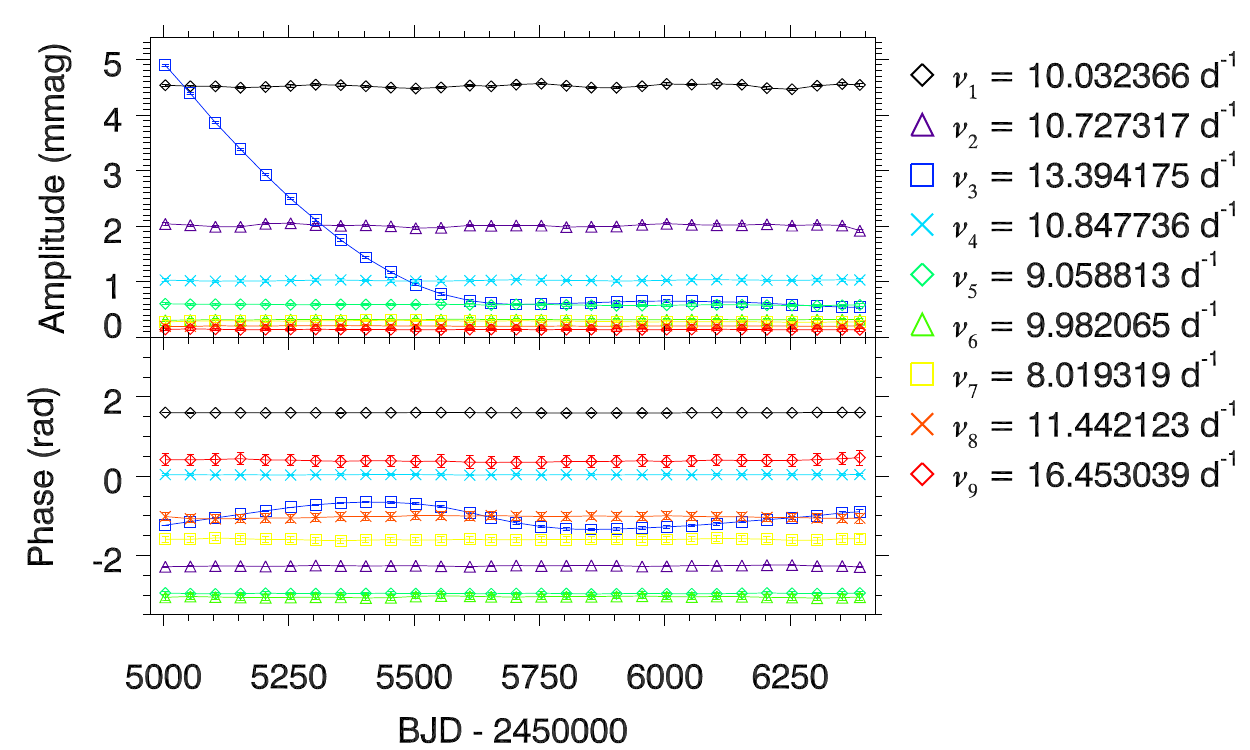}
			\caption{KIC~7106205: the 4-yr amplitude spectrum calculated out to the LC Nyquist frequency is shown in the top panel, and the amplitude and phase tracking plot showing variability in pulsation mode amplitudes and phases over time is shown in the bottom panel.}
			\label{figure: KIC 7106205}
		\end{figure}

		The \dsct star KIC~7106205 was investigated by \citet{Bowman2014} and found to contain only a single pulsation mode with variable amplitude whilst all other pulsation modes remained constant in amplitude and phase between 2009 and 2013. Further work by \citet{Bowman2015a} extended this study back to 2007 using archive data from the WASP project. The amplitude spectrum and tracking plot for the {9} pulsation mode frequencies that have amplitudes higher than {0.10}~mmag are shown in Fig.~\ref{figure: KIC 7106205}. Since no other pulsation modes were observed to vary in amplitude or phase, it was concluded that the visible pulsation mode energy was not conserved in this star \citep{Bowman2014}.
		
\section{Modelling beating}
\label{section: beating model}

{Pure} amplitude modulation of a single pulsation mode may appear as a group of close-frequency peaks in the amplitude spectrum \citep{Buchler1997a}, and thus the presence of multiple peaks does not {\it prove} the existence of multiple independent pulsation modes. {Pure} amplitude modulation of a single pulsation mode excludes phase variability \citep{Breger2006a}, thus it is only by studying the amplitude {\it and} frequency (i.e. phase at fixed frequency) variations of peaks in the amplitude spectrum that it can be determined if the variability is caused by the {pure} amplitude modulation of a single mode, or the beating pattern of multiple close-frequency pulsation modes. 

	\begin{figure}
		\centering
		\includegraphics[width=0.99\columnwidth]{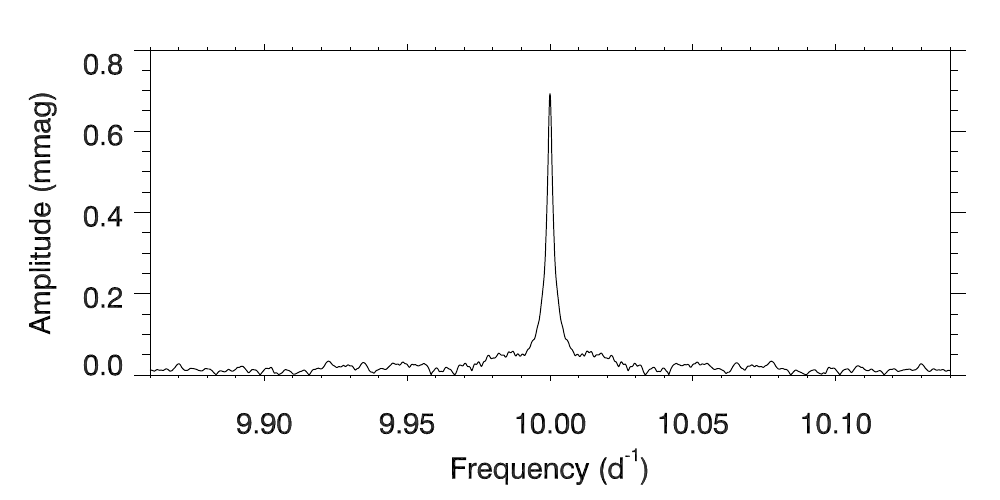}
		\includegraphics[width=0.99\columnwidth]{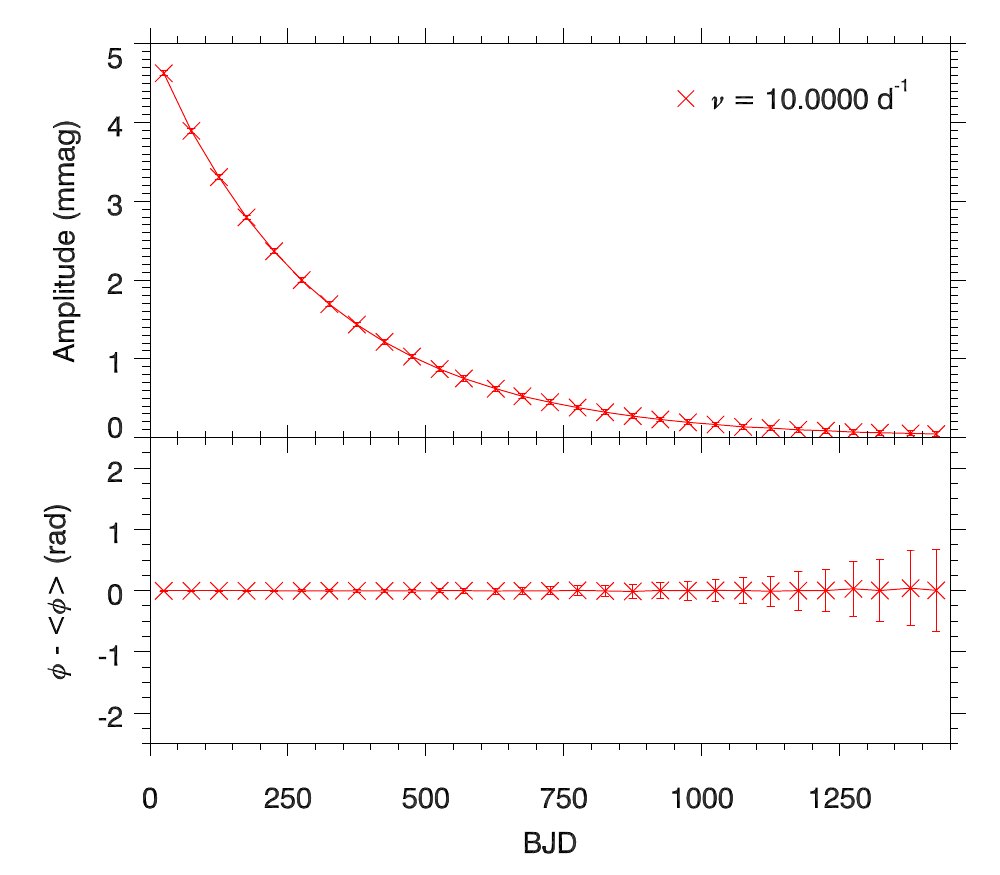}
		\caption{{Pure} amplitude modulation for an exponentially-decaying amplitude and fixed frequency and phase cosinusoid using synthetic \Kepler data. The top panel shows the Lorentzian profile peak produced in the amplitude spectrum and the bottom panel is the amplitude and phase tracking plot for {pure} amplitude modulation with no associated phase variation.}
		\label{figure: exp model}
	\end{figure}

We tested this concept using synthetic data, specifically 4-yr of \Kepler time stamps with calculated magnitudes using 

\begin{equation}
	y = Ae^{{-t}/{\tau}} \cos( 2\pi \nu (t-t_0) + \phi) ~ ,
\label{equation: exp decay}
\end{equation}

\noindent where $\tau$ is an exponential decay time of 300~d, $A$ is an amplitude of 5.0~mmag, $\nu$ is a frequency of 10.0~d$^{-1}$ and $\phi$ is a phase of 0.0 relative to the centre of the data set (i.e. $t_0 = 2\,455\,688.770$~BJD). Calculating the amplitude spectrum of an exponentially decaying signal produces a Lorentzian profile peak in the Fourier domain, which is shown in the top panel of Fig.~\ref{figure: exp model}. We then applied our amplitude and phase tracking method to this input frequency with the results shown in the bottom panel of Fig.~\ref{figure: exp model}. The cosinusoidal signal calculated using eq.~\ref{equation: exp decay} has a constant frequency and phase but a decaying amplitude\footnote{A similar result is obtained if a $\tanh(-t/\tau)$ factor is used instead of $e^{-t/\tau}$ as an amplitude modulation factor.}, which is made clear by the {pure} amplitude modulation and no phase variation shown in the bottom panel of Fig.~\ref{figure: exp model}. 

As previously discussed in section~\ref{subsection: Beating}, the relative amplitudes and separations in close-frequency pulsation modes governs the amplitude and phase modulation of the beating signal. In observations of \dsct stars, close-frequency pulsation modes are not uncommon \citep{Breger2002d} and unlikely to be explained by rotation. Also, one cannot attribute a pair of resolved close-frequency pulsation modes to being caused by the {pure} amplitude modulation of a single pulsation mode, as this leads to a more complex structure in the amplitude spectrum, and not two resolved peaks. From previous studies of \dsct stars, for which mode identification was possible, it was found that non-radial modes were commonly found to cluster around radial modes because of mode trapping \citep{Breger2009a}. 

We constructed beating models of the two pairs of close-frequency pulsation modes in KIC~4641555 and KIC~8246833 using synthetic data, specifically using 4-yr of \Kepler time stamps and magnitudes calculated containing only white noise and the two resolved frequency cosinusoids causing the beating pattern. The frequencies, amplitudes and phases of the two pulsation mode frequencies can be calculated because the peaks are resolved using 4-yr of \Kepler data. These models were successfully matched to observations of amplitude modulation in KIC~4641555 and KIC~8246833, yielding beat periods of {$1166 \pm 1$}~d and {$1002 \pm 1$}~d, respectively. The observed amplitude modulation of the highest amplitude pulsation mode frequency is shown as diamonds and the beating model is shown as crosses in the bottom panels of Fig.~\ref{figure: beating AMod stars} for each star. This demonstrates that it is possible for a \dsct star to pulsate with low-degree p-mode frequencies that lie very close to the Rayleigh resolution criterion in frequency for the 4-yr \Kepler data set and yet maintain their independent identities. 

It is plausible that many more AMod \dsct stars can be explained by the beating of unresolved close-frequency pulsation modes. In a similar way, beating models of multiple unresolved frequencies could be constructed to explain amplitude modulation in many \dsct stars, but this requires the number of frequencies to be known {\it a priori}. Theoretically, it is possible for many non-radial pulsation mode frequencies to exist in a frequency range of less than $0.00068$~d$^{-1}$ and maintain their independent identity over several years (Saio, {\it priv. comm.}). If this is the case for many of the AMod \dsct stars in this study, it would explain the non-sinusoidal modulation cycles because of the complicated beating pattern of multiple unresolved close-frequency pulsation modes.


\section{Modelling {nonlinearity}}
\label{section: coupling model}

In the mode coupling hypothesis, it is required that all three members of a family are variable in amplitude so that the child mode can be identified and a model predicting its behaviour can be constructed as a function of the product of the amplitudes of the parent modes. In section~\ref{subsection: resonance}, it was discussed how families of frequencies must satisfy the frequency criterion given in eq.~\ref{equation: freq coupling} and the phase criterion given in eq.~\ref{equation: phase coupling}, but also how eq.~\ref{equation: amp coupling} can be used to distinguish among possible causes of nonlinearity in a star. By trying different values of the coupling coefficient $\mu_{\rm c}$, the {\it strength} of coupling and nonlinearity among pulsation mode frequencies can be estimated. In this way, families of frequencies that satisfy eq.~\ref{equation: freq coupling} can be distinguished as coupling between a child and two parent modes or combination frequencies. Small values of $\mu_{\rm c}$ imply weak-coupling and favour the non-linear distortion model producing combination frequencies, whereas large values of $\mu_{\rm c}$ imply strong coupling and favour resonant mode coupling \citep{Breger2014}. 

Note that a large value of $\mu_{\rm c}$ is defined by the amplitudes of the parent modes and thus is specific to each family. For example, using eq.~\ref{equation: amp coupling}, if $A_2 = A_3 = 2$~mmag, then to achieve a similar child mode amplitude of $A_1 \simeq 2$~mmag, a value of $\mu_{\rm c} \simeq 0.25$ is required. Using the same parent mode amplitudes, a small value of $\mu_{\rm c} \simeq 0.01$ would produce a child mode amplitude of $A_1 \simeq 0.04$~mmag. Therefore, in this hypothetical example, $\mu_{\rm c} \geq 0.1$ is considered strong coupling and $\mu_{\rm c} \leq 0.01$ is considered weak coupling.

\begin{figure*}
	\centering
		
	\includegraphics[width=0.49\textwidth]{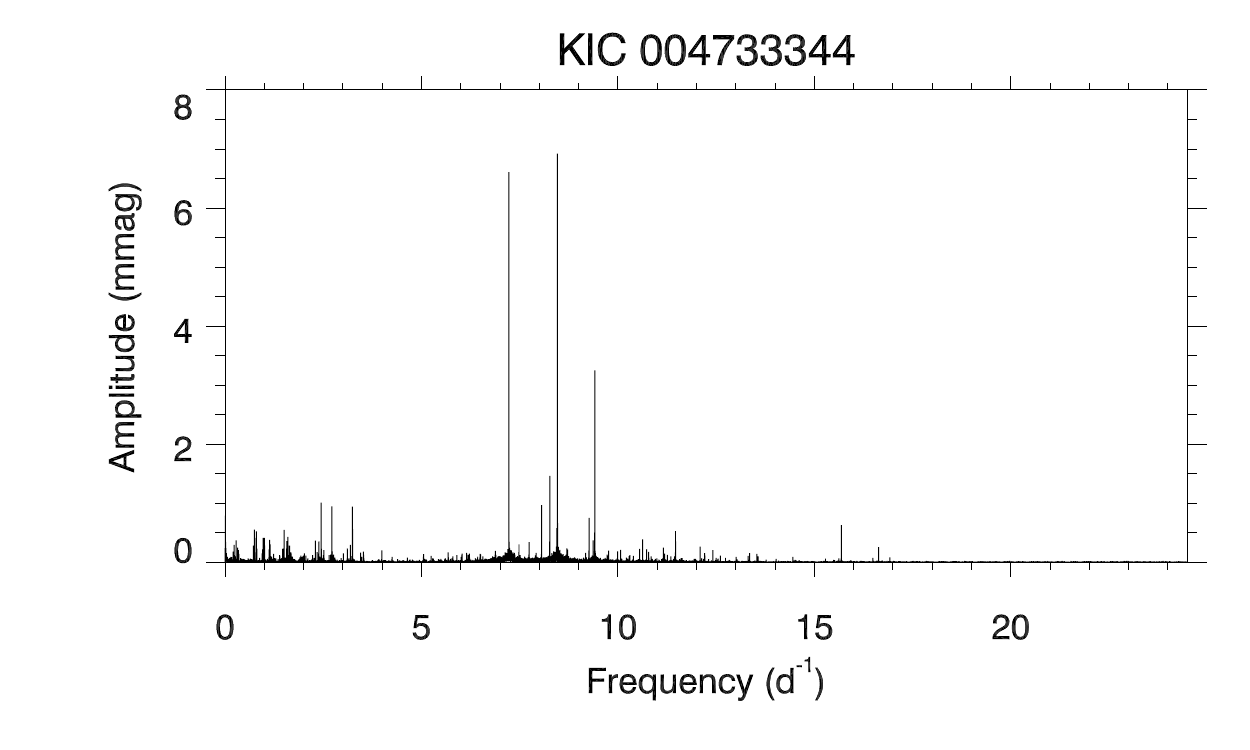}
	\includegraphics[width=0.49\textwidth]{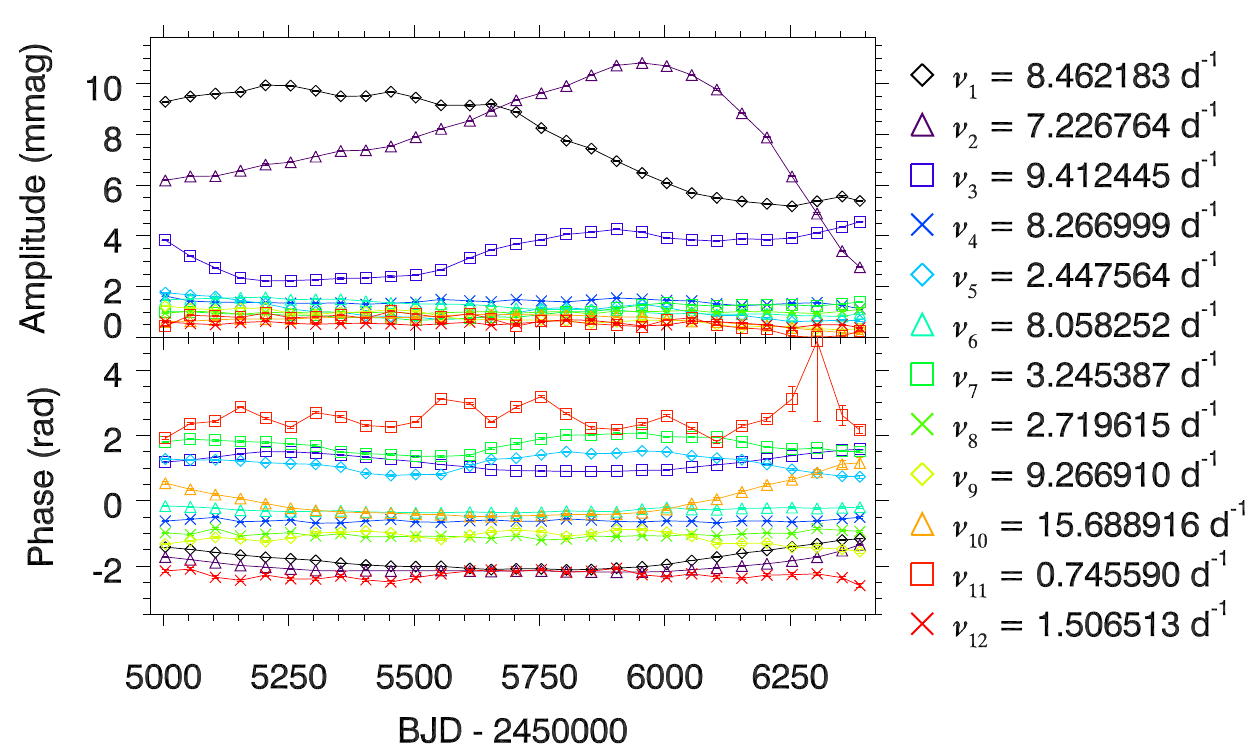}
		
	\includegraphics[width=0.49\textwidth]{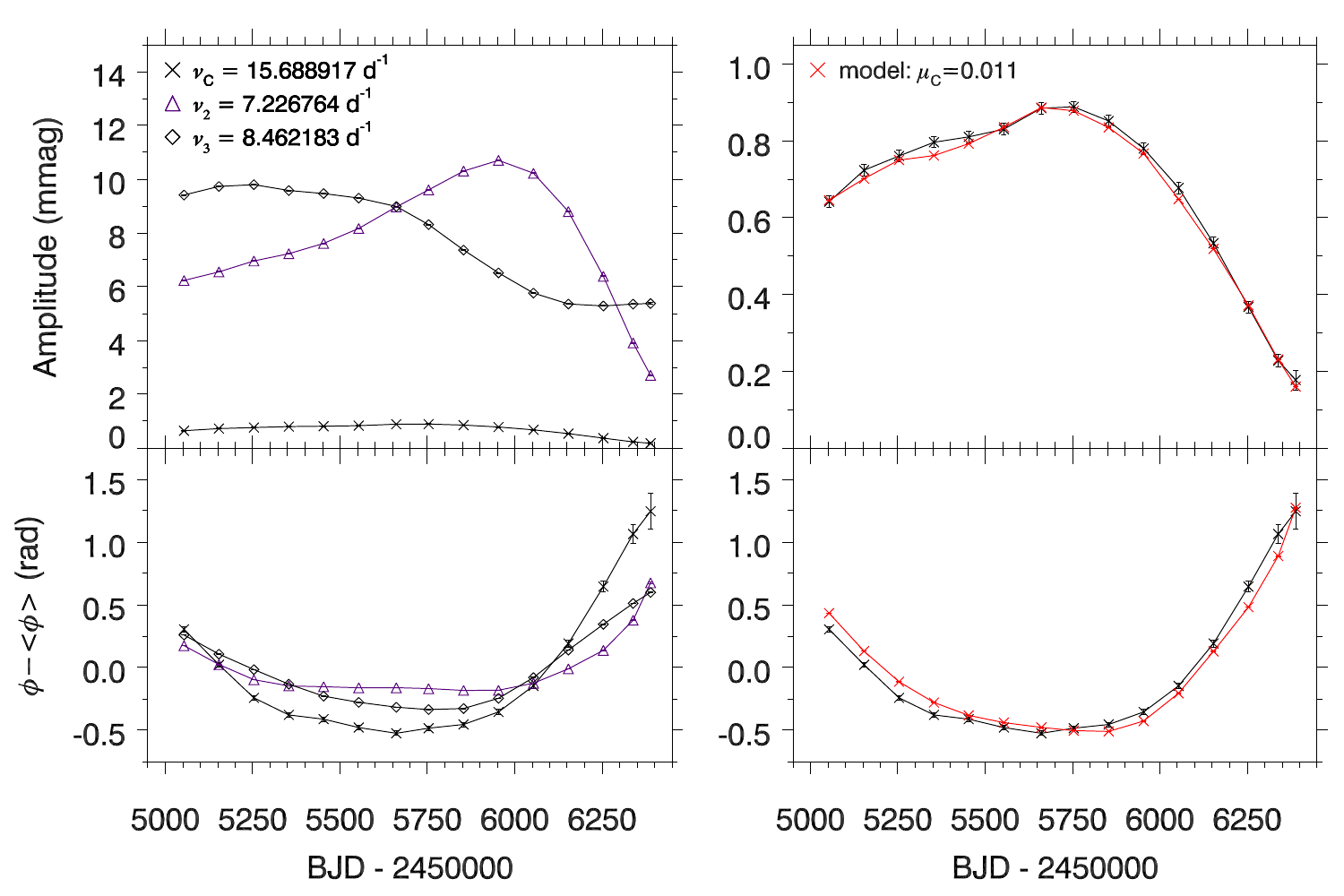}
	\hspace{0.2cm}
	\includegraphics[width=0.49\textwidth]{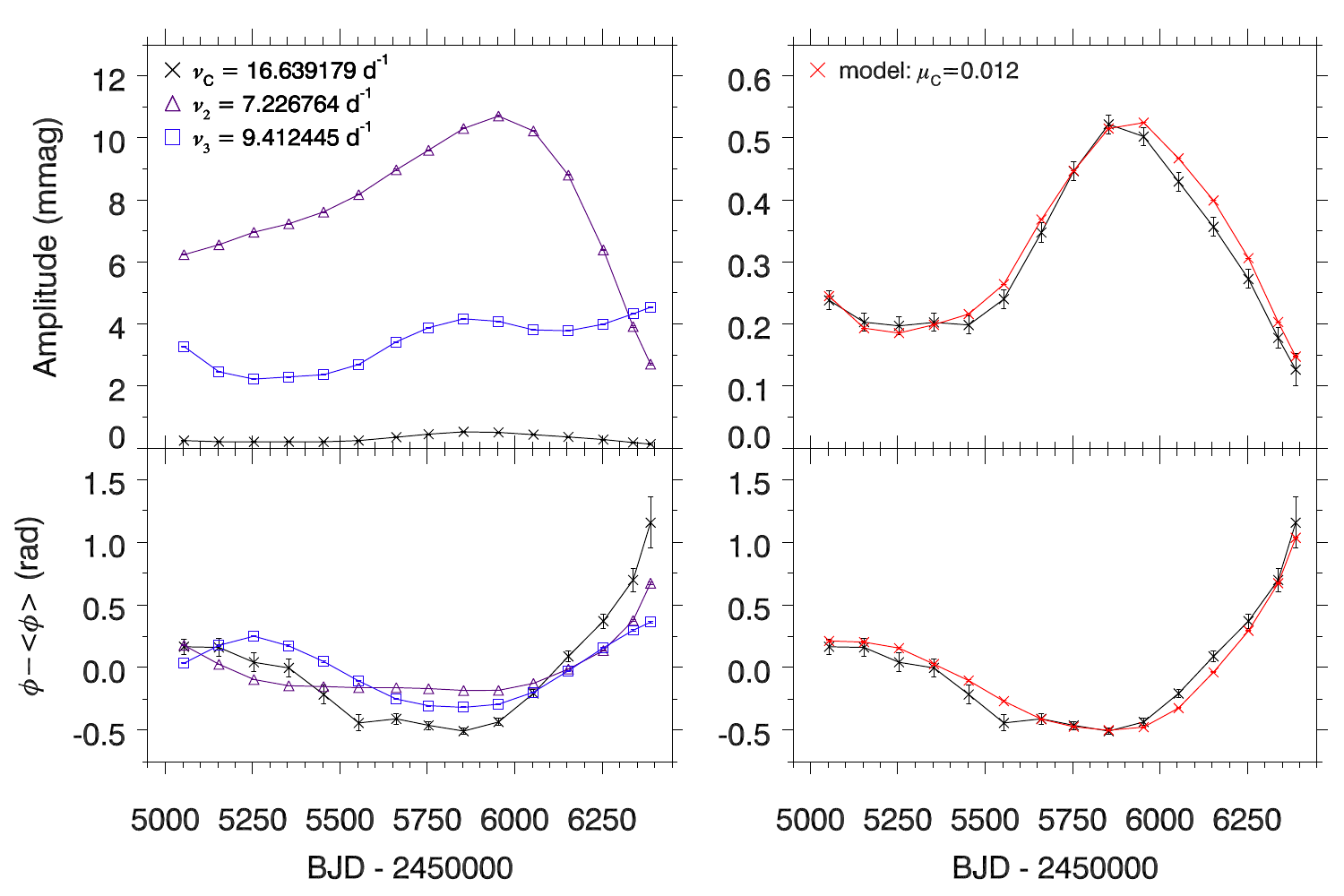}
		
	\caption{The AMod \dsct (hybrid) star KIC~4733344 has variable pulsation amplitudes and phases over the 4-yr \Kepler data set, which can be explained by {nonlinearity}. The top-left panel is the 4-yr amplitude spectrum calculated out to the LC Nyquist frequency and the top-right panel shows the amplitude and phase tracking plot which demonstrates the variability in pulsation amplitudes and phases over 4~yr. The bottom panels show the coupling models as crosses, which are consistent with observations of the child mode variability for two families of frequencies taken from the tracking plot, specifically $\nu_1 + \nu_2$ (left) and $\nu_2 + \nu_3$ (right).}
	\label{figure: coupling AMod stars}
	\end{figure*}

Coupling models for two families of frequencies in KIC~4733344 are shown in Fig.~\ref{figure: coupling AMod stars}. These two families have similar small values of the coupling coefficient $\mu_{\rm c} \simeq 0.01$, which imply nonlinearity in the form of combination frequencies from the non-linear distortion model. For example, since the parent modes have amplitudes of $A_2 \simeq 6$~mmag and $A_3 \simeq 9$~mmag at the start of the data set (see the bottom-left panel of Fig.~\ref{figure: coupling AMod stars}), a coupling coefficient of $\mu_{\rm c} \simeq 0.1$ would produce a child mode amplitude of $A_1 \simeq 5$~mmag, which is an order of magnitude larger than the observed amplitude of the child mode, $A_1 \simeq 0.6$~mmag. Therefore, we conclude that resonant mode coupling is unlikely the cause of nonlinearity and amplitude modulation in KIC~4733344.

\section{Ensemble study statistics}
\label{section: ensemble}

Our ensemble comprised {983} \dsct stars that lie in the KIC effective temperature range of $6400 \leq T_{\rm eff} \leq 10\,000$~K, pulsate in p-mode frequencies between $4 \leq \nu \leq 24$~d$^{-1}$ and were observed continuously by the \Kepler Space Telescope for 4~yr. As previously described in Section~\ref{section: method}, we flagged the number of peaks that have amplitudes greater than $0.10$~mmag (up to a maximum number of 12) in each star that exhibit significant amplitude modulation, with each star labelled as either NoMod or AMod. The criterion of significant amplitude modulation was chosen as at least half of a frequency's time bins being greater than {$\pm5\sigma$} in amplitude from the mean, which is shown graphically in Fig.~\ref{figure: significant amod}. It is important to note that our method for studying amplitude modulation in \dsct stars does not automatically determine if an extracted frequency is a combination frequency, or the cause of the observed amplitude modulation, such as beating or nonlinearity.

\begin{figure*}
	\includegraphics[width=0.4\textwidth]{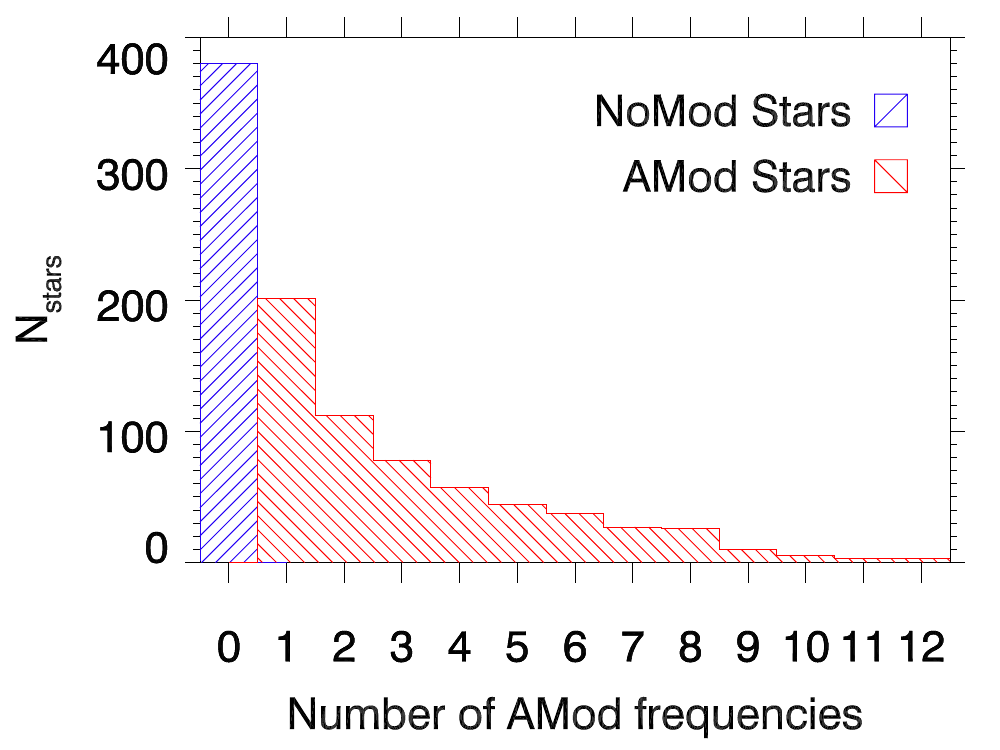}
	\hspace{1cm}
	\includegraphics[width=0.4\textwidth]{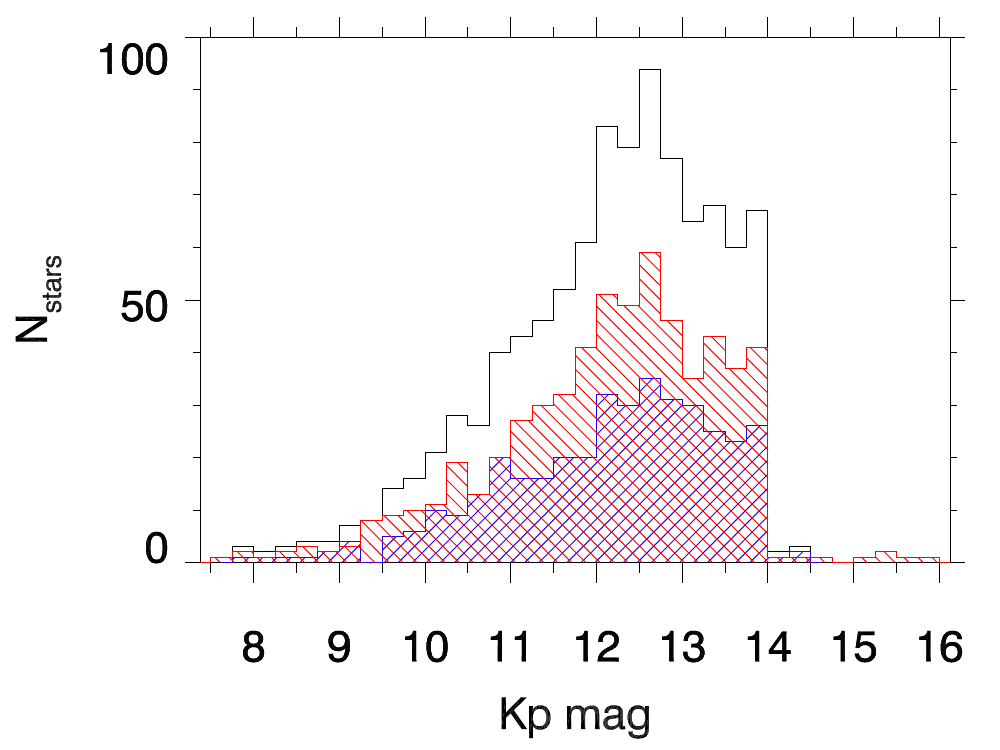}
	
	\vspace{1.2cm}
	
	\includegraphics[width=0.4\textwidth]{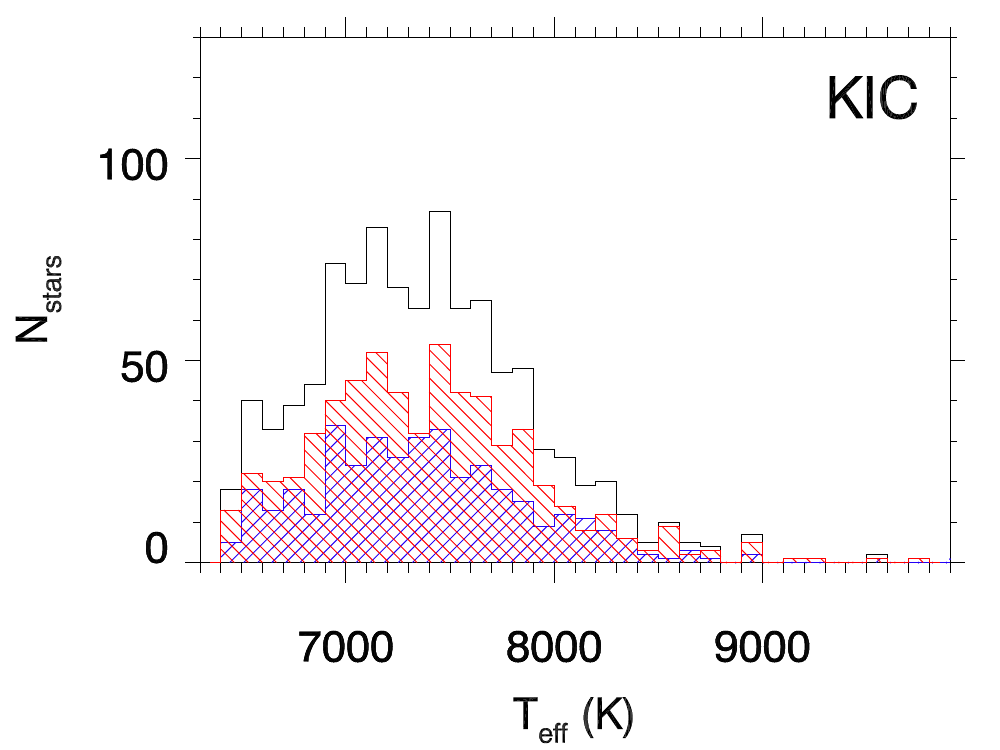}
	\hspace{1cm}
	\includegraphics[width=0.4\textwidth]{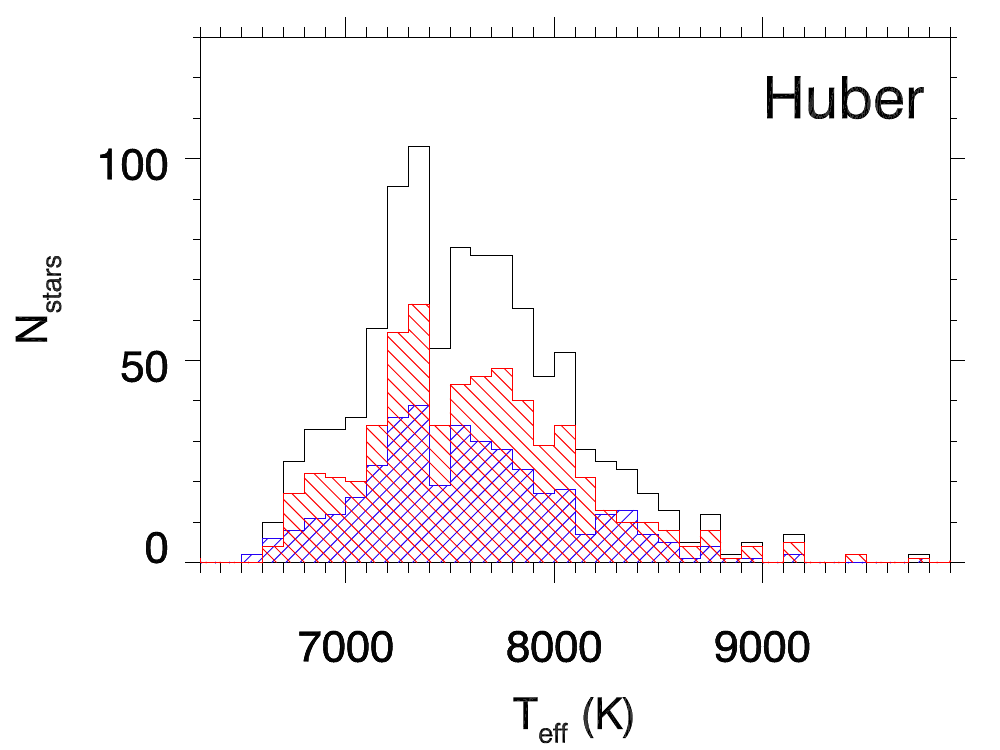}
	
	\includegraphics[width=0.4\textwidth]{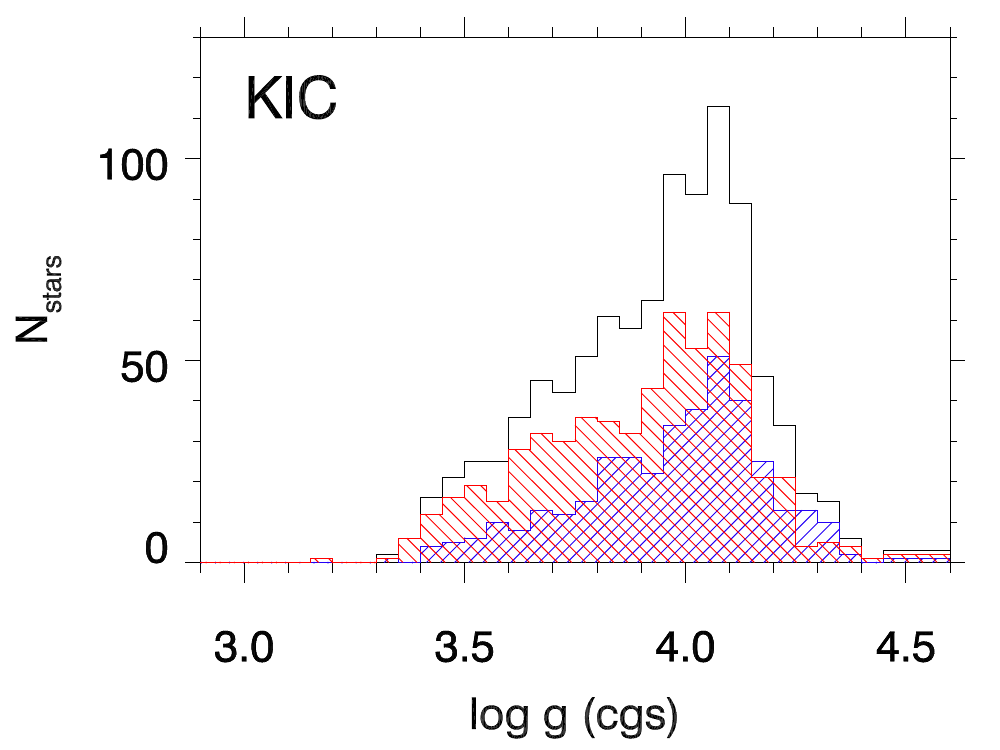}
	\hspace{1cm}
	\includegraphics[width=0.4\textwidth]{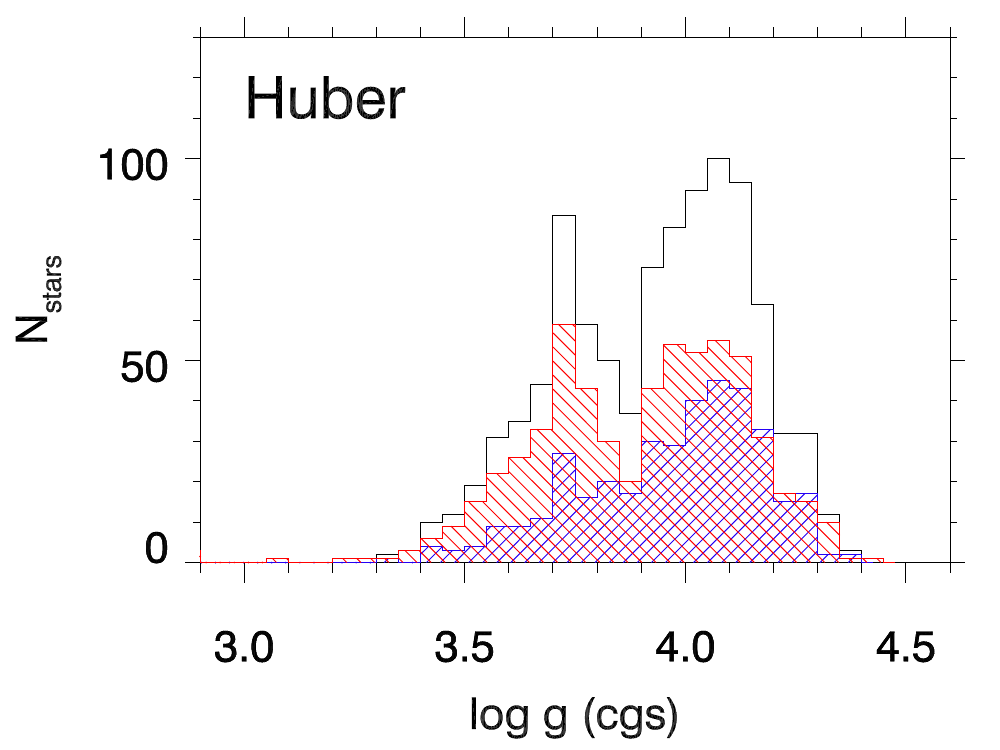}
	
	\includegraphics[width=0.4\textwidth]{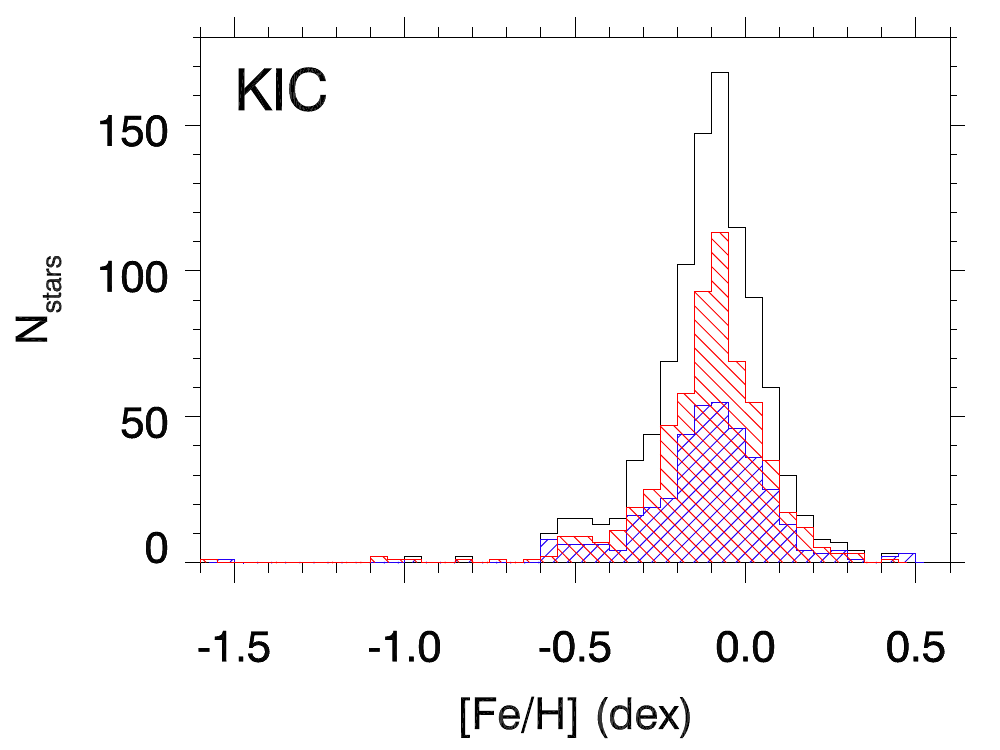}
	\hspace{1cm}
	\includegraphics[width=0.4\textwidth]{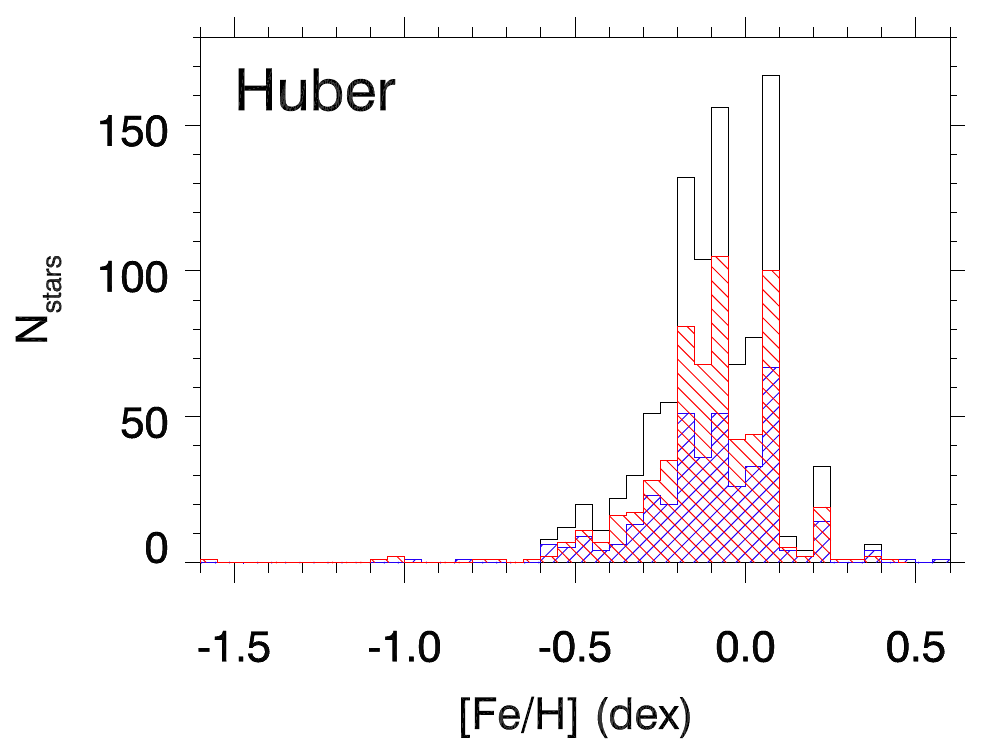}
	
	\caption{The histogram of the distribution of the number of stars against the number of AMod peaks in our ensemble is shown in the top-left panel. Histograms for the number stars against $T_{\rm eff}$, $\log\,g$, $[{\rm Fe}/{\rm H}]$ and Kp mag, in which black represents all {983} stars in our ensemble, the blue hatched region represents the {380} NoMod stars and red hatched region represents the {603} AMod stars. Histograms for $T_{\rm eff}$, $\log\,g$ and $[{\rm Fe}/{\rm H}]$ using the original KIC values are shown in the left panels, and the revised values from \citet{Huber2014} are shown in the right panels.}
	\label{figure: histograms}
\end{figure*}

We found that {380} stars ({38.7}~per~cent) were classed as NoMod and {603} stars ({61.3}~per~cent) had at least a single AMod peak. The histogram for the distribution of the number of stars against the number of AMod frequencies in a star is shown in the top-left panel of Fig.~\ref{figure: histograms}. A significant conclusion from this study is that the majority of \dsct stars have at least a single AMod pulsation mode. More interestingly, {201} stars ({20.4}~per~cent) have only a single AMod frequency, with the other 11 frequencies remaining constant in amplitude. This has previously been demonstrated for KIC~7106205, but the discovery that this behaviour is common among \dsct stars has not been demonstrated before; it is a new result from this work.

We created histograms of various stellar parameters for all stars, which are shown in Fig.~\ref{figure: histograms}, for both the original KIC values \citep{Brown2011} and the revised values given in \citet{Huber2014}. In these histograms, the unhatched region represents all stars and the hatched regions show the NoMod and AMod stars in the various panels. The distributions of $T_{\rm eff}$, $\log\,g$, $[{\rm Fe}/{\rm H}]$ and Kp~magnitude for the NoMod and AMod stars are similar, demonstrating that amplitude modulation is common across the classical instability strip. Therefore, we conclude that the physics that determines which modes have variable amplitudes does not simply depend on the fundamental stellar parameters. Our histograms also show the $200$~K systematic offset in $T_{\rm eff}$ that \citet{Huber2014} found in hot \Kepler stars. This is most clear when comparing the KIC and \citet{Huber2014} $T_{\rm eff}$ histograms in Fig.~\ref{figure: histograms}. 

One might expect the more evolved \dsct stars (i.e. stars with lower $\log\,g$ values) to have variable pulsation mode amplitudes, because they likely contain mixed modes or because the structure of the star is changing in a relatively small period of time. Even whilst on the main-sequence the convective core can increase or decrease in mass depending on the initial mass (see figure~3.6 from \citealt{ASTERO_BOOK}). However, the NoMod and AMod $\log\,g$ distributions in Fig.~\ref{figure: histograms} are centred on approximately the same value. There is, however, a bimodality in the \citet{Huber2014} $\log\,g$ values for the AMod stars compared to the NoMod stars. This supports the above argument that evolved \dsct stars are more likely to be AMod, but since this bimodality is not seen in the KIC $\log\,g$ values, it is likely an artefact of the \citet{Huber2014} method.

\begin{figure}
	\includegraphics[width=0.95\columnwidth]{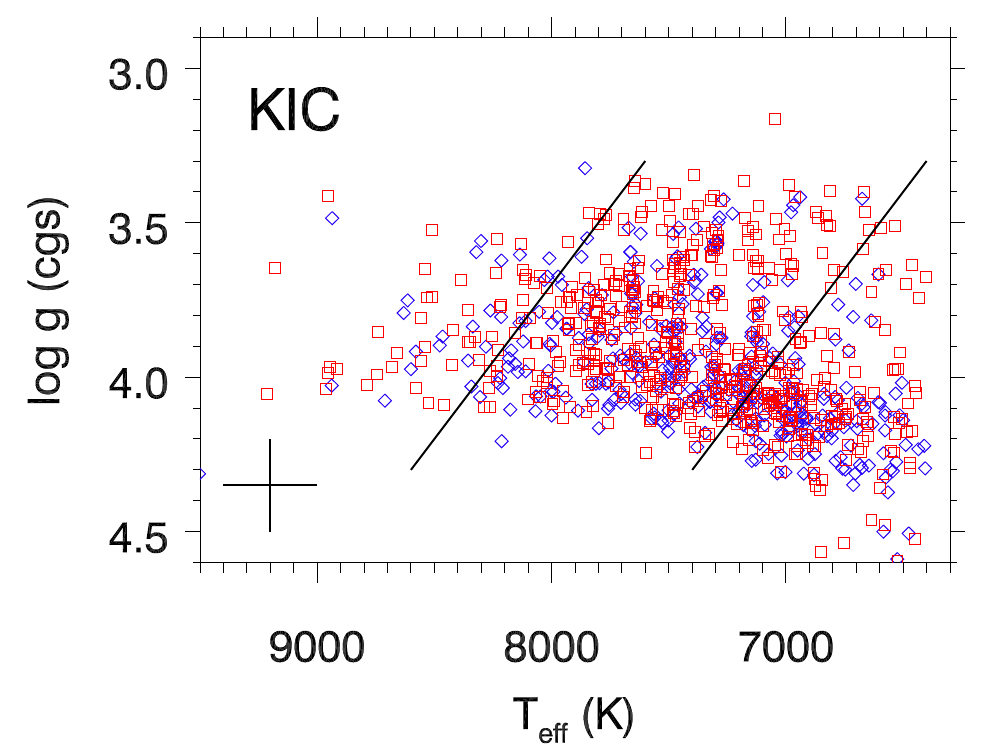}
	\includegraphics[width=0.95\columnwidth]{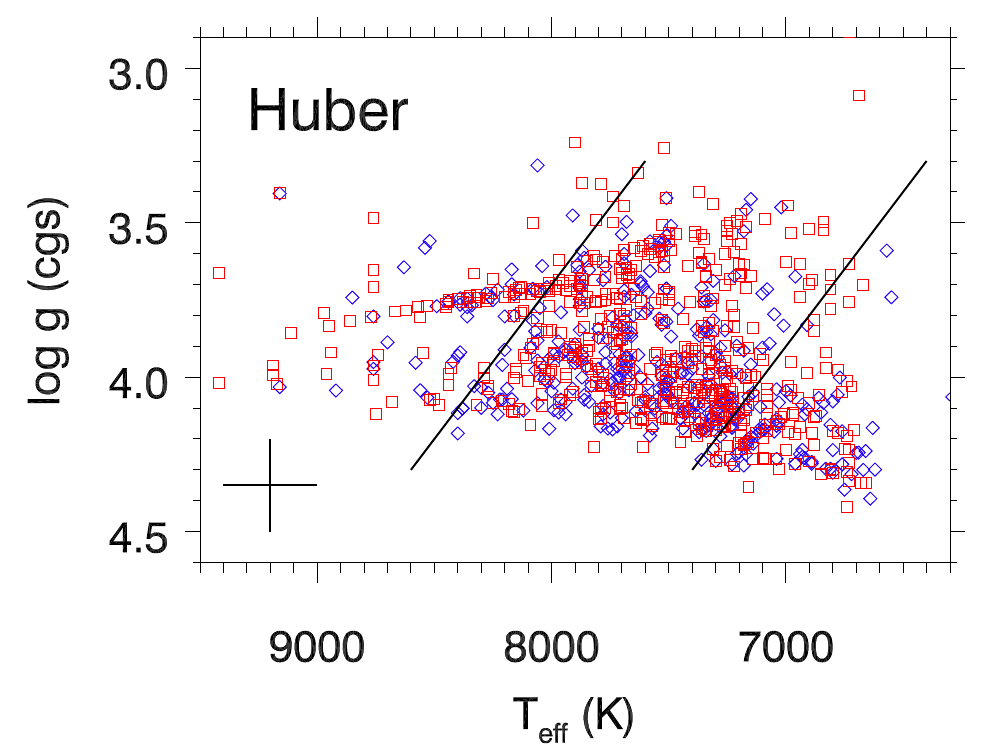}
	\caption{$T_{\rm eff}-\log\,g$ diagrams for the stellar parameters listed in the KIC (top panel) and the revised values given by \citet{Huber2014} (bottom panel). The diamonds represent the {380} NoMod stars and the squares represent the {603} AMod stars. The solid lines are the observational blue and red edges of the classical instability strip from \citet{Rod2001}, and the cross represents the typical uncertainty for each point.}
	\label{figure: HR}
\end{figure}

We also constructed $T_{\rm eff}-\log\,g$ diagrams using the original and revised KIC values, which are shown in Fig.~\ref{figure: HR}, for the NoMod and AMod stars. Observational blue and red edges of the instability strip, taken from \citet{Rod2001}, are also plotted in Fig.~\ref{figure: HR}. There is no obvious correlation between the physical mechanisms that cause amplitude modulation and stellar parameters such as $T_{\rm eff}$ or $\log\,g$ in our ensemble. An inference from this study is that amplitude modulation is not directly dependent on the fundamental stellar parameters of a star, but intrinsically related to the pulsation excitation mechanism itself. Further investigation and theoretical work is needed to address this question. 

\section{Conclusions}
\label{section: conclusions}

In this paper, we have presented the results from a search for amplitude modulation in {983} \dsct stars that were continuously observed by the \Kepler Space Telescope for 4~yr. The \Kepler data set provides extremely high frequency and amplitude precision, which we used to track amplitude and phase at fixed frequency in {100}-d bins with a {50}-d overlap, for a maximum number of 12 peaks with amplitudes greater that $0.10$~mmag in each star. We collated our results into an amplitude modulation catalogue and have presented a selection of case study stars to demonstrate the diversity in pulsational behaviour. A total of {603} \dsct stars ({61.3} per~cent) exhibit at least one pulsation mode that varies significantly in amplitude over 4~yr, and so amplitude modulation is common among \dsct stars.

The 380 NoMod \dsct stars comprise 38.7~per~cent of our ensemble and represent the stars for which the highest precision measurements of amplitude and frequency are possible. This is extremely important in the search for planets orbiting these stars when using the FM method \citep{Shibahashi2012}. The upcoming TESS mission \citep{Ricker2015} will provide short time-scale observations of a large area of the sky, and continuous observations near the polar regions. These NoMod stars are the ideal targets for determining precise frequencies and amplitudes for asteroseismic modelling. They also represent a subset of {\it ideal} \dsct stars that can be compared to observations of \dsct stars in the continuous viewing zones with the TESS mission data.

The causes of variable pulsation amplitudes in \dsct stars, which we termed AMod stars, can be categorised into those that are caused by extrinsic or intrinsic causes. The extrinsic causes of phase variability include binarity (or multiplicity) in the stellar system which acts as a perturbation to the pulsation mode frequencies observed \citep{Shibahashi2012}. A pulsating star can also be easily recognised as being part of a multiple system as its pulsation frequencies will all be phase modulated by the orbital period \citep{Murphy2014}. Similarly, super-Nyquist aliases are easily identifiable because they are periodically phase modulated by a variable Nyquist frequency. This was demonstrated graphically for KIC~5950759 in Fig.~\ref{figure: sNa peaks}. 

The sub-group of stars that exhibit amplitude modulation with no phase change is particularly interesting. In these stars, the amplitude change is non-periodic and is often monotonically variable over many years. Approximately the same number of stars with linearly increasing and decreasing amplitudes are seen, but also non-linearly increasing and decreasing amplitudes in this sub-group. We are possibly observing these stars undergoing slow changes in the relative depths of pulsation cavities driven by stellar evolution. For example, stellar evolution was suggested as the cause of the observed amplitude modulation in the $\rho$~Pup star KIC~3429637 \citep{Murphy2012b}. On the other hand, {pure} amplitude modulation in \dsct stars may be observations of changes in driving and/or damping within a star. These stars remain a challenge to understand and so we can only speculate.

Beating effects from pairs (or groups) of close-frequency pulsation modes are not uncommon in \dsct stars \citep{Breger2002d, Breger2006a}. A resolved beating pattern is most recognisable from the periodic amplitude modulation, but most importantly, from a phase change occurring at the epoch of minimum amplitude. This phase change is $\pi$~rad for two equal-amplitude cosinusoids, and tends to zero as the amplitudes get significantly different from each other. We successfully constructed beating models for two pairs of close-frequency modes separated by less than 0.001~d$^{-1}$ in KIC~4641555 and KIC~8246833, shown in Fig.~\ref{figure: beating AMod stars}, resulting in beating periods of {$1166 \pm 1$}~d and {$1002 \pm 1$}~d, respectively.

Stars that exhibit nonlinearity are evident from the non-sinusoidal shape of the light curve and the presence of harmonics and combination frequencies in the amplitude spectrum. The frequency, amplitude and phase of a coupled or combination frequency are a function of the two parent modes, and so we used a coupling coefficient, $\mu_{\rm c}$, to distinguish between these two forms of nonlinearity within a star. Small values of $\mu_{\rm c}$ imply combination frequencies from a non-linear distortion model that mimic any variability in the parent modes \citep{Brickhill1992a, Wu2001c, Breger2008c}, whereas large values of $\mu_{\rm c}$ imply resonant mode coupling with mode energy being exchanged among similar-amplitude family members \citep{Breger2014}. We have modelled mode coupling in the \dsct star KIC~4733344 and studied the possible energy exchange among pulsation modes. For two families of frequencies in KIC~4733344, we found $\mu_{\rm c} \simeq 0.01$ implying combination frequencies caused by the non-linear distortion model (i.e. nonlinearities in the pulsation waves of the parent mode frequencies), and not strongly-coupled modes. For many \dsct stars, the visible pulsation mode energy is not conserved in 4~yr of \Kepler observations. For example, the \dsct star KIC~7106205 has only a single variable pulsation mode frequency, which is shown in Fig.~\ref{figure: KIC 7106205}. This may be caused by mode coupling to invisible high-degree modes \citep{Dziembowski1985a}. Using \Kepler photometry means we are not sensitive to high-degree modes from geometric cancellation \citep{Dziembowski1977c}, and so it is extremely difficult to determine if the amplitude modulation in a low-degree child p-mode is caused by high-degree parent modes.


	Recent work by \citet{Fuller2015d} and \citet{Stello2016a} has shown that many red giant stars have suppressed dipolar modes ($\ell =1$), which can be explained by the scattering of mode energy into high-degree modes as they interact with a magnetic field in a star's core. This effect, termed the Magnetic Greenhouse Effect \citep{Fuller2015d}, essentially traps the mode energy in the magnetised core of a red giant star resulting in low surface amplitudes for the dipole modes. \citet{Stello2016a} demonstrated that not all red giant stars exhibit suppressed dipole modes and that it is a strong function of stellar mass. Among other pulsating stars, \citet{Cantiello2016a*} modelled a 1.6-M$_{\odot}$ main-sequence \gdor star and suggested that it is possible for a dynamo-generated magnetic field to be induced near the core in such as star. This could alter the pulsational behaviour of a star and redistribute mode energy into higher degrees, hence dramatically reduce their visible amplitudes \citep{Cantiello2016a*}. We speculate that a similar mechanism could be the cause for some of the AMod \dsct stars in our ensemble. The A and F stars have small convective cores and if a magnetic field is sustained throughout the transition from post-main-sequence to the red giant branch, it is reasonable to assume that the progenitors of red giant stars with suppressed dipole modes also had magnetic fields near their cores on the main-sequence. The progenitors of such suppressed dipole mode red giant stars could be within our ensemble of \dsct stars.

	There are various theoretical and observational synergies between pulsating A and B stars, such that \bcep stars can be considered analogues of \dsct stars, from the similar pulsation mode frequencies observed. The $\kappa$-mechanism operating in the metal bump (or `Z bump') in opacity, causes low-order p~modes to become unstable \citep{Dziembowski1993e}. Hybrid B stars pulsating in both g- and p-mode frequencies have also been observed \citep{DeGroote2012b}. Further similarity between \bcep and \dsct stars exists, as \citet{Degroote2009a} found evidence for non-linear resonant mode coupling in the \bcep star HD~180642. It would be interesting to investigate the synergy in mode coupling within hybrid stars, between A and B stars.

	Our catalogue of 983 \dsct stars utilizing 4-yr of \Kepler data demonstrates that observations spanning years (and longer) are often needed to study and resolve pulsational behaviour in these stars. Our catalogue will be useful for comparison purposes when studying observations of \dsct stars from K2 \citep{Howell2014} and TESS \citep{Ricker2015}. Eventually, these missions will observe a large area of the sky, but for only a short length of time. Therefore, \Kepler may represent the best data set for studying \dsct stars as its 4-yr length of continuous observations will not be surpassed for some time.


\section*{acknowledgements}
We thank the anonymous reviewer for their feedback that improved this paper. We also thank professor Hideyuki Saio for useful discussions. DMB is supported by the UK Science and Technology Facilities Council (STFC), and wishes to thank the \Kepler science team for providing such excellent data. SJM is supported by the Australian Research Council. Funding for the Stellar Astrophysics Centre is provided by the Danish National Research Foundation (grant agreement no.: DNRF106) and by the ASTERISK project (ASTERoseismic Investigations with SONG and Kepler) which is funded by the European Research Council (grant agreement no.: 267864). Some of the data presented in this paper were obtained from the Mikulski Archive for Space Telescopes (MAST). STScI is operated by the Association of Universities for Research in Astronomy, Inc., under NASA contract NAS5-26555. Support for MAST for non-HST data is provided by the NASA Office of Space Science via grant NNX09AF08G and by other grants and contracts. 


\bibliography{/Users/Dom/Documents/UCLan/PhD/Bibliography/master_bib.bib}


\label{lastpage}
\end{document}